\documentclass[12pt]{article}

\usepackage{color}
\definecolor{darkblue}{rgb}{0.1,0.1,.7}
\usepackage[colorlinks, linkcolor=darkblue, citecolor=darkblue, urlcolor=darkblue, linktocpage]{hyperref}
\usepackage{ifpdf}
\ifpdf
\usepackage{graphicx}
\usepackage{epsfig}
\usepackage[]{latexsym}
\usepackage{slashed,graphicx,color,amsmath,amssymb}
\usepackage[mathscr]{eucal}
\usepackage{mathrsfs}
\usepackage{geometry}
\usepackage[margin=10pt,font=small,labelfont=bf]{caption}
\usepackage{amscd}
\usepackage{bm}
\usepackage{upgreek}
\usepackage[square, comma, sort&compress,numbers]{natbib}
\usepackage[all,cmtip]{xy}
\numberwithin{equation}{section}
\def\s{\sigma}
\def\xp{{x^+ }}\def\xm{{x^-}}
\def\yp{{y^+}}\def\ym{{y^-}}

\def\CH{ \mathcal{H}}
\def\be{\begin{eqnarray}  }
    \def\ee{\end{eqnarray}}
\def\L{\Lambda}
 \def\CG{{\cal G}}
  \def\p{\partial}
  \def\a{\alpha}
 
 \def\Det{ \text{Det}}
 \def\IK{{\mathrm{K}}}

\def\insde{<\!\!\!\!\cdot \  }
\def\insd{<\!\!\cdot \  }
  \def\IPplus{ {\mathrm{P}_{\! +}}}
   \def\IPminus{ {\mathrm{P}_{\! -}}}
  \def\IIK{{\mathbb{K}}}
  \def\Cbbbun{ [\CC^{\bullet\bullet\bullet}_{123}]}

\newcommand{\tr}{\mathrm{Tr}\,}

\newcommand{\rd}{\mathrm{d}}

\newcommand{\rt}{\mathrm{t}}

\newcommand{\rB}{\mathrm{B}}
\newcommand{\rG}{\mathrm{G}}
\newcommand{\rH}{\mathrm{H}}
\newcommand{\rI}{\mathrm{I}}
\newcommand{\rK}{\mathrm{K}}
\newcommand{\rT}{\mathrm{T}}
\newcommand{\rU}{\mathrm{U}}

\newcommand\CC{\mathcal{C}}
\newcommand\CO{\mathcal{O}}
\newcommand\CN{\mathcal{N}}
\newcommand\ep{\epsilon}
\newcommand\Zh{Zhukovsky\ }
\newcommand\no{\nonumber}
 
 \def\CG{{\cal G}}
  \def\p{\partial}
  \def\a{\alpha}
 
 \def\Det{ \text{Det}}
 \def\IK{{\mathrm{K}}}

   \def\ID{{\rm{D}}}
   \def\Li{{\rm{Li}}}
    \def\eg{{\it e.g.\ }}
    \def\ie{{\it i.e.\ }}
     \def\su{{\mathfrak{su}}}
     \def\sl{{\mathfrak{sl}}}

\newcommand{\caA}{{\mathscr A}}

\newcommand{\caY}{{\mathscr Y}}

\newcommand{\caF}{{\mathscr F}}

\def\nested{ \text{nested}}

\def\g{\gamma }

\def\zz{ { { \bf  z} }}
\def\uu{ { {\bf u} }}
\def\xx{ { {\bf x} }}
\def\yy{ { {\bf y} }}
\def\vv{ { \bf v}}

\def\vv { { \bf v  }}
\def\g{\gamma}
\def\albar{ {\bar\alpha}}
\def\period{ \, .}

\def\hh{{h}}
\def\bb  {{b}}
 \def\Li{ \text{Li}_2}

\def\pp{{p}}
\def\ff{{F}}
\def\xx{ { {\bf x} }}
\def\hf{ {\textstyle{1\over 2}} }

 \def\sutwo{{\mathfrak{su}(2)}}
 \def\sltwo{{\mathfrak{sl}(2)}}

 \def\Tr{\text{Tr}}

\begin{document}
\vspace*{-.4in} \thispagestyle{empty}
\vspace{.1in} {\Large
\begin{center}
{\bf Clustering and the Three-Point Function}
\end{center}}

\begin{center}
\vspace{8mm}

Yunfeng Jiang$^a$, Shota Komatsu$^b$, Ivan Kostov$^c$\footnote{\it
Associate member of the Institute for Nuclear Research and Nuclear
Energy, Bulgarian Academy of Sciences, 72 Tsarigradsko Chauss\'ee,
1784 Sofia, Bulgaria}, Didina Serban$^c$

  \vskip 6mm

\small{\textit{$^a$ Institut f{\"u}r Theoretische Physik,
ETH Z{\"u}rich}},\\
\small{\textit{
Wolfgang Pauli Strasse 27,
CH-8093 Z{\"u}rich, Switzerland}


 {\it $^b$ Perimeter Institute for Theoretical Physics,
 \\
 Waterloo, Ontario, Canada}


 {\textit{ $^c$ Institut de Physique Th\'eorique,
DSM, CEA, URA2306 CNRS\\Saclay, F-91191 Gif-sur-Yvette,
France}}

}

\bigskip

 {\tt jiangyf2008@gmail.com,\  skomatsu@perimeterinstitute.ca, \\
 ivan.kostov $\&$ didina.serban@cea.fr
 }

\end{center}

\begin{abstract}
\normalsize{ We develop analytical methods for computing the structure
constant for three heavy operators, starting from the recently
proposed hexagon approach.  Such a structure constant is a
semiclassical object, with the scale set by the inverse length of the
operators playing the role of the Planck constant.  We reformulate the
hexagon expansion in terms of multiple contour integrals and recast it
as a sum over clusters generated by the residues of the measure of
integration.  We test the method on two examples.  First, we compute
the asymptotic three-point function of heavy fields at any coupling
and show the result in the semiclassical limit matches both the string
theory computation at strong coupling and the tree-level results
obtained before.  Second, in the case of one non-BPS and two BPS
operators at strong coupling we sum up all wrapping corrections
associated with the opposite bridge to the non-trivial operator, or
the ``bottom'' mirror channel.  We also give an alternative
interpretation of the results in terms of a gas of fermions and show
that they can be expressed compactly as an operator-valued
super-determinant.  }
\end{abstract}


\newpage

\setcounter{page}{1}
\begingroup
\hypersetup{linkcolor=black}
\tableofcontents
\endgroup

\section{Introduction}
\label{sec:intro}

 In the strongly-interacting system with a large number of degrees of
 freedom, it is often the case that the system exhibits emergent
 collective behaviour, which is entirely different from that of its
 constituents and provides us with a novel physical picture.  The
 examples of such range from various condensed-matter systems realised
 in the laboratory, to the AdS/CFT correspondence, which claims that
 the strongly-coupled CFTs satisfying certain conditions can be
 described by the gravitational theory in the AdS spacetime.

In this paper, we address one simple but intriguing example of such
phenomena in the context of the AdS/CFT correspondence; namely the
emergence of the classical string worldsheet from the three-point
functions in the planar $\mathcal{N}=4$ super Yang-Mills theory (SYM).
On the one hand, a non-perturbative framework to compute the
three-point functions of $\mathcal{N}=4$ SYM, called the hexagon
vertex, was put forward recently in \cite{BKV}.  It describes the
three-point functions in terms of the dynamics of ``magnons'', which
are the elementary fields constituting the gauge-invariant operator.
On the other hand, the AdS/CFT implies that the very same object in
the strong coupling limit admits a totally different description in
terms of the classical string worldsheet and that the three-point
function is given by its area \cite{Janik:2011bd,Komatsu:3pt1,
Komatsu:3pt2,Kazama:2013qsa}.  However, apart from some partial
results given in \cite{BKV}, it is still not clear whether and how
these two descriptions are consistent with each other.

The main purpose of this paper is to explore the general mechanism
which connects these two results.  We claim that the semiclassical
regime is achieved through a mechanism which we call {\it clustering}.
When a large number of magnons are put together in the hexagon vertex,
they form a sort of bound states, which we call clusters.  As we
demonstrate in several examples, this clustering phenomenon is
essential in order to reproduce the string-theory results from the
hexagon vertex.  It is worth noting that these clusters bear some
resemblance with the bound states in the context of the Thermodynamic
Bethe Ansatz.

In order to explain more in detail what we computed with this method,
let us briefly recall the structure of the hexagon vertex and the
result from the classical string.  The hexagon vertex consists of two
parts: The one is the {\it asymptotic structure constant} which is
given by a sum over partitions of the magnons and describes the
three-point functions of long operators.  The other is the {\it
wrapping corrections}, which is given by the sums and the integrals of
the mirror particles and accounts for the finite size effects.  On the
other hand, the result from the string theory is given in terms of
integrals on the spectral curve, where the integration contours are
either around the branch cuts or around the unit circle.

Let us now describe what we achieved in this paper.  First we study
the asymptotic three-point function of long non-BPS operators in the
rank one sectors and show that the result after clustering reproduces
the integrals around the branch cuts in the string-theory prediction.
Second, in the case of the one non-BPS and two BPS correlators, we sum
up the wrapping corrections associated with the edge opposing to the
non-BPS operator, taking into account the clustering effect.  The
result matches nicely with one of the integrals around the unit circle
in the string-theory computation.

Our analysis is based upon yet another important observation that, in
the regimes of our interest, the expression coming from the hexagon
vertex takes the form of the grand-canonical partition function of
free fermions.  This allows to apply the methods developed in
\cite{GSV:Tailoring3} and in \cite{3pf-prl,SL,Bettelheim:Semi} for the tree-level
correlators.  When the number of magnons is infinite, these fermions
become classical and the result is given by the phase-space integral
of this fermion system, which matches the string-theory prediction.
This Fermi gas description allows to reproduce the results obtained by
clustering in an elegant way, shortcutting the tedious combinatorics.
Furthermore, it reveals that the sum over mirror particles on the
bottom edge can be nicely re-expressed as the operatorial
superdeterminant.  However the derivation based on the Fermi gas is
not, at the present stage, sufficiently rigorous.  Therefore, for the
most parts of the paper, we stay on the safe ground of the clustering
method and only briefly sketch the Fermi gas approach.

The applicability of these two approaches is not limited to the
three-point functions.  For instance, the clustering method has proven
to be useful for various other problems such as the strong coupling
limit of the scattering amplitudes in $\mathcal{N}=4$ SYM
\cite{Basso:2013vsa}, which was otherwise obtained by different
methods \cite{Belitsky2015,Fioravanti:2015dma}.
Clustering-like methods were used to compute the partition functions
in $\mathcal{N}=2$ gauge theories in the Nekrasov-Shatashvilli limit
\cite{Nekrasov:2009aa, JEB-Mayer, Mayer-MY}, and the integrable models
describing non-equilibrium processes
\cite{2011arXiv1111.4408B,2013JPhA...46T5205P,2015CMaPh.339.1167B}.
On the other hand, the Fermi gas approach is used extensively to study
the M-theoretic large $N$ limit of the ABJM and related theories as
well as the super-conformal index in four dimensions
\cite{M-P-ABJM,Drukker:2010ab}.  Our analysis indicates that these
approaches are deeply connected.

The rest of the paper is structured as follows.  In section
\ref{sec:3pthexagon} we review the computation of the three-point
function and the hexagon vertex and summarise our results, as well as
the string-theory prediction at strong coupling \cite{Kazama:2016cfl}.
Then in section \ref{sec:asympA} we study the asymptotic structure
constant for the three-point function of one non-BPS and two BPS
operators.  For this purpose, we first re-express the
sum-over-partitions formula in the hexagon proposal as a multiple
contour integral.  We then explain the basic idea of the clustering
using the tree-level example and show that the method can be applied
at finite coupling.  Next, in section \ref{IIII} we generalise it to
the case of three non-BPS fields and reproduce the string-theory
prediction.  In section \ref{sec:bottommirror}, we turn to the
wrapping corrections and summarise expressions for the basic
quantities at strong coupling.  Using such expressions, we analyse the
clustering of the mirror particles and obtain the expression
consistent with the string theory.  Lastly in section
\ref{subsec:fermitree}, we show that these results can be computed
alternatively using the Fermi gas approach and the Fredholm
determinant.  We in particular show that the summation over the mirror
particles can be expressed as the generalised Fredholm
determinant\footnote{The generalised Fredholm determinant is
introduced originally in the context of topological strings
\cite{CGM}.}, which can be further converted into an operator-valued
superdeterminant.  We conclude in section \ref{sec:conclusion}.
Several appendices are provided in order to explain technical details
and elucidate the relation between the clustering and other methods:
In Appendix \ref{ap:abjm}, we study the ABJM matrix model using the
clustering method, and in Appendix \ref{ap:SoV}, we relate the hexagon
vertex and the separation of variables at tree-level using the
clustering.

 \section{The three-point function and the hexagon proposal}\label{sec:3pthexagon}

 The three-point function of operators in the $\CN=4$ planar SYM
 theory is fixed up to a constant by the conformal invariance,
 \begin{align}
 \langle \CO_1(x_1)\CO_2(x_2)\CO_3(x_3)\rangle
 =\frac{C_{123}(g)}{|x_{12}|^{\Delta_{12}}
 |x_{13}|^{\Delta_{13}}|x_{23}|^{\Delta_{23}}}\,,
 \end{align}
 with $x_i$ vectors in the $3+1$ dimensional Minkowski space,
 $\Delta_i$ the conformal dimension of the operator $\CO_i$ and
 $\Delta_{ij}=\Delta_{i}+\Delta_{j}-\Delta_k$.  The constant $C_{123}$
 is given in terms of the initial data of the three operators, namely
 the charges of the global symmetry group $PSU(2,2|4)$ and the charges
 of the infinite symmetry group associated to integrability.  The
 latter ones, dependent on the coupling constant $g$, can be
 encapsulated, at least in the regime of in the small $g$, by three
 collections of rapidities $\uu_1, \uu_2,\uu_3$, each associated to
 one of the operators $\CO_1(x_1),\ \CO_2(x_2),\ \CO_3(x_3)$.  At
 $g=0$ the three sets of rapidities are determined by Bethe ansatz
 equations for three $PSU(2,2|4)$ spin chains with lengths $L_1$,
 $L_2$ and $L_3$.  At non-zero values of the coupling constant $g$,
 the spin chains acquire long-range interaction and the so-called
 asymptotic Bethe ansatz is not exact anymore.  The long-range
 corrections can be interpreted as coming from virtual particles
 circulating in the so-called mirror channel, where time and space are
 interchanged.  These virtual particles are called mirror particle.
 Their contribution to the spectrum of conformal dimensions
 $\Delta(g)$ can be exactly determined via a set of functional
 equations known under the name of Quantum Spectral Curve, equivalent
 to a system of Thermodynamic Bethe Ansatz equations.  In the large
 volume limit the contribution of the virtual particles is
 exponentially small.

 Through the AdS/CFT correspondence \cite{Maldecena:AdS-CFT}, the
 three-point function is dual to a three-string interaction connecting
 three strings with energies $\Delta_1,\ \Delta_2,\ \Delta_3$.  The
 rapidities can be then associated to the momenta of excitation modes,
 or magnons, propagating on the 1+1 dimensional worldsheet.  For a
 particular subset of the operators, the BPS operators, the conformal
 dimensions do not depend on the coupling constant $g$ and the
 associated rapidities are trivial (\ie infinite).  We are going to
 use a bullet to symbolise a non-BPS operator and an empty circle to
 denote the BPS one with the same global charges.  To remove some
 trivial combinatorial factors we are dividing the three-point
 function by the three-point function of the corresponding BPS
 operators, \eg
 \begin{align}
 \label{unnormsc}
 \CC_{123}^{\bullet \bullet \circ}\equiv\frac{ C_{123}^{\bullet
 \bullet \circ}}{C_{123}^{\circ \circ \circ}} {\sqrt{\CN_1\CN_2}}
 \end{align}
denotes the three-point function of two non-BPS and one BPS operator.
In the above formula, $\sqrt{\CN_i}$ are the normalisation of the
three incoming states, which can be expressed in terms of the Gaudin
determinants.  In this work we are not considering the explicit
expression of the norms, and prefer considering the unnormalised
structure constants $\CC_{123}$ defined in \eqref{unnormsc} instead of
the normalised structure constants $C_{123}$.  The semiclassical limit
of the norms in the absence of mirror correction was taken in
\cite{EGSV,3pf-prl}.

\begin{figure}[h!]
\begin{center}
\includegraphics[scale=0.5]{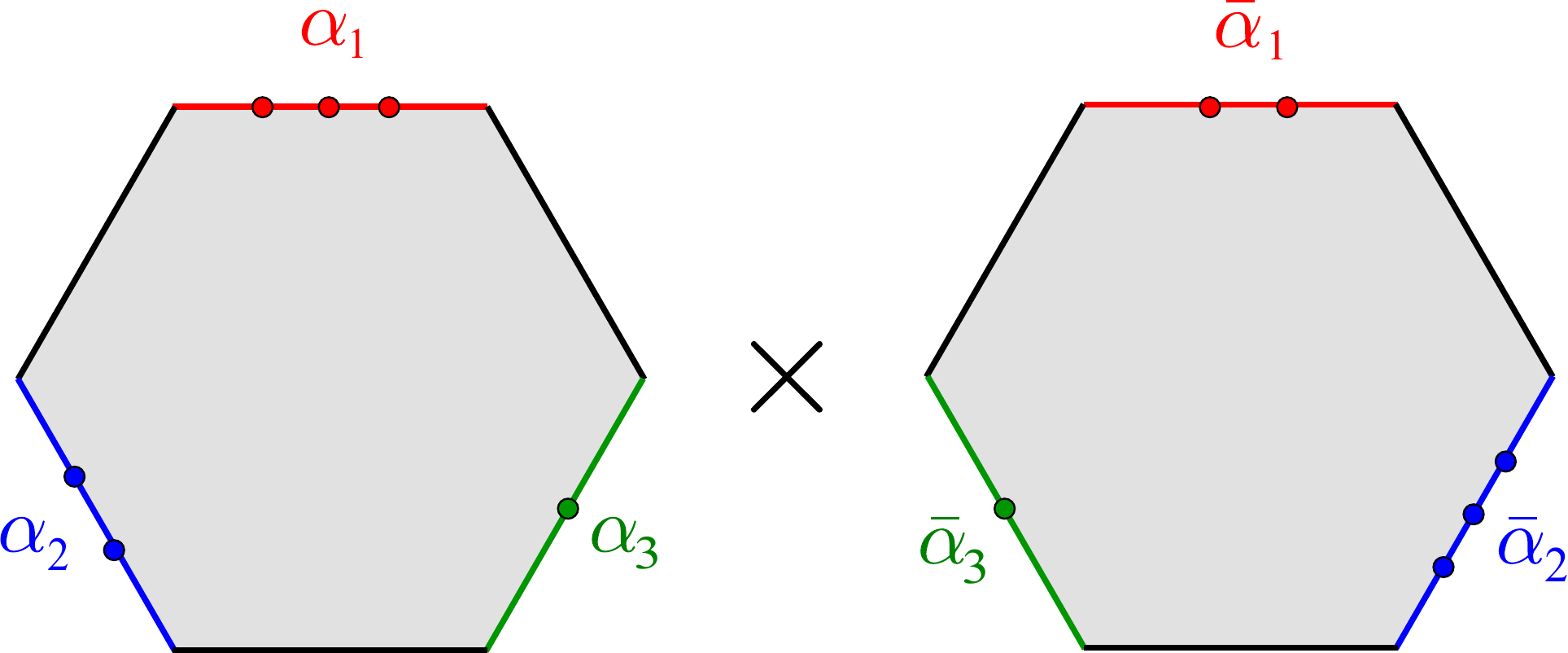}
 \caption{\small\ A possible arrangement of excitations for the
 hexagon form factors.}
\label{fig:CH}
\end{center}
\end{figure}

  An all-loop prescription to compute the three-point function was
  given in \cite{BKV}.  The guiding principle of the proposal is to
  split the worldsheet of the three interacting strings into two
  overlapping hexagons, and then sum over all possible ways of
  distributing the magnon excitations between the two hexagons,
  $\uu_1=\alpha_1\cup \bar \alpha_1, \uu_2=\alpha_2\cup \bar
  \alpha_2,\uu_3=\alpha_3\cup \bar \alpha_3$ as illustrated in figure
  \ref{fig:CH}.  In the absence of the mirror corrections (asymptotic
  limit) the answer is
\begin{align}\label{eq:Chh}
[\mathcal{C}_{123}^{\bullet\bullet\bullet}]^{\rm asympt}
&=\sum_{\substack{\alpha_i \cup\bar{\alpha_i }=\uu_i
 }} \prod_{i=1}^3 (-1)^{| \a_1| +| \a_2|+| \a_3| }\ w_{\ell_{31}}(
 \a_1, \bar\a_1) \, w_{\ell_{12}}(\a_2, \bar\a_2)\, w_{\ell_{23}
 }(\a_3, \bar\a_3) \no\\
&\times \qquad \rH(\a_1| \a_3| \a_2) \rH(\bar \a_2| \bar \a_3| \bar
\a_1)\;.
\end{align}
Explicit expressions for transition factors $w_{\ell_{i-1,i}}(\a_i,
\bar\a_i)$ and hexagon form factors $\rH(\a_1| \a_3| \a_2)$ were
proposed in \cite{BKV} and will be given later.  The building blocks
of the hexagon form factors are the bi-local hexagon amplitudes
$h(u,v)$ proposed in \cite{Basso:2015eqa} and the elements of the
Beisert's scattering matrix \cite{Beisert:2006qh}.  Here we are going
to consider only structure constants of operators from the rank-one
sectors $\su(2)$ and $\sl(2)$ and we are therefore not going to use
the matrix structure of the hexagon form factors.

\begin{figure}[h!]
\begin{center}
\includegraphics[scale=0.35]{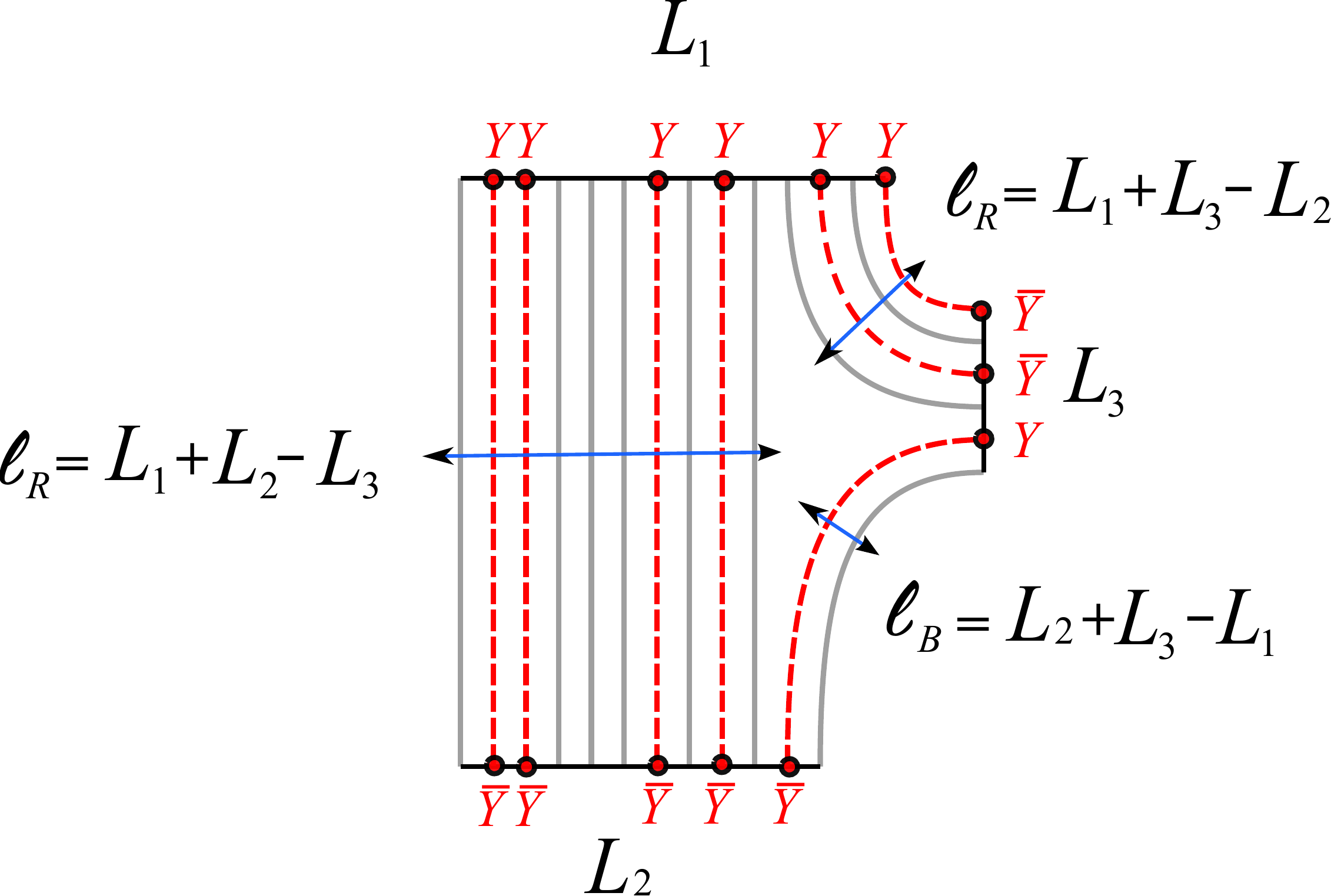}
 \caption{\small\ Vacua and $\su(2)$ excitations in the reservoir
 picture of BKV \cite{BKV}.  }
\label{fig:reservoir}
\end{center}
\end{figure}

To connect with the weak-coupling picture and the corresponding
notations, it is useful to represent the three-point function we
consider in the reservoir picture of \cite{BKV} represented in
\ref{fig:reservoir}.  In this picture, the first operator $\CO_1$ is
of the form $\Tr(Z^{L_1-M_1}Y^{M_1})+ \dots$, the second operator
$\CO_2$ is of the form $\Tr (\bar Z^{L_2-M_2}\bar Y^{M_2})+\dots$, and
the third operator $\CO_3$, the reservoir, is built as $\Tr(Z+\bar
Z+Y-\bar Y)^{L_3-M_3}(\bar Y-\bar Z)^{M_3}+\ldots$.  This type of
structure constant is called type I-I-II in \cite{Kazama:2014sxa},
since two operators belong to the ``left'' $\su(2)$ sector and one
belongs to the ``right'' $\su(2)$ sector in the sense that the
operator $\CO_2$ can be obtained from $\Tr ( Z^{L_2}Y^{M_2})+\dots$
and $\CO_3$ from $\Tr ( Z^{L_3-M_3}\bar Y^{M_3})+\dots$ by one of the
twisted translation defined in \cite{Drukker:2009sf} and used in
\cite{BKV}.  A similar definition works in the $\sl(2)$ sector.

\begin{figure}[h!]
\begin{center}
\includegraphics[scale=0.5]{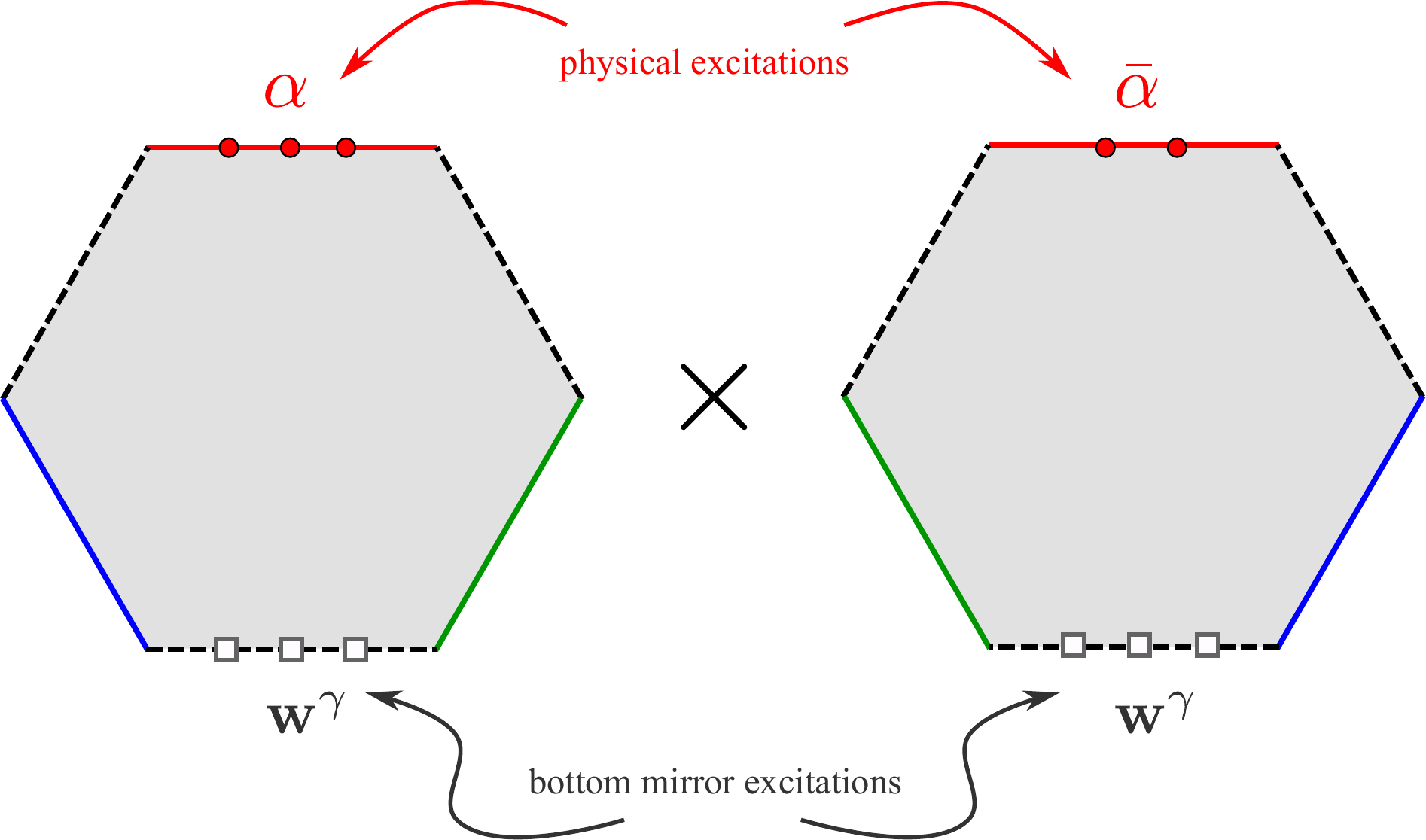}
 \caption{\small\ The physical and bottom mirror excitations.}
\label{fig:bottom}
\end{center}
\end{figure}

The inclusion of wrapping corrections to equation (\ref{eq:Chh}) is
done by including an infinite tower of excitations, as well as their
bound states, circulating in the three mirror channels denoted by
black edges in figure \ref{fig:CH}.  The summation is done over their
rapidities and their polarisations.  The general expression is too
complicated to be reproduced here; instead, we can illustrate the type
of contribution on the case of a single non-BPS operator.  We consider
only the mirror particles in the channel opposed to that operator, as
showed in figure \ref{fig:bottom}.  Following \cite{Basso:2015eqa} we
call this channel the bottom channel.  In this case, the asymptotic
and mirror contributions conveniently factorise,
\begin{align}
\CC^{\bullet \circ \circ}=[\CC^{\bullet \circ \circ}]^{\rm bottom} \;
[\CC^{\bullet \circ \circ}]^{\rm asympt}\;.
\end{align}
Schematically, given in terms of only the fundamental excitations, the
expression of the wrapping corrections is given by \cite{BKV,
Basso:2015eqa}
\begin{align}
\label{eq:integrand}
[\CC^{\bullet \circ \circ}]^{\rm bottom}=\int_{-\infty}^\infty
d\mathbf{w}\, \mu(\mathbf{w}^\gamma)\,
e^{ip(\mathbf{w}^\gamma)\ell_B}\, T(\mathbf{w}^\gamma)\
\hh^{\neq}(\mathbf{w}^\gamma,\mathbf{w}^\gamma) \ \hh(\uu
,\mathbf{w}^{-3\gamma})\,,
\end{align}
with $\ell_B={1\over 2}(L_2+L_3-L_1)$ the length of the bottom bridge
of the correlator, opposed to the operator $\CO_1$ and $T(w)$ the
$\su(2|2)$ spin chain transfer matrix \cite{Beisert:2006qh}.  The full
result takes into account all the bound states and will be given in
the corresponding section.  Here and below the index $\gamma$ stands
for the mirror transformation and we use the shorthand notations
\begin{align}
h(\uu,\vv)\equiv \prod_{i,j}h(u_i,v_j)\;, \qquad
h^{\neq}(\uu,\uu)=\prod_{i\neq j}h(u_i,u_j)\;.
\end{align}

\subsection{Results and comparison with strong coupling}

In the case when the incoming operators correspond to semiclassical
strings, the lengths $L_1,L_2,L_3$ of the three chains and the numbers
of the magnon excitations $M_1,\ M_2,\ M_3$ are large.  The
semiclassical limit is controlled by a small parameter $\ep$ such that
$\ep L_i $ and $\ep M_i $ remain finite when $\ep\to 0$.  This limit
exists for any value of the 't Hooft coupling $g$.  In addition to the
semiclassical limit, one can take the strong coupling limit where the
effective coupling $g' = \ep g$ remains finite when $\ep\to 0$.  Based
on the experience with the spectrum \cite{Gromov:2011ac}, we may
expect that, for $\sl(2)$, the results for the semiclassical strings
can be applied safely to small values of $\ep L_i$.

The summation over the different ways of partitioning the rapidities
in equation (\ref{eq:Chh}), as well as the summation over the mirror
particles remains an open problem in general.  Here we report some
modest progress in taking the sum and the semiclassical limit in three
particular cases when the operators belong to the rank-one sectors
$\su(2)$ and $\sl(2)$:
\begin{itemize}
\item
 the expression of the asymptotic part of the structure constant for
 one non-BPS and two BPS operators, $[\CC^{\bullet \circ \circ}]^{\rm
 asympt}$ for any value of the coupling constant,
\item
 the expression of the asymptotic part of the I-I-II structure
 constant\footnote{ The I-I-I type structure constant remains out of
 reach of our method for the moment.} for three non-BPS operators
 belonging to two different $\su(2)$ or $\sl(2)$ sectors,
 $[\CC^{\bullet \bullet \bullet}]^{\rm asympt}$, for any value of the
 coupling constant,
 \item
the expression of the bottom mirror contribution for one non-BPS and
two BPS operators, $[\CC^{\bullet \circ \circ}]^{\rm bottom}$ in the
strong coupling limit.
\end{itemize}

The first case is a relatively simple generalisation of the result
obtained by \cite{GSV:Tailoring3, 3pf-prl, Bettelheim:Semi} at tree-level.  Here
we use a slightly different method of taking the semiclassical limit,
based on an integral representation of the sums in (\ref{eq:Chh})
which has already appeared in \cite{Bettelheim:Semi}.  This method is
alternative to the Fredholm determinant method used there and it is
easily adaptable to situation when the structure constant cannot be
written exactly as a determinant.  Finally, the structure of the
integrals in the third case ressemble strongly that from the first two
cases, and we are able to take the sum over bound states exactly in
the strong coupling limit.

The answer for the semiclassical structure constants is given in terms
of quasi-momenta associated to the three operators, which encode the
corresponding rapidities.  For operators duals to semiclassical
strings, the rapidities are distributed on a set of cuts, which
connect different sheets of the quasi-momenta.  We are going do denote
by $\tilde \pp^{(k)}$ the sphere part and by $\hat \pp^{(k)}$ the AdS
part of the quasi-momentum associated to the operator $\CO_k$.  The
definition of the quasi-momenta will be given in the main text.  The
results for the $\su(2)$ and $\sl(2)$ sectors are
\begin{align}
\label{C123I-I-IIsu2int}
\!\!\!\!\log [\mathcal{C}_{123}^{\bullet\bullet\bullet}]^{\rm
asympt}_{\su(2)} &=- {1\over \epsilon} \oint_{\CC_{\uu_1 \cup \uu_2}}
{du\over 2\pi} \ \Li \left[ e^{i\tilde \pp^{(1)}_L+i \tilde
\pp^{(2)}_L - i \tilde \pp^{(3)}_R}\right] -{1\over \epsilon}
\oint_{\CC_{\uu_3 } } {du\over 2\pi} \ \Li\left[e^{i\tilde
\pp^{(3)}_R+ i \tilde \pp^{(2)}_L-i \tilde \pp^{(1)}_L} \right], \\
\label{C123I-I-II-sl2int}
\!\!\!\!  \log [\mathcal{C}_{123}^{\bullet\bullet\bullet} ]^{\rm
asympt}_{\sl(2)} &={1\over \epsilon} \oint_{\CC_{\uu_1 \cup \uu_2}}
{du\over 2\pi} \ \Li \left[ e^{i\hat \pp^{(1)}_L+i \hat \pp^{(2)}_L -
i \hat \pp^{(3)}_R}\right] +{1\over \epsilon} \oint_{\CC_{\uu_3 } }
{du\over 2\pi} \ \Li \left[ e^{i\hat \pp^{(3)}_R+ i \hat \pp^{(2)}_L-i
\hat \pp^{(1)}_L} \right].
\end{align}
%
where $\CC_{\uu_k}$ is a contour encircling counterclockwise the
support of the rapidities $\uu_k$.  The result for
$[\mathcal{C}_{123}^{\bullet\circ\circ}]^{\rm asympt}$ is the
particular case where $\uu_2=\uu_3=\emptyset$.  We would like to
emphasise that the expression above are valid when the length of the
three operators $L_1$, $L_2$ and $L_3$ are large and the supports of
$\uu_1$, $\uu_2$ and $\uu_3$ are well separated.  The so-called
heavy-heavy-light diagonal limit, when the length of one of the
operators, say $L_3$, is small and in addition $\uu_1=\uu_2$ was
studied in \cite{2015arXiv151106199J,2016arXiv160106926J}.

A surprisingly similar form is taken by the result of the resummation
of the virtual particles.  Here we succeeded to take the sum only of
the mirror particles for the structure constant with one non-BPS
operator in the channel opposed to the non-trivial operator,
\begin{align}
\log [\mathcal{C}_{123}^{\bullet\circ\circ} ]^{\rm
bottom}_{\su(2)}&=\frac{1}{\ep}\oint_U\frac{du}{2\pi}\left(\Li
\left[e^{i(\hat p^{(2)}+\hat p^{(3)}-\hat p^{(1)})}\right]-\Li \left[
e^{i(\tilde p^{(2)}+\tilde p^{(3)}-\tilde
p^{(1)}(x))}\right]\right)\;,\\
\log [\mathcal{C}_{123}^{\bullet\circ\circ} ]^{\rm
bottom}_{\sl(2)}&=-\frac{1}{\ep}\oint_U\frac{du}{2\pi}\left(\Li
\left[e^{i(\hat p^{(2)}+\hat p^{(3)}-\hat p^{(1)})}\right]-\Li \left[
e^{i(\tilde p^{(2)}+\tilde p^{(3)}-\tilde
p^{(1)}(x))}\right]\right)\;,
\label{eq:mircontintsu}
\end{align}
with the contour of integration $U$ encircling now the \Zh cut with
$u$ between $-2g\ep$ and $2g\ep$.

 The three-point functions at strong coupling admit a completely
 different description, namely in terms of the area of the classical
 string worldsheet.  The computation from the string theory side was
 completed recently building on earlier works \cite{Kazama:2016cfl}.
 In both $\mathfrak{su}(2)$ and $\mathfrak{sl}(2)$ sectors, the result
 is composed of three terms,
 \begin{align}
\log {C}_{123}^{\bullet\bullet\bullet} = \log
[\mathcal{C}_{123}^{\bullet\bullet\bullet} ]^{\rm asympt}+ \log
[\mathcal{C}_{123}^{\bullet\bullet\bullet} ]^{\rm wrapping} + {\rm
Norm}\,.
\end{align}
For the type I-I-II three-point functions in the $\mathfrak{su}(2)$
sector, the asymptotic part and the wrapping part are given on the
string theory side by\footnote{In the convention of this paper.}
\begin{align}
\label{strongsu2as}
\!\!\!\!  \log [\mathcal{C}_{123}^{\bullet\bullet\bullet} ]^{\rm
asympt}_{\su(2)} =& - \frac{1}{\epsilon}\oint_{\mathcal{C}_{{\bf
u}_1\cup {\bf u}_2}}\frac{du}{2\pi}{\rm Li}_2 \left[
e^{i\tilde{p}^{(1)}_L+i\tilde{p}^{(2)}_L-i\tilde{p}^{(3)}_R}\right]
-\frac{1}{\epsilon}\oint_{\mathcal{C}_{{\bf
u}_3}}\frac{du}{2\pi}{\rm Li}_2 \left[
e^{i\tilde{p}^{(3)}_R+i\tilde{p}^{(2)}_L-i\tilde{p}^{(1)}_L}\right]\,,
 \end{align}
 \begin{align}
 \label{strongsu2wr}
\log [\mathcal{C}_{123}^{\bullet\bullet\bullet} ]^{\rm
wrapping}_{\su(2)}&=\frac{1}{\epsilon}\oint_U\frac{du}{2\pi}\left(\Li
\left[e^{i(\hat p^{(1)}+\hat p^{(2)}-\hat p^{(3)})}\right]-\Li
\left[e^{i(\tilde p_L^{(1)}+\tilde p_L^{(2)}-\tilde
p_R^{(3)})}\right]\right)\\ \no
&+\frac{1}{\epsilon}\oint_U\frac{du}{2\pi}\left(\Li \left[e^{i(\hat
p^{(2)}+\hat p^{(3)}-\hat p^{(1)})}\right]-\Li \left[e^{i(\tilde
p_L^{(2)}+\tilde p_R^{(3)}-\tilde p_L^{(1)})}\right]\right)\\ \no
&+\frac{1}{\epsilon}\oint_U\frac{du}{2\pi}\left(\Li \left[e^{i(\hat
p^{(3)}+\hat p^{(1)}-\hat p^{(2)})}\right]-\Li \left[e^{i(\tilde
p_R^{(3)}+\tilde p_L^{(1)}-\tilde p_L^{(2)})}\right]\right)\\ \no
&+\frac{1}{\epsilon}\oint_U\frac{du}{2\pi}\left(\Li \left[e^{i(\hat
p^{(3)}+\hat p^{(1)}+\hat p^{(2)})}\right]-\Li \left[e^{i(\tilde
p_R^{(3)}+\tilde p_L^{(1)}+\tilde p_L^{(2)})}\right]\right)\,.
\end{align}
As is clear from the above expressions,
$[\mathcal{C}_{123}^{\bullet\bullet\bullet} ]^{\rm asympt}$ precisely
matches the result of our analysis \eqref{C123I-I-IIsu2int} and
\eqref{C123I-I-II-sl2int}.  Furthermore, when restricting to the one
non-BPS and two BPS correlators, we can see that the first term in
$[\mathcal{C}_{123}^{\bullet\bullet\bullet} ]^{\rm wrapping}$
coincides with our result of the resummation of the bottom mirror
particles $[\mathcal{C}_{123}^{\bullet\circ\circ} ]^{\rm bottom}$ in
\eqref{eq:mircontintsu} and \eqref{Bottomqccontour}.  Similar match
can be seen also in the $\mathfrak{sl}(2)$ sector, where the result
from the string theory reads
\begin{align}
\label{strongsl2as}
 \log [\mathcal{C}_{123}^{\bullet\bullet\bullet} ]^{\rm
 asympt}_{\sl(2)} =& \frac{1}{\epsilon}\oint_{\mathcal{C}_{{\bf
 u}_1\cup {\bf u}_2}}\frac{du}{2\pi}{\rm Li}_2 \left[
 e^{i\hat{p}^{(1)}_L+i\hat{p}^{(2)}_L-i\hat{p}^{(3)}_R}\right]
 +\frac{1}{\epsilon}\oint_{\mathcal{C}_{{\bf
 u}_3}}\frac{du}{2\pi}{\rm Li}_2 \left[
 e^{i\hat{p}^{(3)}_R+i\hat{p}^{(2)}_L-i\hat{p}^{(1)}_L}\right]\,,
  \end{align}
 \begin{align}
 \label{strongsl2wr}
\log [\mathcal{C}_{123}^{\bullet\bullet\bullet} ]^{\rm
wrapping}_{\sl(2)}=&\frac{1}{\epsilon}\oint_U\frac{du}{2\pi}\left(\Li
\left[e^{i(\hat p_L^{(1)}+\hat p_L^{(2)}-\hat p_R^{(3)})}\right]-\Li
\left[e^{i(\tilde p^{(1)}+\tilde p^{(2)}-\tilde
p^{(3)})}\right]\right)\\ \no
&+\frac{1}{\epsilon}\oint_U\frac{du}{2\pi}\left(\Li \left[e^{i(\hat
p_L^{(2)}+\hat p_R^{(3)}-\hat p_L^{(1)})}\right]-\Li \left[e^{i(\tilde
p^{(2)}+\tilde p^{(3)}-\tilde p^{(1)})}\right]\right)\\ \no
&+\frac{1}{\epsilon}\oint_U\frac{du}{2\pi}\left(\Li \left[e^{i(\hat
p_R^{(3)}+\hat p_L^{(1)}-\hat p_L^{(2)})}\right]-\Li \left[e^{i(\tilde
p^{(3)}+\tilde p^{(1)}-\tilde p^{(2)})}\right]\right)\\\no
&+\frac{1}{\epsilon}\oint_U\frac{du}{2\pi}\left(\Li \left[e^{i(\hat
p_R^{(3)}+\hat p_L^{(1)}+\hat p_L^{(2)})}\right]-\Li \left[e^{i(\tilde
p^{(3)}+\tilde p^{(1)}+\tilde p^{(2)})}\right]\right)\,.
\end{align}
The remaining factors in $[\mathcal{C}_{123}^{\bullet\bullet\bullet}
]^{\rm wrapping}$ supposedly come from other mirror channels.  It
would be an important future problem to reproduce those remaining
terms by resumming the mirror particles in other channels.

\section{Asymptotic structure constant for two  BPS and one non-BPS
 operator
 \label{sec:asympA}
 }

 In this section we are computing the structure constant for the case
 of a single non-BPS operator.  Although this can be considered as a
 particular case of the one treated in the next section, we prefer to
 work out in detail the clustering method on the simpler case, and
 then have a result ready to use for to the more complicated case.
 Since the $\su(2)$ and $\sl(2)$ sectors are largely similar, we treat
 only the former in detail, and just give the results and point out
 the main difference for the latter.

\subsection{From sum-over-partition to multiple contour integral}

 In the definition of the structure constant, the three operators are
 represented by on-shell states of three different spin chains of
 lengths $L_1, L_2, L_3$.  Only the first chain of length $L\equiv
 L_1$ has non-trivial excitations (magnons) with momenta $p_1, \dots,
 p_M$, $M\equiv M_1$.  The momenta are parametrised by the
 corresponding rapidities $\uu= \{u_1, \dots, u_{M}\}$ according to
\begin{align}
e^{ip(u)}=\frac{x(u+i\ep/2)}{x(u-i\ep/2)} \,.
\end{align}
Above, we have rescaled the rapidity variables by $\epsilon$ which
will be set at the typical value for the rapidities $\uu$.  In the
regime dual to semiclassical strings, this overall scale is
$\epsilon\sim 1/L_1$.  The semiclassical limit is $\epsilon\to 0$.
The \Zh variable $x(u)$ is defined as
\begin{align}
x(u)=\frac{{u}+\sqrt{{u}^{2}-{(2g\epsilon)}^{2}}}{2g\epsilon}\;.
\end{align}
The rapidities $\uu$ satisfy the Bethe equations
\be
\label{BAE}
e^{i\phi_j}=1\,, \qquad j=1, \dots, M,
\ee
where $\phi_j$ is the total scattering phase for the $j$-th magnon
\be
\label{Betheq}
e^{\phi_j} = e^{-i p(u_j) L} \prod_{k(\ne j)} {S(u_j, u_k)},
\ee
 $S(u,v)$ being the scattering matrix, which can be represented as the ratio
 \be
 S(u,v) =  {h(v, u)\over h(u,v)}\,.
 \ee
 The function $h(u,v)$, which is given in our case by $h(u,v)_\sutwo
 \equiv h_{YY}(u,v)$, is the building block for the hexagon expansion
 in the configuration described above.  It is given by the product of
 three factors,
\begin{align}
\label{eq:h}
\hh (u,v)_\sutwo=&\,\frac{u-v}{u-v+i\ep}\;\frac{1}{s(u,v) \sigma(u,v)}\, ,
\end{align}
where $s(u,v)$ is the symmetric part,
\begin{align}
\label{htildef}
s(u,v) = {(1- 1/\xp\yp)\, (1- 1/\xm\ym)\over (1- 1/\xp\ym)\, (1-1/\xm\yp)}
\end{align}
and $\sigma(u,v)=1/\sigma(v,u)$ is the square root of the BES dressing
phase \cite{BHL,BES}.  The reason to split $h(u,v)$ as above is that
at tree ($g=0$) level, $s(u,v)=\sigma(u,v)=1$.  It will be important
in the following that neither $s(u,v)$ nor $\sigma(u,v)$ has
singularities close to $u=v$.  We use the notation $x^\pm=x(u\pm
i\ep/2)$ and $y^\pm=x(v\pm i\ep/2)$.
The unnormalised structure constant, is defined as a sum over
partitions of the rapidities $\uu$ into two subsets, $\uu=
\a\cup\bar\a$,
 \begin{align}
\label{eq:altC123}
[\CC^{\bullet\circ\circ}_{123}]^{\rm asympt}\equiv
\caA=\sum_{\alpha\cup\bar{\alpha} =\mathbf{u}}(-1)^{|{\bar
\alpha}|}\;\prod_{j\in{\alpha}}e^{ip(u_j)\ell_R}
\prod_{j\in\alpha,k\in\bar{\alpha}}\frac{1}{h(u_k,u_j)}\;,
\end{align}
where $\ell_R = \hf(L_1+L_3-L_2)$ is the length of the bridge between
the first operator (on the top) and the third one.  In order to have a
complete match with the original tree-level result reported in
\cite{EGSV, GSV:Tailoring3}, we will work with an equivalent representation,
\begin{align}
\label{eq:A}
 \caA=\sum_{\alpha\cup\bar{\alpha}
 =\mathbf{u}}(-1)^{|{\alpha}|}\;\prod_{j\in{\alpha}}e^{-ip(u_j))\ell}
 \prod_{j\in\alpha,k\in\bar{\alpha}}\frac{1}{h(u_j,u_k)}, \quad
 \ell\equiv \ell_L,
\end{align}
where $ \ell_L={1\over 2}(L_2+L_1-L_3)$ is the length of the bridge
connecting the first and the second operator.  The equivalence of the
two expression can be shown by using the Bethe ansatz equations
(\ref{BAE}) with $L_1=\ell_L+\ell_R$.  Formally, at tree-level, the
two expressions (\ref{eq:A}) and (\ref{eq:altC123}) can be obtained
from each other by exchanging $\ell_L$ and $\ell_R$ and sending
$\ep\to-\ep$.

Extending the tree-level observation in \cite{Bettelheim:Semi}, the
sum over partitions (\ref{eq:A}) can be written as a multiple contour
integral
\begin{align}
\label{C123int}
\caA=\sum_{n=0}^N\frac{1}{n!}\, \oint_{\mathcal{C}_{\uu
}}\prod_{j=1}^n\frac{dz_j}{2\pi\ep}\;\ff (z_j)\,\prod_{j<k}^n h(z_j,
z_k) \, h(z_k, z_j) \, ,
\end{align}
where the integration contour $\CC_\uu$ closely encircles the
rapidities $\uu =\{u_1\ldots u_N\}$ counterclockwise, the function
$\ff (x)$ is given by
\begin{align}
\ff (z)=\frac{e^{-ip(z)\ell}\ \mu(z)}{\hh (z,\uu )}\,, \qquad
h(z,\mathbf{u}) \equiv \prod_{j=1}^N h(z,u_j)\,,
\end{align}
and the measure
\begin{align}
\mu(z)
 = \frac{(1-1/x^+x^-)^2}{(1-1/(x^+)^2)(1-1/(x^-)^2)}
\end{align}
 is defined so that $h(z,u) \simeq i \ep \, \mu(z) (z-u)$ at $z=u$.

In the all loop pairwise interaction
\begin{align}
\Delta^{\text{all}}(z_j,z_k)\equiv \hh(z_j,z_k)\ \hh (z_k,z_j)
=\Delta(z_j,z_k)\ s(z_j,z_k)^2\,
\end{align}
the dressing factor drops out due to the anti-symmetry of the dressing
phase.  In the semiclassical limit $\ep\to 0$, the deviation of the
interaction $\Delta^{\text{all}}(u,v)$ with respect with its tree
level value $\Delta(u,v)$ is subleading,
\begin{align}
\Delta^{\text{all}}(u,v)=\,\Delta(u,v)\left(1-c (u,v,g\ep)^2\,
\ep^2+\mathcal{O}({\ep}^3) \right)\,,
\end{align}
where $c(u,v,g\ep)$ is some function of the rapidities $u$ and $v$ and
the effective coupling $g'=g\ep$.  It is important that even at strong
coupling, where $g'$ is finite, the correction to the interaction is
subleading.  A similar property is valid for the measure $\mu(u)$
\begin{align}
\mu(u)= 1-c(u,u,g\ep)\,\ep^2+\mathcal{O}({\ep}^3) \,.
\end{align}
This will allow us to take the semiclassical limit of the asymptotic
contribution for any value of the coupling constant, including strong
coupling.  The main steps of the derivation can be understood on the
tree-level example, which can be treated exactly and will be worked
out in detail in the following.  The clustering procedure explained
below works exactly as in the tree-level, as long as the integration
contours are kept as distance from the cuts of the dressing phase
$\sigma(u,v)$, that is out of the region
$-2g'<{\textrm{Re}}(z_k)<2g'$.  This is certainly the case for
semiclassical strings.
%

\subsection{Tree-level revisited}

 The structure constant of one non-BPS and two BPS operators at three
 level $\caA$ was first studied thoroughly in \cite{EGSV} and
 \cite{GSV:Tailoring3}.  In this section, we revisit the tree-level result by a
 different method which allows direct generalisations to all loops.

The starting point is the multiple integral contour integral
\eqref{C123int}
\begin{align}
\label{eq:multiC}
\caA=\sum_{n=0}^N\frac{1}{\,n!}\;
\oint_{\mathcal{C}_{\mathbf{u}}}\;\prod_{j=1}^n\frac{dz_j}{2\pi \ep
}\;\ff(z_j)\;\prod_{j<k}^n \Delta(z_j,z_k),
\end{align}
where the different ingredients take their tree-level
values\footnote{To avoid proliferation of symbols we keep the same
notations as for all-loop case for most objects.}
\begin{align}
\ff(z)=\frac{e^{-ip(z)\ell}}{h(z,\mathbf{u})},\qquad h(z,u)
=\frac{z-u}{z-u+i\ep}\;.
\end{align}
  The sum (\ref{eq:multiC}) is given by single integrals coupled by
  the pairwise interaction $\Delta(u,v)$ defined as
\begin{align}
\Delta(u,v)=&{h(u,v)\,h(v,u)}=\frac{(u-v)^2}{(u-v)^2+\ep^2}\\\nonumber
=&\,1+ \frac{i\ep/2}{u-v-i\ep}- \frac{i\ep/2}{u-v+i\ep}\,.
\end{align}
In other words, the function $\Delta(u,v)$ differs from $1$ only when
$|u-v|\sim \ep$, and it has two poles at $u=v\pm i\ep$.  For later
convenience, we define a generalisation of this function
\begin{align}
\label{eq:delta}
\Delta_{mn}(u,v)=&\,\frac{(u-v)(u-v+i(m-n)\ep)}{(u-v+im\ep)(u-v-in\ep)}\\\nonumber
=&\,1-\frac{mn}{m+n}\left(\frac{i\ep}{u-v+im\ep}+
\frac{i\ep}{u-v-in\ep}\right)\, ,
\end{align}
so that $\Delta(u,v)=\Delta_{1,1}(u,v)$.  The summation limit in
(\ref{eq:multiC}) can be extended to infinity, since the result of
integration is zero if there are more integrals than rapidities in the
set $\uu$.  The multiple contour integral representation
(\ref{eq:multiC}) is our starting point.  Similar integrals have
appeared recently in the context of integrable probabilities for
example in \cite{2011arXiv1111.4408B, 2013JPhA...46T5205P} and
\cite{2015CMaPh.339.1167B}.

{\bf Semiclassical limit.}
 The rhs of (\ref{eq:multiC}) can be viewed as a grand canonical
 partition function of a matrix model.  This matrix model appeared
 when computing the partition function of dimensionally reduced SYM
 with four supercharges \cite{Moore:1998et}.  The semiclassical limit
 (large $N$ or large chemical potential for the grand canonical
 partition function) was found in \cite{Kazakov:1998ji} using the
 standard matrix model techniques.  The spectral curve of the matrix
 model is associated with an elliptic Riemann surface with two
 parallel cuts at distance $\ep$ from each other.

The semiclassical limit we are interested in is more subtle.  It
consists of taking the limit $L,M\to \infty $ so that $M/L\sim 1$, or
taking $\ep\to 0$ so that $ M\ep$ remains finite.\footnote{This is the
large $L$ limit of a solution of the Bethe equations, characterised by
one or more Bethe strings with mode numbers $n_k$ and filling
fractions $\a_k=M_k/L_1$.  When $L_1\to\infty$, the distribution of
the magnon rapidities along each Bethe string converges to a
continuous linear density.  This limit of the spin chain has been
first studied in \cite{PhysRevLett.74.816} and then rediscovered in
the context of AdS/CFT in \cite{Beisert:2003xu}.} In this limit, which
is very similar to the Nekrasov-Shatashvili limit
\cite{Nekrasov:2009aa}, the standard matrix model techniques do not
work.  The leading and the subleading term of the partition function
were evaluated in \cite{Bettelheim:Semi} by representing the partition
function as a Fredholm determinant and resolving the corresponding
Riemann-Hilbert problem.  A shorter, although less rigorous derivation
used the mapping to a system of chiral fermions.

Here we will give a rigorous derivation of the semiclassical limit
based on an exact evaluation of each term in the sum in
(\ref{eq:multiC}) and then taking the limit.  We will observe a
formation of bound states in close analogy to the bound states of
instantons appearing in the Nekrasov-Shatashvili limit.

\subsection{Deformation of contours and clustering\label{subsec:deformation}}

\begin{figure}[h!]
\begin{center}
\includegraphics[scale=0.35]{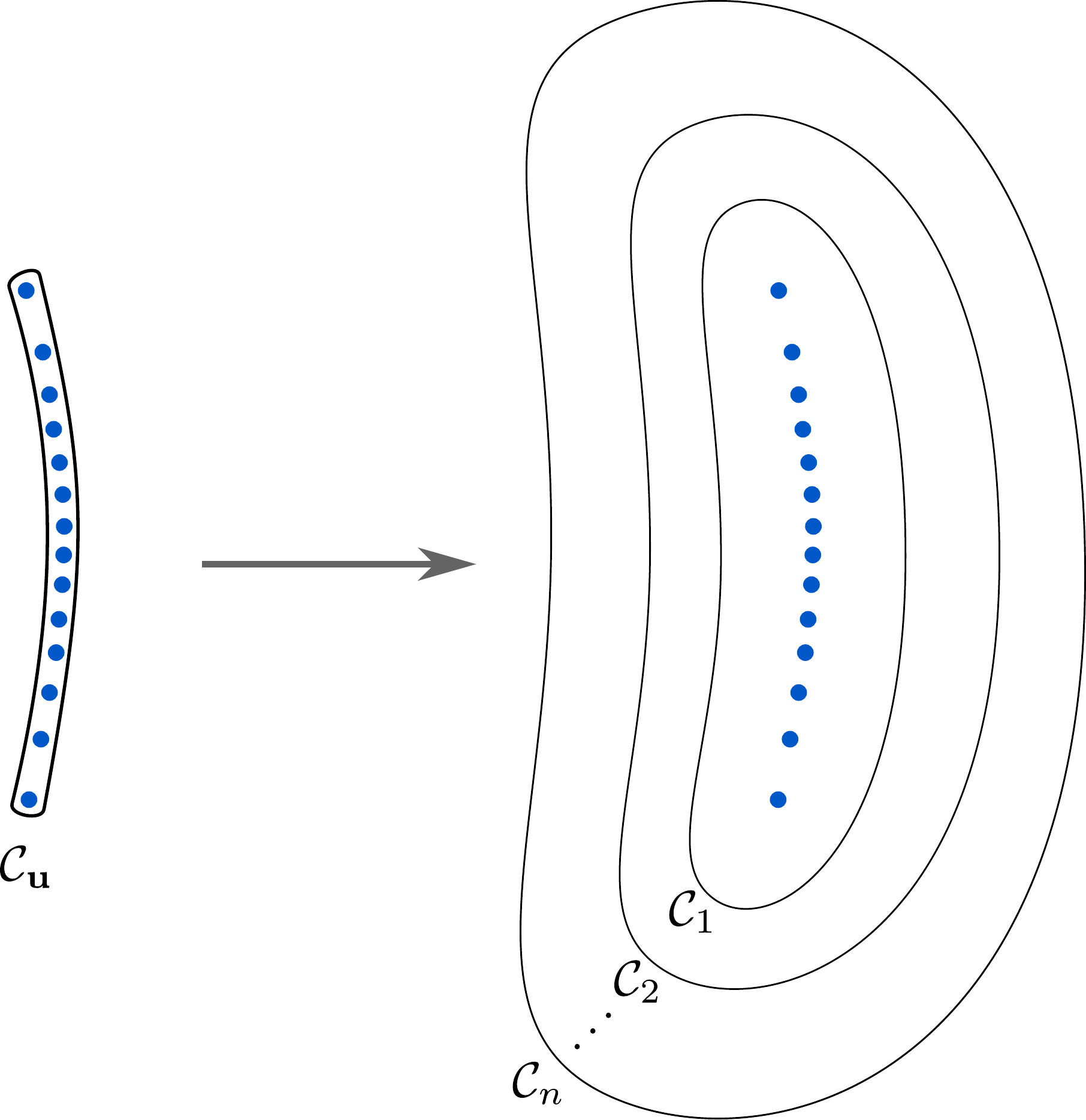}
 \caption{\small\ Deformation of the integration contours.  Here
 $\mathcal{C}_k$ is the deformed contour of the integration variable
 $x_k$, which is situated at a distance larger than $\ep$ from all the
 other contours.}
\label{fig:deformation}
\end{center}
\end{figure}

\begin{figure}[h!]
\begin{center}
\includegraphics[scale=0.4]{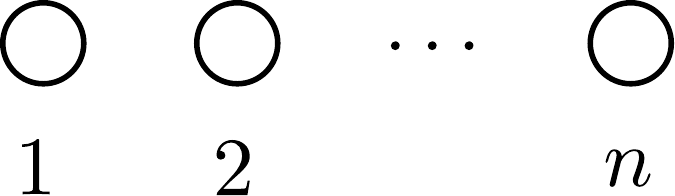}
 \caption{\small\ A diagrammatic representation of the integrand of
 $\mathrm{I}_n$.}
\label{fig:particles}
\end{center}
\end{figure}

  Here we will set up a procedure which allows to perform an expansion
  in the parameter $\ep$ around the semiclassical limit $\ep \to 0,\ L
  \sim M\sim 1/\ep\to\infty$ of the functional $\caA$ in
  (\ref{eq:multiC}).  Namely, we deform the integration contours
  sequentially so that they become widely separated and far way from
  the support of $\mathbf{u}$, as is shown in figure
  \ref{fig:deformation}.
After the contour deformation, we have $|z_j-z_k|\gg \ep$ and the
singularities in the multiple integrals are removed.  In the procedure
of deformation of contours, one has to take into account the residues
of the poles in the interaction terms $\Delta(z_i,z_j)$ in
(\ref{eq:multiC}).  This leads to a phenomenon we call
\emph{clustering} which was considered in various forms in
\cite{Moore:1998et,Nekrasov:2009aa}, in \cite{JEB-Mayer,Mayer-MY} and
in \cite{2011arXiv1111.4408B,2015CMaPh.339.1167B} and which is
reminiscent of the formation of bound states as solutions of the Bethe
equations.  A similar procedure was suggested in \cite{Basso:2013vsa}
in order to take the strong coupling limit of the scattering
amplitudes for gluons.
Let us consider in more detail an integral of the type
\begin{align}
\label{defIn}
\mathrm{I}_n=\oint_{\mathcal{C}_{\mathbf{u}}}\prod_{j=1}^n
\frac{dz_j}{2\pi \ep }\ \ff(z_j)\prod_{j<k}^n \Delta(z_j,z_k),
\end{align}
which corresponds to the $n$-th term in the sum in (\ref{eq:multiC}).
The integrand is a product of functions $\ff(z)$ and
$\Delta(z_i,z_j)$.  We can imagine a collection of $n$ particles, each
particle $z_i$ is associated with a function $\ff(z_i)$ and between
any two particles $z_i$ and $z_j$, there are interactions described by
the function $\Delta(z_i,z_j)$.  Then the integrand can be represented
by the diagram shown in figure\,\ref{fig:particles}.

\vskip 0.8cm

\begin{figure}[h!]
\begin{center}
\includegraphics[scale=0.40]{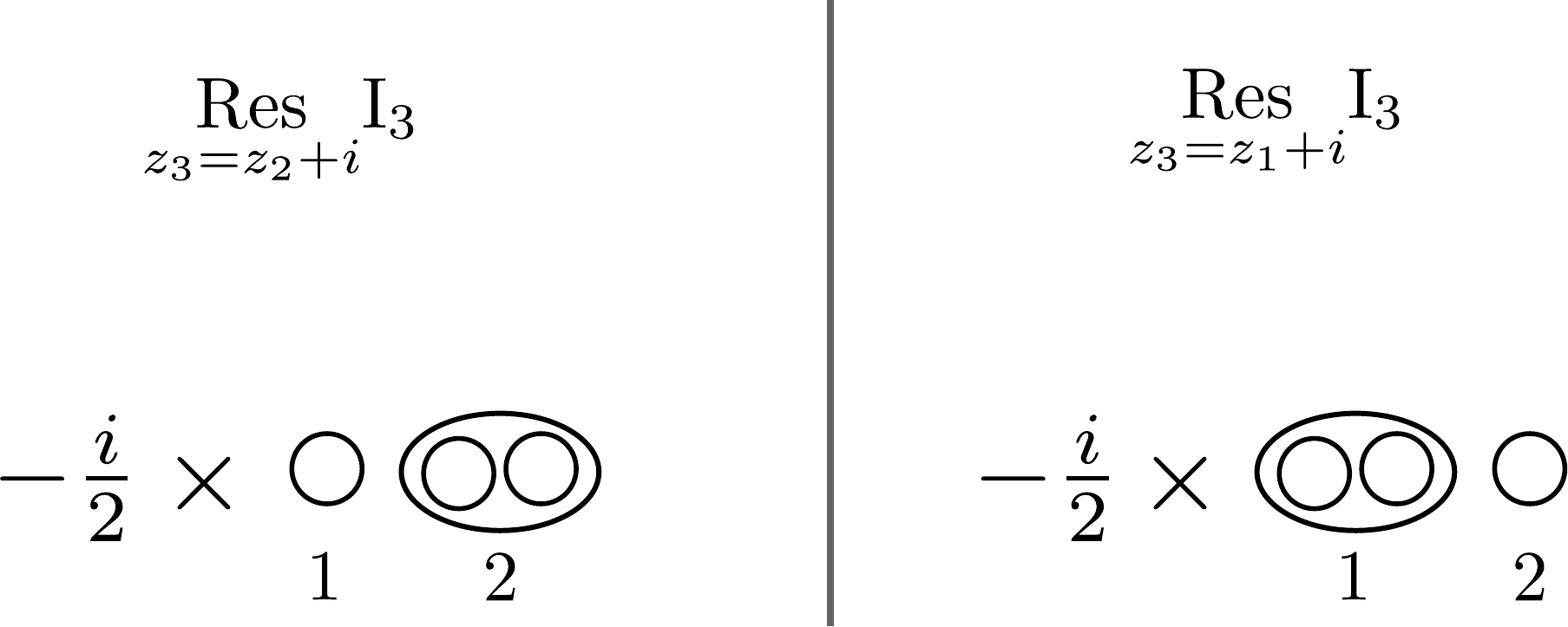}
 \caption{\small The clustering of fundamental particles into bound
 states.}
\label{fig:residue}
\end{center}
\end{figure}

In order to illustrate the idea, we analyse an example for $n=3$
explicitly.  We start with
\begin{align}
\mathrm{I}_3=\oint_{\mathcal{C}_{\mathbf{u}}}\frac{dz_1 dz_2
dz_3}{(2\pi \ep)^3} \ff(z_1)\ff(z_2)\ff(z_3)
\Delta(z_1,z_2)\Delta(z_1,z_3)\Delta(z_2,z_3)
\end{align}
We first deform the contour of integration for $z_3$ from
$\mathcal{C}_{\mathbf{u}}$ to a contour $\mathcal{C}_3$ which is
situated outside $\mathcal{C}_{\mathbf{u}}$ at a distance larger than
$\ep$. There are poles at $z_3=z_2\pm i \ep$ and $z_1\pm i\ep$ due
to the interaction $\Delta(z_1,z_3)$ and $\Delta(z_2,z_3)$,
respectively.  If we take the pole at $z_3=z_2-i\ep$, the residue is
proportional to the following integral
\begin{align}
\oint_{\mathcal{C}_{\mathbf{u}}}\frac{dz_1 dz_2}{(2\pi \ep)^2}
\ff(z_1)\ff(z_2)\ff(z_2-i\ep) \Delta(z_1,z_2)\Delta(z_1,z_2-i\ep).
\end{align}
Because
\begin{align}
\ff(z_2)\ff(z_2-i\ep)=\left(\frac{z_2-3i\ep/2}{z_2+i\ep/2}\right)^\ell
\frac{z_2-\mathbf{u}+i\ep}{z_2-\mathbf{u}-i\ep}
\end{align}
is analytic inside the contour $\mathcal{C}_{\mathbf{u}}$, the
integration over $z_2$ gives zero.  The same argument works for
$z_3=z_1-i\ep$.  This implies that we only need to consider the poles
$z_3=z_2+i\ep$ and $z_3=z_1+i\ep$.  If we take the pole
$z_3=z_2+i\ep$, the result reads
\begin{align}
\label{eq:residue}
\frac{1}{2}\oint_{\mathcal{C}_{\mathbf{u}}}\frac{dz_1 dz_2}{(2\pi
\ep)^2}\ff(z_1)\ff(z_2)\ff(z_2+i\ep)\Delta(z_1,z_2)\Delta(z_1,z_2+i\ep)\,.
\end{align}
We have taken here into account that, while deforming the
counter-clockwise contour $\CC_\uu$ into $\CC_k$, the contours
surrounding the poles will be oriented clockwise.  Let us define the
functions $F_1, F_2, F_3 $, etc.  by\begin{align}
\label{eq:wavef}
F_n(z)=\ff(z)\ff(z+i\ep)\cdots \ff(z+(n-1)i\ep).
\end{align}
Using the fact that
\begin{align}
\Delta(u,v)\Delta(u,v+i\ep)=\Delta_{1,2}(u,v)
\end{align}
where $\Delta_{1,2}(u,v)$ is defined in (\ref{eq:delta}), we can write
the residue (\ref{eq:residue}) as
\begin{align}
\frac{1}{2}\oint_{\mathcal{C}_{\mathbf{u}}}\frac{dz_1 dz_2}{(2\pi
\ep)^2}\ff(z_1)F_2(z_2)\Delta_{1,2}(z_1,z_2).
\end{align}
This result can be interpreted as the following.  Taking the residue
gives rise to a cluster, or bound state, of length 2.  The function
associated to this cluster is given by $F_2(z)$ and its interaction
with a fundamental particle at the point $z'$ is described by
$\Delta_{1,2}(z,z')$.  This is symbolised graphically in figure
\ref{fig:residue}, left.

\begin{figure}[h!]
\begin{center}
\includegraphics[scale=0.35]{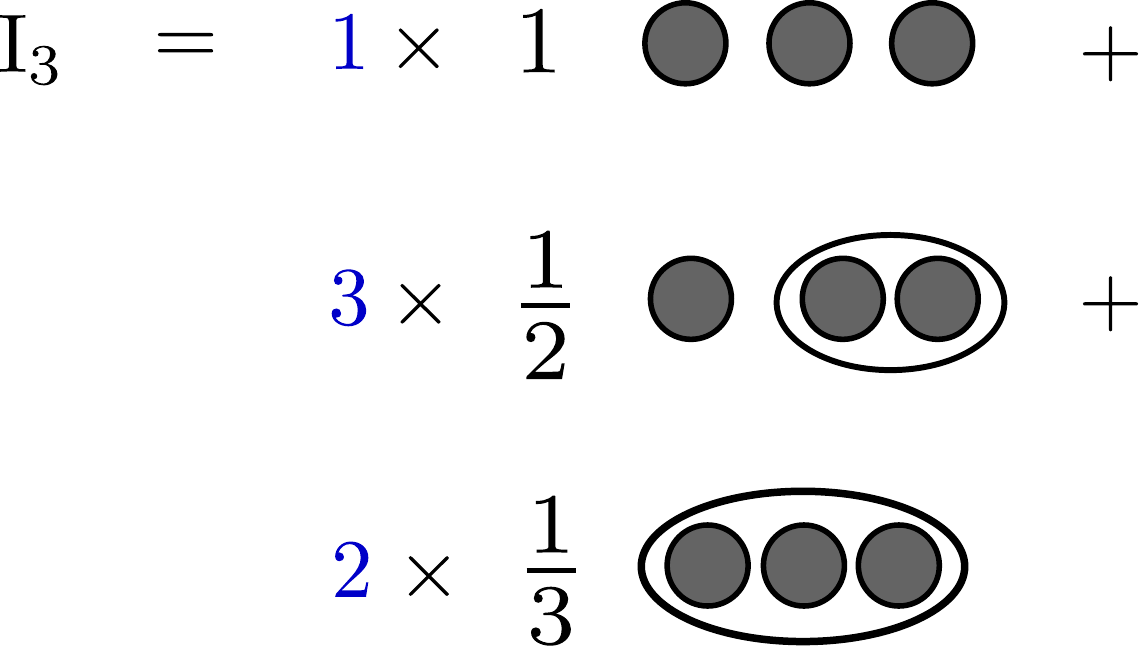}
 \caption{\small\ The final result of $\mathrm{I}_3$ after deforming
 the contours.  Here the black dots mean the integration contour for
 $x_j$ is $\mathcal{C}_j$.  The numbers in blue represent the
 multiplicities of clusters and they are given by equation
 \ref{eq:Cn}, for example $C_3^{1,1,1}=1$, $C_3^{1,2}=3$ and
 $C_3^{3}=2$.}
\label{fig:I3}
\end{center}
\end{figure}

When moving their integration contours from $\CC_\uu$ to $\CC_j$, the
bound states themselves undergo further clustering and form larger
bound states.  A length $n$ bound state is associated to the wave
function $F_n(z)$ defined in (\ref{eq:wavef}) and the interaction
between bound states of length $m$ and $n$ is described by
$\Delta_{mn}(z,w)$.  The full result of our example $n=3$, which
illustrates the origin of the combinatorial factors, is given in
appendix\,\ref{sec:n3}.  In terms of diagrams, it is given in figure
\ref{fig:I3}.

\vskip 1.5cm

As we can see, the result is given by the sum of all possible bound
states, each bound state of length $n$ multiplied by a factor $1/n$.
To see that this is true in general, let us consider the integral with
a bound state of length $m$ and length $n$
\begin{align}
\oint_{\mathcal{C}_\mathbf{u}}\frac{dz_j}{2\pi
\ep}\frac{dz_k}{2\pi\ep}\cdots
\frac{F_m(z_j)}{m}\times\frac{F_n(z_k)}{n} \
\Delta_{mn}(z_j,z_k)\cdots
\end{align}
Suppose we now want to deform the contour $z_k$ to $\mathcal{C}_k$ and
pick the pole $z_k=z_j+im\ep$.  The extra contribution from the pole
is
\begin{align}
\label{eq:FF}
&\oint_{\mathcal{C}_{\mathbf{u}}}\frac{dz_j}{2\pi \ep}\cdots
\frac{F_m(z_j)}{m}\times\frac{F_n(z_j+im)}{n}\times\left(\frac{mn
}{m+n}\right)\cdots\\\nonumber
&=\oint_{\mathcal{C}_{\mathbf{u}}}\frac{dz_j}{2\pi
\ep}\cdots\left(\frac{F_{m+n}(z_j)}{m+n}\right)
\end{align}
where we have used that
\begin{align}
F_m(z)F_n(z+im)=F_{m+n}(z),\qquad
\underset{v=u+im\ep}{\text{Res}}\Delta_{mn}(v,u)=\frac{imn\ep}{m+n}\,.
\end{align}
In what follows we will denote the fusion rules like (\ref{eq:FF})
simply as
\begin{align}
\frac{F_m(z_j)}{m}\times\frac{F_n(z_j+im\ep)}{n}\rightarrow
\frac{F_{m+n}(z_j)}{m+n}.
\end{align}
The fusion rules ensure that the final result is a sum over all
possible bound state configurations.  Each configuration comes with a
combinatorial factor.  We will derive these factors and write down an
exact expression for $\mathrm{I}_n$ in the next section.

\subsection{The exact result and semiclassical   limit}
\label{sec:combas}

In this section, we give an exact expression for $\mathrm{I}_n$ and
$\caA$ and then take its semiclassical limit.  As discussed above,
while deforming the contour we need to pick up poles which lead to the
formation of bound states.  The final result is a sum over all
possible configurations of bound states
\begin{align}
\label{eq:exact}
\mathrm{I}_n=\sum_{k=1}^n\sum_{q_1+\cdots
q_k=n}C_n^{q_1,\cdots,q_k}\;\prod_{j=1}^k\oint_{\mathcal{C}_j}\frac{dz_j}{2\pi
\ep}\frac{F_{q_j}(z_j)}{q_j}\;\prod_{i<j}^n\Delta_{q_i,q_j}(z_i,z_j).
\end{align}
Here $k$ is the number of bound states in a given configuration and
$q_1\le\cdots\le q_k$ are the lengths of the bound states.  They
should satisfy $q_1+\cdots+q_k=n$.  $F_{q_j}(z_j)$ is the wavefunction
for the bound state defined in (\ref{eq:wavef}) and
$\Delta_{q_i,q_j}(z_i,z_j)$ is defined in (\ref{eq:delta}).  The
combinatorial factor $C_n^{q_1,\cdots,q_k}$ counts the number of the
bound state configuration with lengths $\{q_1,\cdots,q_k\}$.  In what
follows, it is convenient to represent the bound state configuration
in a different way.  Suppose among the bound state configurations
$\{q_1,q_2,\cdots,q_k\}$, $d_l$ of them have length $l$
($l=1,2,\cdots$), then we can represent the configuration by a vector
$\vec{d}=\{d_1,d_2,\cdots\}$:
\begin{align}
\label{eq:vec}
\{q_1,\cdots,q_k\}=\{\underbrace{1\cdots
1}_{d_1},\cdots,\underbrace{l,\cdots,l}_{d_l},\cdots\}\mapsto
\vec{d}=\{d_1,d_2,\cdots\}.
\end{align}
We will use the two notations interchangeably.  The following two
obvious identities will be useful
\begin{align}
\sum_{j=1}^k \ff(q_j)=\sum_{l=1}^\infty d_l\,\ff(l),\qquad
\prod_{j=1}^k \ff(q_j)=\prod_{l=1}^\infty \ff(l)^{d_l}.
\end{align}
In particular, the constraint $\sum_{j=1}^k q_j=n$ can be rewritten as
$\sum_l d_l\,l=n$.  We have
\begin{align}
\label{eq:Cn}
C_n^{q_1,\ldots,q_k}&=\frac{1}{d_1!d_2!\cdots}{n\choose
q_1}{n-q_1\choose q_2}\cdots {q_k\choose q_k}(q_1-1)!\cdots (q_k-1)!\\
\nonumber&=\frac{1}{d_1!  d_2!\cdots}\frac{n!}{q_1\cdots
q_k}=\frac{n!}{\prod_l l^{d_l}d_l!}\,.
\end{align}
Let us explain briefly how to obtain the first line of the expression
above.  It is constituted from three different blocks: the middle one
is the way to make $k$ clusters of lengths $n_1\leq \ldots\leq n_k$
out of $n$ variables, while the first block insures that clusters with
the same number of elements are indistinguishable.  The last block
gives the number of different ways to arrange the objects inside each
cluster.  For a cluster with $n_1$ elements, one can choose the label
of the surviving integration variable at will, while the number of
different possible orders of clustering for the other variables is
$(n_1-1)!$.  Inserting (\ref{eq:Cn}) into (\ref{eq:exact}) and summing
over $n$ we obtain the exact result
%
\begin{align}
\label{eq:exactC}
 \caA=&\,\sum_k\sum_{q_1\leq \cdots\leq
 q_k}\frac{1}{d_1!d_2!\cdots}\prod_{j=1}^k\oint_{\mathcal{C}_j}\frac{dz_j}{2\pi
 \ep}
 \frac{F_{q_j}(z_j)}{q_j^2}\prod_{i<j}^k\Delta_{q_i,q_j}(z_i,z_j)\\\nonumber
 =&\,\sum_k\frac{1}{k!}\;\sum_{q_1,\ldots\,,q_k}\prod_{j=1}^k
 \oint_{\mathcal{C}_j}\frac{dz_j}{2\pi \ep}
 \frac{F_{q_j}(z_j)}{q_j^2}\prod_{i<j}^k\Delta_{q_i,q_j}(z_i,z_j)\;.
\end{align}
In the last line the summation over $q_j$ is unrestricted.  This exact
expression can be taken as the starting point for a systematic
semiclassical expansion.  There are two sources of $\ep$ corrections,
from the wavefunction $F_n(z)$ and from the interaction
$\Delta_{mn}(z_i,z_j)$,
\begin{align}
\label{FDeltaexp}
F_n(z)=&\, \ff(z)^n+\epsilon\frac{n(n-1)}{2} \ff(z)^{n-1}\partial_z
\ff(z)+\mathcal{O}(\epsilon^2)\\\nonumber
\Delta_{mn}(z_i,z_j)=&\,1-\frac{mn}{(z_i-z_j)^2}
\epsilon^2+\mathcal{O}(\epsilon^3)\,.
\end{align}
If we are interested in the leading order of $\epsilon$ expansion of
(\ref{eq:exactC}) we can replace $F_n(z)$ by $F^n(z)$ and
$\Delta_{mn}(z_i,z_j)$ by 1, which simplifies (\ref{eq:exactC})
drastically.  The multiple integrals decouple and the result
exponentiates,
\begin{align}
\label{leadingA}
\caA \simeq &\,\sum_k\frac{1}{k!}\;\prod_{j=1}^k
\sum_{q_j}\oint_{\mathcal{C}_j}\frac{dz_j}{2\pi \ep}
\frac{\ff(z_j)^{q_j}}{q_j^2}\;=\exp \oint_{\CC_\uu}\frac{dz}{2\pi \ep}
\sum_q \frac{\ff(z)^{q}}{q^2}\;.
\end{align}
Here the integration contour is far way from the support of
$\mathbf{u}$, but now we can deform it back to encircle closely the
support of the rapidities $\mathbf{u}$.  We recognise in the expression above
  the expansion of the dilogarithm.  Taking into
account the subleading corrections from \eqref{FDeltaexp} we obtain
the first two terms from \cite{Bettelheim:Semi}
\begin{align}
\label{weak-semiclassical-all}
\log\caA=\oint_{\mathcal{C}_{\uu}}
\frac{dz}{2\pi\ep}\text{Li}_2\left[\ff(z)\right]-
\frac{1}{2}\oint_{\mathcal{C}_{\uu}^{\times 2}}\frac{dz dz'}{(2\pi)^2}
\ { \log\left[1-F(z)\right] \log\left[1-F(z')\right]\over (z-z')^2}
+\ldots\,.
\end{align}
To avoid the singularity when $z$ and $z'$ coincide in the double
contour integration above, the two contours can be separated, which is
equivalent to taking the principal value integral.  More terms in the
expansion \eqref{weak-semiclassical-all} can in principle be obtained
by a cluster expansion of \eqref{eq:exactC}.

\subsection{The semiclassical limit for the $\su(2)$ sector}

We now specialise the expression in \eqref{weak-semiclassical-all} to
the particular case of the $\su(2)$ sector
\begin{align}\label{weak-semiclassical}
\log\caA\simeq \oint_{\mathcal{C}_{\mathbf{u}}}\frac{dz}{2\pi\ep}\text{Li}_2
\left[ e^{-ip(z)\ell+iG_{\uu}(z)}\right]\,,
\end{align}
which agrees with the results in \cite{GSV:Tailoring3} and \cite{SL}.  Above, we
denoted with $p(z)$ and $G_{\uu }(z)$ the momentum and resolvent at
tree level, in the semiclassical limit $\epsilon \to 0$,
\begin{align}
p ^{\rm{tree}}(z)=\frac{\ep}{z}\,, \qquad G^{\rm tree} _{\uu }(z)
=\sum_{i=1}^N \frac{\ep }{z-u_i}\;.
\end{align}

The all-loop result has exactly the same structure, but with the
quasi-momentum replaced by its full expression, which contains now the
dressing phase,
\begin{align}
\label{alloopres}
G_{\uu }(z)=\sum_{i=1}^N \left[\frac{\ep}{z-u_i}- i \log
\s(z,u_i)\right] \,.
\end{align}
In the physical regime the dressing phase can be expressed as
\begin{align}
-i\log
\sigma(u,v)=\chi(u^+,v^-)+\chi(u^-,v^+)-\chi(u^+,v^+)-\chi(u^-,v^-)\simeq
\ep^2\partial_u\partial_v\chi(u,v)\,,
\end{align}
with $\chi(u,v)$ given by an integral representation
\cite{Dorey:2007xn}.
Defining the sphere all-loop quasi-momenta $\tilde \pp^{(k)}(z)$ by
\begin{align}
\tilde \pp^{(2,3)}(z)=\ep \frac{x'(z)L_{2,3}}{2x(z)}\,, \quad \tilde
\pp^{(1)}(z)=\ep \frac{x'(z)L_1}{2x(z)}-{{G}}_{\mathbf{u}}(z)\,,
\end{align}
the semiclassical limit of the asymptotic all-order contribution is
given by
\begin{align}
\log\caA=\oint_{\mathcal{C}_{\mathbf{u}}}\frac{dz}{2\pi\ep}\text{Li}_2\left[e^{i(\tilde
\pp^{(3)}(z) -\tilde \pp^{(2)}(z)-\tilde \pp^{(1)}(z))}\right]=
-\oint_{\mathcal{C}_{\mathbf{u}}}\frac{dz}{2\pi\ep}\text{Li}_2\left[e^{i(\tilde
\pp^{(3)}(z) -\tilde \pp^{(2)}(z)+\tilde \pp^{(1)}(z))}\right]
\end{align}
The last expression is the semiclassical limit of (\ref{eq:A}).  We
used that
\begin{align}
\label{BAEcl}
(e^{-i\tilde p^{(1)}(z)}) _{\text{on the first sheet}} = (e^{i\tilde
p^{(1)}(z)}) _{\text{on the second sheet}}
\end{align}
 which is a consequence of the classical limit of the Bethe equations
 (\ref{Betheq}),
\begin{align}
\tilde p(u+i0)+\tilde p(u-i0)= L p(u) -(G_\uu(u+i0)+G_\uu(u-i0)) = 0\
\text{mod}(2\pi),
\end{align}
  and that the contour of integration changes its orientation when
  deformed to the second sheet.

In the strong coupling limit the dressing phase simplifies,
$\chi(u,v)\simeq \frac{u-v}{\ep}\log\left(1-\frac{1}{xy}\right)$.
Since $\frac{1}{u-v}=-\partial_u\partial_v (u-v)\log(u-v)$ the
resolvent becomes
\begin{align}
 G_{\mathbf{u}}(z)=
 \ep\sum_{i=1}^N \frac{x'(u_i)}{x(z)-x(u_i)}- p(x)\, \ep\sum_{i=1}^N
 \frac{x'(u_i)}{x^2(u_i)} \equiv {\cal{G}}_\uu(x)- {\Delta- L\over
 2}\, p(x)\,,
\end{align}
with $\Delta-L$ the anomalous dimension, or the spin-chain energy.
The quasi-momenta $\tilde p^{(k)}(z)$ assume in this limit the simpler
form
\begin{align}
\label{ptilde123}
\tilde \pp^{(2,3)}(z)=\ep \frac{x'(z)L_{2,3}}{2x(z)}\,, \quad \tilde
\pp^{(1)}(z)=\ep
\frac{x'(z)\Delta}{2x(z)}-{\cal{G}}_{\mathbf{u}}(x(z))\,.
\end{align}

\subsection{The semiclassical limit for the $\sl(2)$ sector}

The expression for the three-point function with a single non-BPS
operator in the $\sl(2)$ sector is the same as \eqref{eq:A} with
$h(u,v)=h_{\su(2)}(u,v)$ replaced with the corresponding $\sl(2)$
quantity
  \begin{align}
  \label{hsl2}
h_{\sl(2)}(u,v)=\frac{x^+-y^-}{x^--y^+}h_{\su(2)}(u,v)
=\frac{u-v}{u-v-i\ep}\,\frac{1-1/x^-y^+}{1-1/x^-y^-}\,
\frac{1-1/x^-y^+}{1-1/x^+y^+}\,\frac{1}{\sigma(u,v)}
\,.
\end{align}
At tree-level, $\caA_{\sl(2)}$ can be obtained from $\caA_{\su(2)}$
just by sending $\ep \to -\ep$ and $\ell \to -\ell$.  At higher loop,
the change comes from changing the expression of $h(u,v)$ as in
\eqref{hsl2}, which affects the expression of the quasi-momenta in the
semiclassical limit,
 \begin{align}
 f_{\sl(2)}(u)\to e^{i(\hat \pp^{(3)}(z) -\hat \pp^{(2)}(z)-\hat
 \pp^{(1)}(z))}\,.
\end{align}
The quasi-momenta appearing in the asymptotic part of the $\sl(2)$
structure constant correspond now to the AdS part of the spectral
curve \cite{2006CMaPh.263..659B,2012LMaPh..99..169S},
 \begin{align}
 \label{phat123}
 \hat \pp^{(2,3)}(z) =\ep \frac{x'(z)L_{2,3}}{2x(z)}\,, \quad \hat
 \pp^{(1)}(z)=\ep \frac{x'(z)L_1}{2x(z)}+{{\CG}}_{\mathbf{u}}(x(z))\,.
\end{align}
The slightly different appearance of \eqref{phat123} with respect to
\eqref{ptilde123} is due to the extra factor in the second member of
\eqref{hsl2}.

We can therefore write the semiclassical limit of the asymptotic
all-order contribution in the $\sl(2)$ sector as
\begin{align}
\log\caA_{\sl(2)}=-\oint_{\mathcal{C}_{\mathbf{u}}}\frac{dz}{2\pi\ep}\text{Li}_2\left[e^{i(\hat
\pp^{(3)}(z) -\hat \pp^{(2)}(z)-\hat \pp^{(1)}(z))}\right]=
\oint_{\mathcal{C}_{\mathbf{u}}}\frac{dz}{2\pi\ep}\text{Li}_2\left[e^{i(\hat
\pp^{(3)}(z) -\hat \pp^{(2)}(z)+\hat \pp^{(1)}(z))}\right]\,.
\end{align}
Upon permutation of indices $2$ and $3$, which is possible due to
symmetry, this expression coincides with the strong coupling result
\eqref{strongsl2as}.

\section{Asymptotic   structure constant for three non-BPS fields }
\label{IIII}

 Here we consider the all-loop prediction for a configuration
 equivalent to that studied in \cite{EGSV} where two of the operators
 belong to the left sector and the third operator belongs to the right
 sector of $\mathfrak{so}(4) =\mathfrak{su}(2)_L\oplus
 \mathfrak{su}(2)_R$.  The excitations for the three operators are
 chosen to be the longitudinal scalars
\begin{align}
&\mathcal{O}_1\in\sutwo_L :\, \text{vacuum} \ Z^{L_1}, \ \ M_1 \
\text{excitations}\ Y=\Phi_{1\dot{2}}\, , \no\\
& \mathcal{O}_2\in\sutwo_L :\, \text{vacuum} \ Z^{L_2}, \ \ M_2 \
\text{excitations}\ Y = \Phi_{1\dot{2}}\, , \no\\
& \mathcal{O}_3 \in\sutwo_R:\, \text{vacuum} \ Z^{L_3 } ,\ \ M_3 \
\text{excitations}\ \bar Y=\Phi_{2\dot{1}} \, .  \no
\end{align}
 After the twisted rotation the three operators are mapped to
 operators type $\{ Z,Y\}$ at the origin, $\{ \bar Z, \bar Y\}$ at
 infinity, and $\{ \tilde Z, \tilde Y\}$ at some finite point, say
 $\vec x = (0, 0, 1, 0)$, where $\tilde Z=\hf \left(Z + Y + \bar Z -
 \bar Y\right), \ \ \tilde Y = \textstyle{{1\over \sqrt{2}}}(\bar Y
 -\bar Z ).$ This corresponds, in the conventions of \cite{BKV}, to
 excitations $\chi_{\text{top} }=\chi_{\text{bottom} } =
 \chi_{\text{reservoir} }= Y$.

 To compute such three-point functions using the hexagon, we first
 collect all the scalar excitations to one of the edges by performing
 the mirror transformation $\gamma$ several times \cite{BKV}.  (See
 figure \ref{fig:CH} for the configuration of the excitations before
 performing the mirror transformations.)  After collecting them on the
 second edge $(\mathcal{O}_2)$ on the left hexagon and on the first
 edge $(\mathcal{O}_1)$ on the right hexagon, we obtain the hexagons
 with $\{\alpha_1^{4\gamma}, \alpha_3^{2\gamma},\alpha_2 \}$ and with
 $\{\bar{\alpha}_2^{4\gamma},\bar{\alpha}_3^{2\gamma},\bar{\alpha}_1\}$.
 There are of course several other ways to collect the excitations to
 one of the edges.  However, the advantage of the choice described
 here is that all the excitations become $Y$ after the transformation
 owing to the transformation property of the excitations clarified in
 \cite{BKV}:
    \begin{align}
    Y \overset{2\gamma}{\to} -\bar{Y}\,,\quad \bar{Y}
    \overset{2\gamma}{\to} -Y\,.
    \end{align}
Then, since all the excitations are of the $Y$ type, the hexagon form
factor factorises into two-particle form factors\footnote{Here we
included the matrix part $A(u,v)$ is included in the definition of
$h(u,v)$ for the $\mathfrak{su}(2)$ sector, as we did in section
\ref{sec:asympA}.} $h(u,v)$.

 We also study an analogous configuration in the $\mathfrak{sl}(2)$
 sector, where $\mathcal{O}_1$ and $\mathcal{O}_2$ contain $D$
 excitations and $\mathcal{O}_3$ contains $\bar{D}$ excitations.  The
 hexagon form factor for this configuration can be computed in a
 similar way, namely by collecting all the excitations to the one of
 the edges by using the mirror transformations.

\subsection{Formulation in terms of multiple   contour integrals}
\label{sec:sumint}

    The asymptotic part of the un-normalised structure constant with
    three non-BPS operators, which we denote by $\Cbbbun$, is given by
    a sum over the partitions of all the three sets of Bethe roots
    into left and right subsets, $\uu_i= \alpha_i \cup\bar{\alpha_i
    }$:
\begin{align}\label{eq:Chh1}
 \Cbbbun&=\sum_{\substack{\alpha_i \cup\bar{\alpha_i }=\uu^{(i)}
 }} \prod_{i=1}^3 (-1)^{| \a_1| +| \a_2|+| \a_3| }\ w_{\ell_{31}}(
 \a_1, \bar\a_1) \, w_{\ell_{12}}(\a_2, \bar\a_2)\, w_{\ell_{23}
 }(\a_3, \bar\a_3) \no\\
&\times \qquad \rH(\a_1| \a_3| \a_2) \rH(\bar \a_2| \bar \a_3| \bar
\a_1)
\end{align}
with the splitting factors given by 
 \be w_{\ell}(\a, \bar\a)= e^{- i p_{\a} \ell} \
 \frac{\hh^<(\bar{\alpha},\alpha) }{\hh^>(\alpha,\bar{\alpha})}\,,
 \qquad \hh^{\buildrel >\over <}(\uu,\vv)\equiv \prod_{j{\buildrel
 >\over <}k} \hh(u_j,v_k)\,.  \ee
 The hexagon form factor can be computed by performing crossing
 transformation on all the excitations to bring them on the same edge
\begin{align}
\label{eq:phase12}
\rH(\a_1|\a_3|\a_2)=&\,\texttt{phase}_1\,\rH(\a_1^{4\gamma};\a_3^{2\gamma};{\a_2})
\\\nonumber
\rH(\bar{\a}_2|\bar{\a}_3|\bar{\a}_1)=&\,\texttt{phase}_2\,
\rH(\bar{\a}_2^{4\gamma};\bar{\a}_3^{2\gamma};\bar{\a}_1)
.
\end{align}
A subtle point is the definition of the crossing-transformed factors.
For fields from the $\sl(2)$ sector it is sufficient to change the
argument $x^\pm\to 1/x^\pm$.  In the general case the crossing
transformation is more complicated.  It is computed by going to string
frame, perform the analytic continuation and transforming back to the
spin frame, {\it cf.} appendix F of \cite{BKV}.  In general the
hexagon form factor contains a matrix and a scalar part, {\it cf.}
equation (2) of \cite{BKV}.

As we mentioned above, in the $\sltwo$ case the matrix part of the
hexagon form factors is trivial and the weights in the sum over
partitions are products of scalar factors:
\be\begin{split}
\label{eq:sop112}
 \Cbbbun^{\rm asympt}=&\sum_{\substack{\alpha_i
 \cup\bar{\alpha}_i=\uu_i
 }} (-1)^{| \a_1| +| \a_3| +| \a_3| }\ e^{- i p(\a_1) \ell_{3 1}}\
 e^{- i p(\a_2)\ell_{1 2}}\ e^{- i p(\a_3) \ell_{2 3}} \\
 & {\hh(\a_1^{4\gamma},\a_2) \ \hh(\a_1^{4\gamma},\a_3^{2\gamma}) \
 \hh(\a_3^{2\gamma},\a_2 ) \ \hh(\bar\a_2 ^{4\gamma},\albar_1) \
 \hh(\bar\a_2 ^{4\gamma},\bar\a_3 ^{2\gamma}) \ \hh(\bar\a_3
 ^{2\gamma},\albar_1) \over \hh(\a_1,\albar_1)\hh(\a_2 ,\bar\a_2 )
 \hh(\a_3,\bar\a_3 )}\times {\tt phase} \, .
\end{split}
\ee
 For fields from the $\sl(2)$ sector the crossing transformation is
 done analytically continuing $x^\pm\to 1/x^\pm$ and {\tt phase}$=1$.
 For $\sutwo$ fields the phase factors are derived in in Appendix
 \ref{app:phases}.  The explicit forms of the hexagon amplitudes in
 the two sectors is given in \eqref{hsl2}.
and the factors $h(u^{4\g},v)$ and $h(u^{2\g}, v^{2\g})$ are related
to $h(u,v)$ in a simple way:
\be
\label{h4gamma}
\hh(u^{4\gamma},v)=1/\hh(v,u), \qquad \hh(u^{2\gamma},v^{2\gamma})=
\begin{cases}
    \hh(u,v) & \text{for } \ \sltwo, \\
    \hh(u,v) \, e^{i p(u) -ip(v)} & \text{for}\ \sutwo\, .
\end{cases}
\ee
The unnormalised structure constant takes the same form for $\sutwo$
and $\sltwo$ if we define \be \bb (u,v) \ = \begin{cases} { h(u^{2\g},
v)= \s(u,v) /A(u,v)}& \text{for } \ \sltwo, \\
 e^{-ip(v)}h(u^{2\g}, v)= \s(u,v) & \text{for}\ \sutwo\,, \end{cases}
 \quad {\rm with}\quad A(u,v) ={1- {1\over x^-y^+}\over 1- {1\over x^+
 y^-}}.  \ee
so that
\begin{align}
\label{C123beh}
\!\!\!\![\mathcal{C}_{123}^{\bullet\bullet\bullet}]^{\rm asympt}=
\sum_{\a_i\cup \bar \a_i = \uu_i} \prod_{i=1}^3 (-1)^{|\bar \a_i|} \
{e^{- p(\a_i) \ell_{i-1, i}} \over \hh (\a_i,\bar\a_i)}
&\times \frac{1}{h ({\a_2},\a_1)h (\bar{\a_1},\bar{{\a_2}})} \ \
\frac{\bb({\a_3} ,{\a_2})}{\bb({\a_3} ,\a_1)}\frac{\bb(\bar{{\a_3}}
,\bar{\a_1})}{\bb(\bar{{\a_3}},\bar{{\a_2}})} .
\end{align}
%
 %
 The next step is to convert the sum over partitions to a multiple
 contour integral, a generalisation of (\ref{C123int}):
 %
\begin{align}
\label{IntC123}
\!\![{\CC_{123}^{\bullet\bullet\bullet}}]^{\rm asympt}\; &{\propto
\sum_{m,n,r}^\infty\frac{1}{m!\,n!\,r!} \oint_{\mathcal{C}_{\uu
_1}}\prod_{j=1}^m\frac{\mu(z_{1,j})\,dz_{1,j}}{2\pi \ep}
\oint_{\mathcal{C}_{\uu
_2}}\prod_{k=1}^n\frac{\mu(z_{2,k})\,dz_{2,k}}{2\pi \ep}
\oint_{\mathcal{C}_{\uu
_3}}\prod_{l=1}^r\frac{\mu(z_{3,l})\,dz_{3,l}}{2\pi \ep}}\nonumber\\
\ &\qquad \times \,\frac{h^{\ne}(\zz _1,\zz _1)}{h(\zz _1,\uu _1)}
\times\frac{h(\zz _1,\uu _2)}{\bb(\uu_3 ,\zz _1)}e^{-ip(\zz
_1)\ell_{13}} \\\nonumber &\qquad \qquad \times \frac{h^{\neq}(\zz
_2,\zz _2)}{h(\zz _2,\uu_2 )} \times \frac{h(\uu ,\zz _2)}{\bb(\zz
_2,\uu_3 )}e^{-ip(\mathbf{z_2})\ell_{12}}\\
\nonumber &\qquad \qquad \qquad \times \frac{h^{\neq}(\zz _3,\zz
_3)}{h(\zz _3,\uu_3 )} \times \frac{1}{\bb(\uu_2 ,\zz _3) \bb(\zz
_3,\uu )} \ e^{-ip(\zz _3 )\ell_{23}}
\\
\nonumber &\qquad \qquad \qquad \qquad\times\frac{\bb(\zz _1,\zz
_3)\bb(\zz _3,\zz _1)\bb(\zz _2,\zz _3)\bb(\zz _3,\zz _2)} {h(\zz
_1,\zz _2)h(\zz _2,\zz _1)}.
 \end{align}
%
 where the last line describes the interactions between different sets
 of variables $\zz_1 $ and $\zz_2$.  The numerator in the last line is
 equal to one due to the property $b(u,v)b(v,u)=1$,
 and thus the integration over the third set of variables $\zz_3 $
 completely decouples.  This is what is expected, since the left and
 the right $\sutwo$ fields do not feel each other perturbatively.

 The integral (\ref{IntC123}) splits into three independent integrals of the type
    already studied in \cite{Bettelheim:Semi}, if it were not for the
    bi-local factor entangling the groups $\zz_1 $ and $\zz_2$
    of variables.
Remarkably, in the semiclassical limit $\epsilon\to 0$  and $\ell \ep$ finite,
 the
integration contours for the variables $\zz _1 $ and $\zz_2$
are at macroscopic distance
and $h(\zz  _1,\zz  _2)h(\zz  _2,\zz  _1)= 1+ o(\ep)$.

   In conclusion, the asymptotic coupling constant is given in the
   semiclassical limit again by a product of determinants.  This can
   be used to work out a systematic quasi-classical expansion, which
   is however out of the scope of this paper.  Our goal here is to
   compute the leading term and compare it with the result obtained on
   the string theory side \cite{Kazama:2013qsa}.  In the semiclassical
   limit the structure constant factorises as
\be
\label{C123factor}
[\mathcal{C}_{123}^{\bullet\bullet\bullet}]^{\rm asympt}
\propto
\caA_1\times \caA_2\times \caA_3.
\ee
where the integrals $\caA_1$, $ \caA_2$ and $\caA_3$, defined as
  \begin{align}
\mathscr{A}_1=&\,\sum_{n=0}^\infty\frac{1}{n!}
\oint_{\mathcal{C}_{\uu_1 }}\prod_{j=1}^n\frac{\mu(z_j)\,dz_j}{2\pi \ep}
\times\frac{h^{\ne}(\zz  ,\zz  )}{h(\zz  ,\uu_1 )}
\times\frac{h(\zz  ,\uu_2 )}{\bb (\uu_3  ,\zz  )}e^{-ip(\zz  )  \ell_{13}}\\\nonumber
\mathscr{A}_2=&\,\sum_{n=0}^\infty\frac{1}{n!}
\oint_{\mathcal{C}_{\uu_2 }}\prod_{j=1}^n\frac{\mu(z_j)\,dz_j}{2\pi \ep}\times
\frac{h^{\neq}(\zz  ,\zz  )}{h(\zz  ,\uu_2 )}\times
\frac{h(\uu_1 ,\zz  )}{\bb(\zz ,\uu_3  )}e^{-ip(\zz  ) \ell_{12}}\\\nonumber
\mathscr{A}_3=&\,\sum_{n=0}^\infty\frac{1}{n!}
\oint_{\mathcal{C}_{\uu _3}}\prod_{j=1}^n\frac{\mu(z_j)\,dz_j}{2\pi \ep}\times
\frac{h^{\neq}(\zz   ,\zz  )}{h(\zz  ,\uu_3  )}\times
\frac{1}{\bb(\uu_2 ,\zz   )\bb(\zz   ,\uu_1 )}e^{-ip(\zz  ) \ell_{23}}.
\end{align}
Neglecting the subleading factors in the product of the scalar factors, we
can approximate the functionals $\caA_k$ by the objects we have already
computed in the previous section,
\be\label{integr-var}
\begin{aligned}
\caA_k \propto&
\sum_{n =0}^\infty {1 \over
n  !}\ \oint_{\CC_{\uu_k  }} \prod_{j=1}^{n  } {dz_{ j}
\ \L_k  (z_{ j})\over 2\pi i} \prod_{i<j}^n\Delta(z_i,z_j)
  \end{aligned}
\ee
 where the functions $\L_1 , \L_2, \L_3 $ assemble the local factors for
 the three groups of integration variables:
\be
\label{localV123}
\begin{aligned}
\L_1 (z)&={e^{ -i\ell_{31}\,p(z) }\ \hh(z, \uu_2)\over \hh(z,
\uu_1  )\ \bb  ( \uu_3 , z) } ,  \qquad
 \L_2(z)={e^{ -i\ell_{12}\,p(z) }\ \hh(\uu_1 , z)\over \hh(z,
 \uu_2 )\ \bb  ( z, \uu_3 ) },
\\
& \qquad \L_3 (z) ={e^{ -i\ell_{23}\,p(z) }\over \hh(z, \uu_3  )\ \bb  (
 \uu_2, z )\ \bb  ( z, \uu_1 ) }.
 \end{aligned}
\ee
 The three factors in the product (\ref{C123factor})
  are exponentially small and the  exponent of the product is    given by
\be\label{logfac}
\begin{aligned}
\log [\mathcal{C}_{123}^{\bullet\bullet\bullet} ]^{\rm asympt}={1\over \epsilon}
\left( \caY _1  + \caY_2 + \caY_3  + o(\epsilon) \right) .
\end{aligned}
\ee
  where $\caY$ is   a contour integral of a dilogarithm
\be
\label{defdilog}
 \caY_i =\pm \oint_{\CC_{\uu_i } } {du\over 2\pi} \ \Li (
\L_i (u)), \quad i=1,2,3.  \ee
 where the $(+) $ sign is for $\sutwo$ and the $(-)$ sign is for $\sltwo$.

 \subsection{Taking the semiclassical limit, $\sutwo$}
 \label{sec:semiclsu2}

To obtain explicit expressions, we will express the local factors
$\L_a (u)$ in terms of the three quasi-momenta (\ref{pseudomms}).
Consider first the $\sutwo$ case where $b(u,v)=\s(u,v)$.  In the
leading order in $\ep$ we have (see Appendix \ref{app:hslsu})
    \begin{align}\label{limith1}
  \begin{aligned}
 \log \hh(u,v) & \to\
  -\, {i\epsilon\,  y'\over x-y}+    \, { ip(x)\over y^2-1}
 = \ \ i {\epsilon\,x'\over y-x}- i    \, { p(y)\over x^2-1}
,\\
 \log \bb (u,v)
  & \ \to\ \
 -  {i\ep  y' \over 1/x- y}-   {ip(x)\over y^2-1}- i\,  p(y)
  =
\frac{i \epsilon  x'}{{1/y}-x}+\frac{i p(y)}{x^2-1}+i p(x)
 \end{aligned}
    \end{align}
or, after taking the product with
$x_j= x(u_j), \ u_j\in\uu$
\be
  \begin{split}
&\log  \hh (u,\uu ) \to   - i \, \CG_\uu (x)+i  \, {\Delta-
 L\over 2}\, p(x)
 \\
& \log \bb  (u,\uu ) \to  \  -  i \, \CG_\uu(1/x)-i  \, {\Delta-
 L\over 2}\, p(x)
  \\
 &\log   \hh (\uu , u)  \to \  i \, \CG_\uu(x)- i \, {\Delta-
 L\over 2}\, p(x)   , \\
& \log \bb  (\uu, u ) \to \  i \, \CG_{\uu}(1/x)+
 i  \, {\Delta- L\over 2}\, p(x)
 ,
    \end{split}
  \label{resolvents}
   \ee
where the resolvent for the set $\uu $ is defined by\footnote{We have
set the mode numbers to zero for simplicity.}
\be
\CG_\uu(x) = \sum _{j} {x_j'\over x- x_j},
\qquad
  \sum_j {1\over x_j^2-1} ={\Delta-L\over 2} \,.
\ee
%
The next step is to express the measure factors  (\ref{localV123})
 in terms of the  quasi-momenta
 of the three operators
\begin{align}
 \tilde \pp^{(1)}(x)&  =  \hf   \, \Delta_1 \, p(x) -\CG_{\uu_1}(x) ,
\no  \\ \label{pseudomms}
  \tilde \pp^{(2)}(x)&  =  \hf   \, \Delta_2 \, p(x) -\CG_{\uu_2}(x) ,
\\ \no
\tilde \pp^{(3)}(x)&= \hf   \, \Delta_3 \, p(x) - \CG_{ \uu_3}(x) .
\end{align}
Substituting (\ref{resolvents}) in  (\ref{localV123}) we get
\be
\label{Vonetwothree}
\begin{aligned}
\L_1 (x)&\ \ \to\ \ \exp\left(+ i \tilde \pp^{(2)}(x) -i\tilde \pp^{(1)}(x)+ i \tilde
\pp^{(3)}(1/x)  \right) ,
\\
\L_2(z)&\ \ \to\ \ \exp\left( - i\tilde \pp^{(1)}(x)
-i \tilde \pp^{(2)}(x) - i \tilde
\pp^{(3)}(1/x)  \right),
\\
\L_3 (z)&\ \ \to\ \ \exp\left( -i \tilde \pp^{(3)}(x) -i \tilde
\pp^{(1)}(1/x)+ i \tilde \pp^{(2)}(1/x) \right) .
\end{aligned}
\ee
Using the classical Bethe equations on the cut of $\tilde p^{(1)}$,
we can  change the sign of $\tilde p^{(1)}$ in the exponent and
write $\caY_1$, eq. (\ref{defdilog}), as
\be\begin{aligned} \caY_1
&=- \oint_{\CC_{\uu_1 } } {du\over 2\pi} \ \Li(e^{i\tilde
\pp^{(1)} (x)+i \tilde \pp^{(2)}(x) + i \tilde \pp^{(3)}(1/x)}) .
\end{aligned}
\ee
Here we took into account that the contour of integration changes its
orientation when deformed to the second sheet.  We also change the
sign of the exponents in the other two integrals using the functional
equation for the dilogarithm,
$\Li(X^{-1})=-\Li(X)-\frac{\pi^2}{6}-\frac{1}{2}\log^2(-X)$.  This
again leads to a minus sign in front of the integrals:
\be\begin{aligned}
 \caY_2 &= -\oint_{\CC_{\uu_2}} {du\over 2\pi} \ \Li (
 e^{i\tilde \pp^{(1)}(x)+i \tilde \pp^{(2)}(x) + i \tilde \pp^{(3)}(1/x)})
 \\
\caY_3  &=- \oint_{\CC_{\uu_3 }} {du\over 2\pi} \ \Li(e^{-i\tilde
\pp^{(2)}(1/x)+i \tilde \pp^{(1)}(1/x) +i \tilde \pp^{(3)}(x)}) .
\end{aligned}
\ee
The final formula is
\be\begin{aligned}
\label{integr-var-1}
\log [\mathcal{C}_{123}^{\bullet\bullet\bullet} ]^{\rm asympt}_{\su(2)}
&=- {1\over \epsilon} \oint\limits_{\CC_{\uu_1 \cup \uu_2}} {du\over
2\pi} \ \Li ( e^{i\tilde \pp^{(1)}(x)+i \tilde \pp^{(2)}(x) + i \tilde
\pp^{(3)}(1/x)})
\\
&\hskip -1.5cm
-{1\over \epsilon} \oint\limits_{\CC_{\uu_3 } } {du\over 2\pi}
\ \Li(e^{ - i \tilde \pp^{(3)}(1/x)+ i \tilde \pp^{(2)}(x) -i\tilde
\pp^{(1)}(x) } ) + \text{subleading in } \ep.
\end{aligned}
\ee
This expression gives, up to subleading $o(\epsilon ^0$) terms, the
exponent for the all-loop perturbative structure constant for three heavy fields.

Let us interpret this expression from the point of view of the
spectral curves  of the three heavy states
which is  written in termes of the classical monodromy matrix
\begin{align}
\Omega(u) &= \text{Diag} \left( e^{i \hat \pp_1(u)}, e^{i \hat
\pp_2(u)}, e^{i \hat
\pp_3(u)}, e^{i \hat
\pp_4(u)}| e^{i\hat \pp_1(u)}, e^{i\hat \pp_2(u)},
e^{i \hat \pp_3(u)}, e^{i \hat
\pp_4(u)} \right).
\no\end{align}
The finite zone solutions in this sector
are characterised by cuts between 1-4 and 2-3 sheets of the Riemann
surface.
 The Bethe equations give
boundary conditions on these cuts for the combinations $\tilde \pp_L=
\hf (\tilde \pp_1 - \tilde \pp_4)$ and $\tilde \pp_R= \hf(\tilde \pp_2-\tilde
\pp_3)$, representing the quasi-momenta in the left and in the right
$\sutwo$ sectors.
The spectral curve of the $SO(4)$ sector is invariant under the
inversion symmetry $x\leftrightarrow 1/x$, which exchanges $\tilde \pp_L
$ and $\tilde \pp_R $
  \be
  \label{LRsymmetry}
  \tilde \pp _R(x) = - \tilde \pp_L(1 /x) 
  . \ee
This allows to go from the four-sheeted Riemann surface in the
$u$-parametrization to a two-sheet Riemann surface in the
$x$-parametrization
%
%
 \begin{eqnarray}
 \label{plpr}
 \label{pLpRp} \tilde \pp_{R} (x) = -\tilde \pp(1/x) \Big|_{|x|>1},
 \qquad \tilde \pp_{_L}( x) = \tilde \pp(x) \Big|_{|x|>1}.
 \end{eqnarray}
In  the notations $\pp_{L,R}(u)$ via (\ref{plpr}), the
unnormalised structure constant takes the form
\be\begin{aligned}
\label{C123I-I-II}
\log [\mathcal{C}_{123}^{\bullet\bullet\bullet} ]^{\rm
asympt}_{\su(2)}&=- {1\over \epsilon} \oint_{\CC_{\uu_1 \cup \uu_2}}
{du\over 2\pi} \ \Li ( e^{i\tilde \pp^{(1)}_L+i \tilde \pp^{(2)}_L - i
\tilde \pp^{(3)}_R}) -{1\over \epsilon} \oint_{\CC_{\uu_3 } } {du\over
2\pi} \ \Li(e^{i\tilde \pp^{(3)}_R+ i \tilde \pp^{(2)}_L-i \tilde
\pp^{(1)}_L} ) \\
&+ \text{subleading in } \ep.
\end{aligned}
\ee
In the strong coupling limit this expression reproduces exactly the
the result of the string theory computation, eq.  \eqref{strongsu2as}.

 \subsection{Taking the semiclassical limit, $\sltwo$}
 \label{sec:semiclsl2}

 In the case of $\sltwo$ fields the scalar factors $h(u,v)$ have
 asymptotics, {\it cf.} appendix \ref{app:hslsu},
  \begin{align}\label{limith}
  \begin{aligned}
 \log \hh_\sltwo(u,v) & \to\ { i\ep y' \over x-y} - {ip(y)\over x^2-1}
 = { i\ep x' \over x-y} + {ip(x)\over y^2-1} ,\\
 \log \bb _\sltwo (u,v)\ \ \ &\to \ {i\ep y'\over 1/x- y} +
 {ip(y)\over 1- 1/x^2} = \frac{i \epsilon x'}{x-1/y}-\frac{i p(x)}{1-
 1/y^2}
 \end{aligned}
    \end{align}
which gives
\be
  \begin{split}
\log \hh (u,\uu ) &\to \ i \, \CG_\uu(x)
  , \\
  \log \hh (\uu, u ) &\to
  - i\CG_\uu(1/x)  
 \\
\log  \bb  (u,\uu )& \to     i \, \CG_\uu (1/x) 
\, ,
\\
\log \bb (\uu ,u )& \to
- i \CG_\uu (1/x) \,.
    \end{split}
  \label{resolvents1} \ee
Substituting in (\ref{localV123}), we obtain
\be
\label{Vonetwothree1}
\begin{aligned}
\L_1 (x)&\ \ \to\ \ \exp\left(+ i \hat \pp^{(2)}(x) -i\hat
\pp^{(1)}(x)+ i \hat \pp^{(3)}(1/x) \right) ,
\\
\L_2(z)&\ \ \to\ \ \exp\left( - i\hat \pp^{(1)}(x) -i \hat
\pp^{(2)}(x) - i \hat \pp^{(3)}(1/x) \right),
\\
\L_3 (z)&\ \ \to\ \ \exp\left( -i \hat \pp^{(3)}(x) -i \hat
\pp^{(1)}(1/x)+ i \hat \pp^{(2)}(1/x) \right) \,,
\end{aligned}
\ee
  where $\hat p^{(k)}$ are the $\sltwo$ quasi-momenta,
  \begin{align}
  \hat \pp^{(k)}(x)& = \hf L_k\, p(x) + \CG_{\uu_k} (x)\,, 
\qquad k=1,2,3\,.
  \end{align}
  The rest is in complete analogy with the $\sutwo$ sector.  Taking
  into account the opposite sign of the dilogarithm, we write it as
\be\begin{aligned}
\label{C123I-I-II-sl2}
\log [\mathcal{C}_{123}^{\bullet\bullet\bullet}]^{\rm asympt}_{\sl(2)}
&={1\over \epsilon} \oint_{\CC_{\uu_1 \cup \uu_2}} {du\over 2\pi} \
\Li \left[ e^{i\hat \pp^{(1)}_L+i \hat \pp^{(2)}_L - i \hat
\pp^{(3)}_R}\right] +{1\over \epsilon} \oint_{\CC_{\uu_3 } } {du\over
2\pi} \ \Li\left[e^{i\hat \pp^{(3)}_R+ i \hat \pp^{(2)}_L-i \hat
\pp^{(1)}_L} \right]
\\
&+ \text{subleading in } \ \ep.
\end{aligned}
\ee which is what is expected from the strong coupling result
\cite{Kazama:2016cfl} in \eqref{strongsl2as}.
%

\section{Bottom mirror excitations\label{sec:bottommirror}}
The full result of the structure constant requires taking into account
mirror excitations on all the three edges.  The general expression is
too complicated to be treated here; moreover, the interaction of
mirror particles in crossed channels is affected by singularities
which need careful regularisation.  The simplest, tractable case of
mirror contribution is that of the structure constant with only one
non-BPS operatos, in the channel opposed to the on the opposite edge
of the physical excitations, or bottom channel, as shown in figure
\ref{fig:bottom}.
These mirror particles do not enter the sum over partitions and they
do not interact with the other mirror excitations, so they can be
factorised out and considered separately.  Written schematically in
terms of the fundamental excitations, the integrand is given by
\cite{Basso:2015eqa}
\begin{align}
\label{eq:integrand1}
\mu(\mathbf{w}^\gamma)\, e^{ip(\mathbf{w^\gamma})\ell_B}\,
T(\mathbf{w}^\gamma)\ \hh^{\neq}(\mathbf{w}^\gamma,\mathbf{w}^\gamma)
\ \hh(\uu ,\mathbf{w}^{-3\gamma})\,,
\end{align}
with $\ell_B={1\over 2}(L_2+L_3-L_1)$ the length of the bottom bridge
of the correlator, opposed to the operator $\CO_1$.  The last factor
can also be transformed to the same mirror dynamics by using
\eqref{h4gamma}, $\hh(\uu
,\mathbf{w}^{-3\gamma})=1/\hh(\mathbf{w}^{\gamma},\uu)$.  The mirror
transformation $\gamma$ is defined as the analytical continuation
through the branch cut of the variable $x^+$, namely $x^+\to 1/x^+$,
as shown in \ref{fig:thumbrule}.

In \cite{BKV} the contribution of a single mirror particle was
analysed, and shown to reproduce lowest order contribution of the
expected strong coupling answer \cite{Kazama:2013qsa,Kazama:2016cfl}.
The full mirror corrections involve all the bound states, and in the
integrand (\ref{eq:integrand}) all the quantities should be replaced
by their bound state counterparts.  Here we are able to sum all the
bound state contribution, in the strong coupling limit, and to
retrieve part of the strong coupling result.  This imply summing over
all the configurations $\vec n= \{ n_1, n_2, \dots \}$, where $n_a$ is
the number of bound states of $a$ magnons,
\begin{align}
\label{eq:Bottom}
[\CC^{\bullet\circ\circ}]^{\rm bottom}=\sum_{\vec{n}}
\frac{\rB[\vec{n}]}{\prod_a n_a!} \, .
\end{align}
 The contribution of the configuration $\vec n$ is given
 by\footnote{{The $(-1)^n$ factor
 comes from the crossing transformation of the mirror magnons
 $\Phi^{a\dot{a}}\overset{2\gamma}{\to}-\Phi^{\dot{a}a}$.}}
\begin{align}
\label{eq:intb}
\rB[\vec{n}]= (-1)^n\, &
\int_{-\infty}^\infty\prod_a\,\prod_{j=1}^{n_a}\,
\frac{dz_j^a}{2\pi\ep}\,\mu_a^\gamma(z_j^a)\;g^\gamma
_a(z_j^a)\;T_a^\gamma(z_j^a) \\\nonumber &
\times\!\!\!\!\prod_{\substack{a \\ 1\le i<j\le
n_a}}\!\!\!\! H^\gamma_{aa}(z_i^a,z_j^a) \prod_{\substack{a<b \\
1\le i\le n_a\\ 1\le j\le n_b}} \!\! H^\gamma_{ab}(z_i^a,z_j^b)\;
,\qquad n=\sum_a n_a\, a.
\end{align}
%
The integration contour is along the real axis in the mirror regime
shown in \ref{fig:thumbrule}.  The bi-local factors
$H^\gamma_{ab}(z_i^a,z_j^b)$ coupling two bound states of length $a$
and $b$ are given by
\begin{align}
H^\gamma_{ab}(u,v)\equiv\hh _{ab}(u^\gamma,v^\gamma)\, \hh
_{ba}(v^\gamma,u^\gamma)\, ,
\end{align}
where $\hh _{ab}(u,v)$ is the bound state counterpart of $\hh (u,v)$
and is defined in (\ref{eq:hbound}).  The functions $g
_a^{\gamma}(u)\equiv g _a(u^{\gamma})$ and $\mu_a^\g (u)\equiv
\mu_a(u^\g)$ are mirror transforms respectively of the local weight
factor $ g _a(u )$ and the measure $\mu_a(u)$ defined as
\begin{align}
\label{eq:gm}
g _a(u)=\frac{ e^{ip_a(u)\, \ell_B}}{\hh _{a,1}(u,\uu )}\,,\qquad
\mu_a(u) =\frac{ 1
}{a}\ \frac{(1-1/x^{[-a]}x^{[+a]})^2}{(1-1/x^{[+a]}x^{[+a]})\
(1-1/x^{[-a]}x^{[-a]})} \, .
 \end{align}
 %

 Throughout this chapter we are using the notation
 $x^{[k]}=x(u+ik\ep/2)$.
 %
We want to take the semiclassical limit of (\ref{eq:Bottom}) and
(\ref{eq:intb}), focusing on the strong coupling limit $g\to \infty$.
Since the $\sutwo$ and the $\sltwo$ cases are treated in almost
identical way, we will focus on the $\sutwo$ case and will briefly
summarise the $\sltwo$ case at the end.

\subsection{Quantities for bound states at strong coupling
\label{sec:boundstrong}}

In this section, we determine the strong coupling limit expressions of
the various quantities in the integrand (\ref{eq:intb}).  The bound
state counterparts can be obtained by fusing the corresponding
fundamental quantities.  Notice that when we perform strong coupling
expansion, there are different regimes in the complex $x$-plane.
Since the integration contours for the rapidities $\zz$ are situated
on the real axis in the mirror dynamics, it is enough to analyse the
strong coupling limit in this regime \cite{BHL}.
The near-flat-space regime, where $u$ is situated close to the
singularities $x(u)=\pm1$ is not relevant for this case, as we are
concerned with semiclassical strings.  We have mainly to check the
mirror giant magnon regime $|x(u)|>1$ and the mirror BMN regime
$|x(u)|=1$\footnote{
The denomination of the various regimes follows the analytical
continuation of the corresponding ones in the physical dynamics.}.
The integrals over mirror particles contain the factor
\begin{align}
\label{eq:expsuppr}
e^{-E_a(u)\ell_B}\sim \frac{1}{ x^{2\ell_B}}\,, \qquad |x(u)|>1
\end{align}
which strongly suppresses the contribution of the mirror giant magnon
particles for large values of the bridge length $\ell_B$.
We are therefore going to concentrate on the BMN mirror regime
$|x|=1$.

\begin{figure}[h!]
\begin{center}
\includegraphics[scale=0.8]{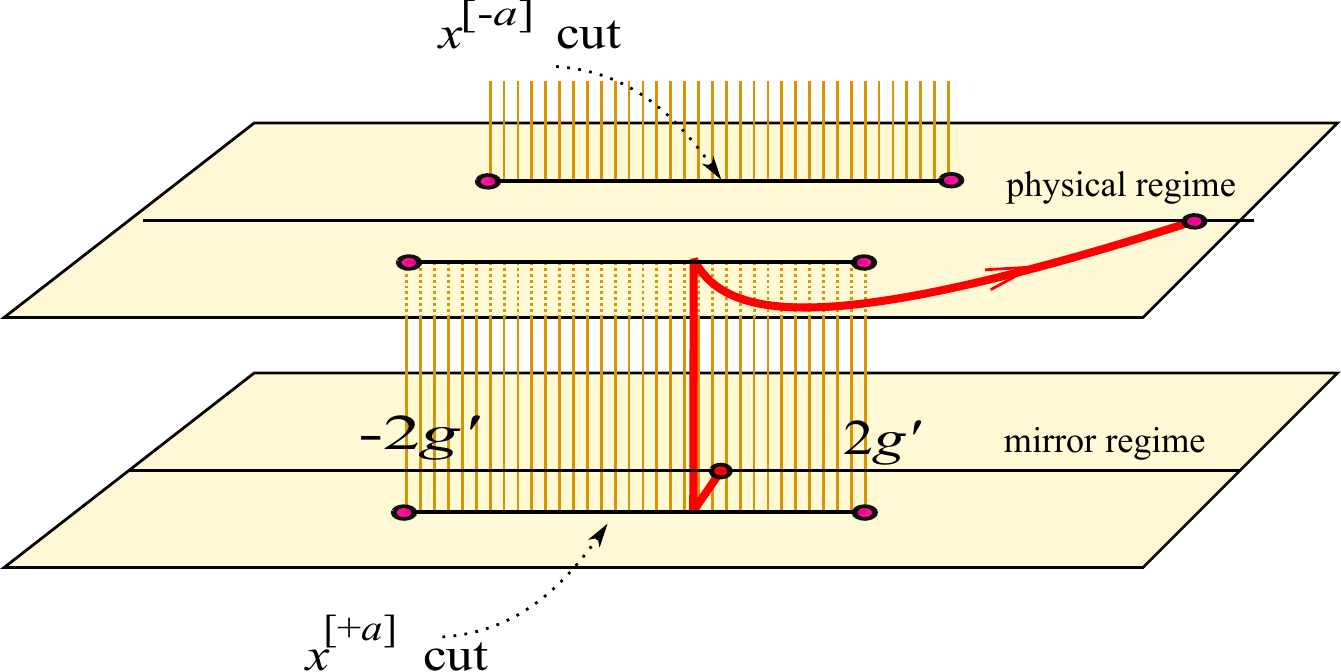}
 \caption{\small\ The rule for analytic continuation from the BMN
 mirror regime to BMN physical regime at strong coupling, when the
 real axis is pinched between the branch cuts of the \Zh variables
 $x^+$ and $x^-$.  }
\label{fig:thumbrule}
\end{center}
\end{figure}

As illustrated in \ref{fig:thumbrule}, the contributions from the
mirror BMN regime at strong coupling can be determined by first taking
the strong-coupling limit of the relevant quantities in the physical
regime and then analytically continuing them to the lower half of the
unit circle $|x| =1$.  This simple rule should be applied with care
for the bound-state quantities, which may have an array of branch
cuts.  In this case, the passage to the mirror regime of an object
associated to a bound state of size $a$ is done by substituting
$x^{[+a]}(u)$ by $1/x^{[+a]}(u)$, that is by analytically continuing
$u$ through the branch cut of the \Zh variable $x^{[+a]}(u)$ and
leaving the other cut untouched.

In the strong coupling limit, the different branch cuts collapse on
each other and on the real axis and the dependence on the rapidities
will be given by variable with a single branch cut $x(u)$.  In the
mirror giant magnon regime, $x^{[+a]}\simeq x^{[-a]}\simeq x$, while
in the mirror BMN regime, where the branch cut is situated,
$x^{[+a]}\simeq 1/x^{[-a]}\to x$.  The net result is that after both
mirror transformation {\it and} strong coupling limit, $x^{[\pm a]}\to
1/x$, which is equivalent to continuing $x$ to the lower half unit
circle $U_-$.

Special care has to be devoted to the continuation of the dressing
phase for bound states to the mirror dynamics, where extra cuts
appear.  As we explain in appendix \ref{app:nopolesinthephysicalstrip}
based on \cite{Arutyunov:2009kf,Vieira:2010aa}, the dressing phase
appears in combination with other functions which cancel exactly the
cuts on the real axis and all the other cuts between those of
$x^{[-a]}(u)$ and $x^{[+a]}(u)$.  The same combination has no branch
cut below that of $x^{[+a]}(u)$ in the physical dynamics, therefore we
are again in the situation represented in figure \ref{fig:thumbrule}
and we can use the analytical continuation from the BMN physical
regime to the BMN mirror regime.


\paragraph{The scalar factor $H_{ab}(u,v)$.} The scalar factor for
scattering of two bound states of length $a$ and $b$ is given by
\begin{align}
\label{eq:hbound}
\hh _{ab}(u,v) =\prod_{k=-\frac{a-1}{2}}^{\frac{a-1}{2}}
\prod_{l=-\frac{b-1}{2}}^{\frac{b-1}{2}} \hh (u^{[2k]},v^{[2l]})\, .
\end{align}
%
%
The symmetric scalar factor $H_{ab}(u,v)=h_{ab}(u,v)h_{ba}(v,u)$ is
then given by
\begin{align}
H_{ab}(u,v)=\frac{x^{[-a]}-y^{[-b]}}{x^{[+a]}-y^{[-b]}} \
\frac{x^{[+a]}-y^{[+b]}}{x^{[-a]}-y^{[+b]}} \
\frac{1-1/x^{[-a]}y^{[+b]}}{1-1/x^{[+a]}y^{[+b]}} \
\frac{1-1/x^{[+a]}y^{[-b]}}{1-1/x^{[-a]}y^{[-b]}}\, .
\end{align}
The dressing factor dropped out from the expression of the symmetric
factor.  In the strong coupling limit in the mirror dynamics
$H^\gamma(u,v)$ takes the simple form
\begin{align}
\label{Habappr}
H^\gamma_{ab}(u,v)\simeq
\frac{u-v-i\ep\frac{a-b}{2}}{u-v-i\ep\frac{a+b}{2}} \ \
\frac{u-v+i\ep\frac{a-b }{2}}{u-v+i\ep\frac{a+b}{2}}\, .
\end{align}
We notice that the pairwise interaction takes in the strong coupling
limit the same form as the interaction of the bound states
(\ref{eq:delta}) in the asymptotic structure constant,
\begin{align}
H^\gamma_{ab}(u,v)\simeq\Delta_{ab}(u^{[-a]},v^{[-b]})
=\Delta_{ab}(u-\hf ia\ep,v-\hf ib\ep)\, ,
\end{align}
 the only difference being that the position of the pole is shifted to
 $v=u\pm i(a+b)\ep/2$.

\paragraph{The measure $\mu_a(u)$.}
The  expression for the measure for a bound state, eq.(\ref{eq:gm}), is
\begin{align}
{\mu_a(u)
=\,\frac{1}{a}\
\frac{(1-1/x^{[-a]}x^{[+a]})^2}{(1-1/x^{[+a]}x^{[+a]})\
(1-1/x^{[-a]}x^{[-a]})}}\,.
\end{align}
%
%
Performing mirror transformation for $\mu_a(u)$ and expanding at
strong coupling in the mirror BMN regime $|x|=1$, we find
\begin{align}
\label{eq:mua}
\mu_a(u^\gamma)\simeq \frac{{1 }}{a}.
\end{align}

\paragraph{The factor $g _a(u)$.}
%
%
Recall that
\begin{align}
\label{defgsutwo}
g _a(u)=
\frac{e^{ip_a (u)\ell_B }}{\hh _{a1}(u,\mathbf{u})}\;,
\end{align}
with
\begin{align}
\hh _{a1}(u,v)=
\frac{u^{[-a]}-v^-}{u^{[+a]}-v^-}
\
\frac{1-1/x^{[-a]}\, y^+}{1-1/x^{[+a]}\, y^+}
\
\frac{1-1/x^{[+a]}\, y^-}{1-1/x^{[-a]}\, y^-}\frac{1}{\sigma_{a,1}}
\;.
\end{align}
After the continuation to the mirror dynamics,
$\sigma_{a,1}(u^\gamma,v)$ has extra cuts between those situated at
$u-ia\ep/2$ and $u+ia\ep/2$ with $u\in [-2g', 2g']$.  In particular,
for even $a$ one of those cuts is situated on the real axis, \ie on
the contour of integration for the mirror particle contribution.
These cuts are compensated by an extra factor coming from the
normalisation of the transfer matrix matrix, as we will show below.
The quantity we have to consider is
\begin{align}
\label{tildega}
\tilde g _a
= g_a(u) \frac{R^{(-)[2-a]}}{R^{(+)[2-a]}}\ldots  \frac{R^{(-)[a]}}{R^{(+)[a]}}\;.
\end{align}
Here and below we use the notation
\be
\begin{split}
R^{(\pm)}(u)&=(x - \xx^{\mp})\equiv \prod_j(x(u)-x^\mp(u_j)),
\\
 B^{(\pm)}(u)&=
 (1/x- \xx^{\mp}) \equiv \prod_j (1/x(u)-x^\mp(u_j)).%
\end{split}
\label{defRpRm}
\ee
where the functions $R^{(\pm)}(u), B^{(\pm)}(u)$ play the role of the
Baxter polynomials in the \Zh plane and encode the rapidities of the
incoming state.  The simplest strategy to take the strong coupling
limit of the quantity above is to compute it first in the BMN physical
dynamics and the analytically continue to the BMN mirror dynamics.
This can be done because there is no singularity to be met on the path
of the analytical continuation.  Taking the strong coupling limit, one
can express $g_a(u)$ in terms of the three quasi-momenta
(\ref{ptilde123})
\begin{align}
\label{asymga}
{g_a(u)\ \to\  e^{iap(x)\ell_B+ia\CG(x)-ia\frac{\Delta-L_1}{2}p(x)}=
e^{ia(\tilde p_2(x)+\tilde p_3(x)-\tilde p_1(x))}}
\end{align}
and, after the continuation to the mirror regime one has
\begin{align}
{\tilde g_a(u^\gamma)=\ \to\
e^{iap(1/x)\ell_B-ia\frac{\Delta-L_1}{2}p(1/x)}= e^{ia(\tilde
p_2(1/x)+\tilde p_3(1/x)-\tilde p_1(1/x))}e^{-ia\CG(1/x)}.}
\end{align}
We have used that at strong coupling we have, in the physical BMN
regime
\begin{align}
f^{[a]}(u)\equiv \frac{R^{(+)[+a]} }{R^{(-)[+a]} }\to e^{i\CG (x)}\,,
\qquad \bar{f}^{[a]}(u)\equiv \frac{B^{(-)[+a]} }{B^{(+)[+a]} } \to
e^{-i\CG (1/x)}\,,
\end{align}
where
\begin{align}
\label{defresx}
 \CG(x)=\frac{1}{i}\sum_j \ln \frac{x-x_j^-}{x-x_j^+}\ \ \to \ \
 \epsilon \sum_j \frac{x'(u_j)}{x-x(u_j)}\,
\end{align}
is the resolvent in the $x$-plane, while in the mirror BMN regime we
have
\begin{align}
\label{fbarfmir}
f^{[a]}(u^\gamma)\to e^{i\CG (1/x)}\,, \qquad \bar{f}^{[a]}(u^\gamma)
\to e^{-i\CG (x)}\,.
\end{align}

%
%

%
%
%

{\bf The transfer matrix at strong coupling.} Another important
element for the integrand is the transfer matrix, arising after
summing over the various polarisations of the mirror particles.  The
transfer matrix in the $\mathfrak{su}(2)$ sector is given, up to a
global factor, by \cite{Tsuboi-SmatGF, Beisert:2006qh, Kazakov:2007fy,
Zabrodin:2007rq,Gromov:2009tq})
 \begin{align}
 \label{genfnl}
&\sum _{a=0}^\infty
 \bar{T}_{a} ^{[a-1]}(u) \ID^{2a}=\left(1-Y_{2,2}\ID^2\right)\left(1-X_{2,1}
 \ID^2\right)^{-1}\left(1-X_{1,1}\ID^2\right)^{-1}\left(1-Y_{0,1}\ID^2\right)\,,\\ 
 \no &\text{with}\quad X_{2,1}=X_{1,1}=1\,,\quad Y_{2,2}=
 \frac{R^{(-)-}}{R^{(+)-}} ={1\over f^-}, \quad Y_{0,1}=
 \frac{B^{(+)+}}{B^{(-)+}} = {1\over \bar f^+} \quad {\rm and } \quad
 \ID^2=e^{i\ep \partial_u}\,.
\end{align}
The bar on the function $T_a$ means complex conjugation, and assuming
the rapidity $u$ to be real this means just changing the sign of the
imaginary shifts.

A change in the normalisation of the transfer matrix can be obtained
by multiplying the shift operator by an arbitrary function, $\ID^2\to
-\bar N (u)\, \ID^2 $ in the generating functional.  {Since we are
using the $\mathfrak{su}(2)$ hexagon form factor as the dynamical
part, we should normalise the transfer matrices in such a way that the
component in the $\mathfrak{su}(2)$ sector is just $1$.  This
corresponds to taking $N (u)$ to be $R^{(-)+}/R^{(+)+}= 1/f^+$.
Expanding \eqref{genfnl} and taking into account the normalization, one
obtains (see {\it e.g.} eq.  (8.67) of \cite{Serban:2010sr})}
\begin{align}\label{newnorm}
T_a(u)=(-1)^a N_a(u)\left[
(a+1)-a\,\frac{R^{(+)[+a]}}{R^{(-)[+a]}}-a\,
\frac{B^{(-)[-a]}}{B^{(+)[+a]}}+(a-1)
\frac{R^{(+)[+a]}}{R^{(-)[+a]}}\frac{B^{(-)[-a]}}{B^{(+)[-a]}} \right]
\end{align}
where $N_a(u)= N^{[1-a]}\, N^{[1-a+2]}\, \dots \,N^{[a-1]}$.
 Notice that the bound-state quantities entering $g _a(u)$ in
 (\ref{eq:gm}) have exactly the same structure as the prefactor
 $N_a(u)$.  Therefore we can absorb $g _a(u)$ into the normalisation
 factor.  This amounts to replacing $N_a (u)$ in \eqref{newnorm} with
\begin{align}
\tilde{g}_a (u) \equiv N_a (u) g_a (u)\,.
\end{align}
%
We have prefered to keep the sign $(-1)^a$ out of the normalisation
factor, since it will exactly compensate the factor $(-1)^n$ in
\eqref{eq:intb}.  Thus the re-normalised transfer matrix $\rT_a$ takes
the form
\begin{align}
 \label{defrTa}
\rT_a(u)\equiv g _a(u)\,T_a(u)= \tilde g_a(u)
\left[(a+1)-a\,f^{[a]}-a\,\bar{f}^{[a]}+(a-1)f^{[a]}\bar{f}^{[a]}\right].
\end{align}
As it was discussed in the beginning of this chapter, the
re-normalised transfer matrix $\rT_a(u)$ does not have any cut beyond
the cuts of $x^{[\pm a]}$ on the physical sheet, and no cuts within
the strip $|\Im u|<\hf a \ep $
 on the mirror sheet.  Therefore, the strong coupling limit of
 \eqref{defrTa} in the BMN mirror dynamics can be obtained by simply
 substituting $x$ with $1/x$ in the physical BMN expression,
\begin{align}
 \label{defrTagamma}
\rT_a(u^\gamma)\to \tilde g^a(u^\gamma)
\left[(a+1)-a\,f-a\,\bar{f}+(a-1)f\bar{f}\right]\,,
\end{align}
with $f$ and $\bar f$ being those from \eqref{fbarfmir}.  Let us
further define a quantity
\begin{align}
\rt_n=\tilde g^n\, (2-f^n-\bar{f}^n)\,.
\end{align}
It is interesting that only these quantities will appear in the final
result of semiclassical limit.  They can be expressed in terms of the
transfer matrix by the following relations\footnote{The relevance of
this type of relations to clustering of mirror bound states was
pointed out to us by Benjamin Basso.  These relations are valid in the
semiclassical limit only, both in the BMN and mirror dynamics,
therefore we have dropped the arguments which specify the dynamics.}
\begin{align}
\label{eq:recursive}
&\rt_1=\rT_1\\\nonumber &\rt_2=2\rT_2-\rT_1^2\\\nonumber
&\rt_3=3\rT_3-3\rT_2\rT_1+\rT_1^3\\\nonumber
&\rt_4=4\rT_4-4\rT_3\rT_1-2\rT_2^2+4\rT_2\rT_1^2-\rT_1^4\\\nonumber
&\rt_5=5\rT_1-5\rT_4\rT_1-5\rT_3\rT_2+5\rT_3\rT_1^2+5\rT_2^2\rT_1-5\rT_2\rT_1^3+\rT_1^5\;,
\end{align}
which can be derived from the generating functionals,
\begin{align}
\label{eq:Tt}
\text{Sdet}(1-z \, \rG)^{-1}=&\,\frac{(1-z y_1)(1-z
y_2)}{(1-zx_1)(1-zx_2)}=\sum_a z^a\,\rT_a\\\nonumber \text{Str}(1-z\,
\rG)^{-1}=&\,z\frac{\rd}{\rd z}\log\text{Sdet}(1-z \, \rG)^{-1}=\sum_a
z^a\,\rt_a.
\end{align}
Here ``Sdet'' and ``Str'' denote super-determinant and super-trace
respectively and $\rG$ a supergroup element with eigenvalues
$(x_1,x_2|y_1,y_2)=\tilde g (1,1|f,\bar{f})$.  By inserting the first
equation of (\ref{eq:Tt}) into the second equation of (\ref{eq:Tt})
and comparing the coefficient of $z^n$, we obtain the relations
(\ref{eq:recursive}).  In general, the result reads
\begin{align}
\label{eq:tn}
\frac{\rt_n}{n}=\sum_{\vec{n}\;:\,\sum_a
n_a\,a=n}(-1)^{k-1}(k-1)!\prod_a\frac{\rT_a^{n_a}}{n_a!},\qquad
k\equiv\sum_a n_a.
\end{align}

\subsection{Clustering the mirror particles
\label{sect:clustmirpart}}

We have now prepared all the ingredients necessary to take the strong
coupling limit of the contribution of mirror particles.  After
substituting the strong coupling quantities in the BMN mirror regime
evaluated in the previous chapter, we obtain
%
\begin{align}
\label{eq:intbstrong}
\rB_{sc}[\vec{n}]=\int_{\rU_-}\prod_a\prod_{j=1}^{n_a}\;\frac{dz_j^a}{2\pi
a \ep}\;\rT_a(z_j^a)\!\!\!\prod_{\substack{a \\ 1\le i<j\le n_a}}
\!\!\!\Delta_{aa}(z_i^a,z_j^a) \times \!\!\prod_{\substack{a<b \\ 1\le
i\le n_a\\ 1\le j\le n_b}}\!  \!\Delta_{ab}(z_i^a-\hf ia\ep,z_j^b-\hf
ib\ep)
\end{align}
Here $\rB_{sc}[\vec{n}]$ stands for the strong coupling limit of
$\rB[\vec{n}]$ in \eqref{eq:intb}.  Since in the strong coupling limit
$\rT_a(z)$ has a single branch cut, the one of the \Zh variable $x(z)$
situated on the real axis, one has to specify the contour of
integration.  By convention we denote by $x$ the determination
$x(z+i0)$ and by $1/x$ the determination $x(z-i0)$, when $z$ is real.
In agreement with the argument on the analytical continuation we have
employed in the previous chapter, the integrals in
\eqref{eq:intbstrong} run on the contour just below the \Zh cut,
$\rU_-=[{-2 g' -i0},{2g'-i0}]$, where the determination of the \Zh
variable is $1/x$.  The resummation of \eqref{eq:Bottom} will employ a
method closely related to the clustering method from the asymptotic
case.

{\bf The mechanism of clustering for mirror particles.}
In the asymptotic case, the contour of integration is closely
surrounding the roots $\uu $.  By deforming the contour, we pick up
poles $x_k=x_j+im\ep$ which leads to clustering and makes it possible
to safely take the semiclassical limit.  Here the situation is
slightly different since the contour is along the real axis in the
complex $z$-plane.  The integrals are independent except for the
factors $\Delta_{ab}$ which become important at $|z_j-z_k|\sim \ep$.
At strong coupling, it is convenient to use as rescaling factor
$\ep=1/2g$.  When $g\to\infty$, the two poles of $\Delta_{ab}$ are
approaching each other and pinch the integral contours on the real
axis \footnote{This is the contour pinching which is alluded to in
\cite{Basso:2013vsa}.}.  In order to avoid this singularity, we can
deform the contours.  Again the deformation of contours will catch
poles and leads to clustering.  Here we are interested in obtaining
only the limit $g\to \infty$ and not the $1/g$ corrections, which are
more involved and which are left for a future work.  A more
straightforward, equivalent way to obtain the result is to notice that
\begin{align}
\label{eq:Deltadelta}
\Delta_{ab}(z_1-ia\ep/2,z_2-ib\ep/2)=1-\delta_{ab}(z_1-z_2)
\end{align}
with $\delta_{ab}(z)$ the contribution of the singularities pinching
the integration contour, which becomes a delta-like function as $g\to
\infty$,
\begin{align}
\label{eq:deltanorm}
\int \frac{dz}{2\pi\ep}\;\delta_{ab}(z)=\frac{ab}{a+b}\,.
\end{align}
Evaluating the contribution from $\delta_{a,b}(z)$ will be equivalent
to the clustering procedure.

\par Let us analyse some simple examples in order to illustrate the
idea.  Let us organise the sum of (\ref{eq:Bottom}) as
\begin{align}
[\CC^{\bullet\circ\circ}]^{\rm bottom}=\sum_{n=1}^\infty \rB_n,\qquad
\rB_n=\sum_{\vec{n}:\sum_a n_a\,a=n}\frac{\rB_{sc}[\vec{n}]}{\prod_a
n_a!}\,,
\end{align}
and consider $\rB_1$, $\rB_2$ and $\rB_3$.  We have
\begin{align}
\rB_1=&\,\int\frac{dz}{2\pi\ep}\rT_1(z)=\int\frac{dz}{2\pi\ep}\rt_1(z),
\\\nonumber
\rB_2=&\,\frac{1}{2}\int\frac{dz_1dz_2}{(2\pi\ep)^2}\rT_1(z_1)\rT_1(z_2)
\Delta_{11}(z_1,z_2)+\frac{1}{2}\int\frac{dz}{2\pi\ep}\rT_2(z).
\end{align}
Using (\ref{eq:deltanorm}) the first term of $\rB_2$ can be re-written
as
\begin{align}
\label{eq:B2}
\frac{1}{2}\int\frac{dz_1dz_2}{(2\pi\ep)^2}\rT_1(z_1)\rT_1(z_2)-
\frac{1}{4}\int\frac{dz_1}{2\pi\ep}\rT_1(z_1)^2
\end{align}
where in the first term of (\ref{eq:B2}), the integrals for $z_1$ and
$z_2$ are independent.  The second term of (\ref{eq:B2}) comes from
taking the pole or clustering.  Using the second line in
(\ref{eq:recursive}), $\rB_2$ can be written as
\begin{align}
\rB_2=\frac{1}{2}\left(\int\frac{dz}{2\pi\ep}
\rt_1(z)\right)^2+\frac{1}{4}\int\frac{dz}{2\pi\ep}\rt_2(z)\,.
\end{align}
In appendix \ref{sec:B3strcp} we show that the third term in the
expansion is equal to
\begin{align}
\rB_3=&\frac{1}{3!}\left(\int\frac{dz}{2\pi\ep}\rt_1(z)\right)^3
+\frac{1}{4}\left(\int\frac{dz}{2\pi\ep}\rt_1(z)\right)
\left(\int\frac{dz}{2\pi\ep}\rt_2(z)\right)+
\,\frac{1}{9}\int\frac{dz}{2\pi\ep}\rt_3(z).
\end{align}
These three terms are consistent with the expansion of the exponential
of a sum of dilogarithms.  In the next section we prove that the full
expression is indeed the exponential of a sum of dilogarithms.

{\bf Combinatorics of clusters of bound states.  }
%
Now we can perform the clustering of bound states in the mirror
channel in full generality following the pattern we have just
explained.  We start with the sum over bound states with the
interactions $\Delta_{ab}$.  After taking into account the
contribution of the poles $\delta_{ab}$, which is equivalent to
clustering, we re-organise the bound states.  The final result is
again a sum over bound states, but without the interactions
$\delta_{ab}$ and with a different combinatorics factor.  It is this
combinatorics factor that we are going to determine next.  \par As a
first step, we consider the problem of clustering $k$ bound states
into a single bound state.  The initial bound state configuration can
be labeled by $\vec{n}=\{n_1,n_2,\cdots\}$ where $n_a$ is the number
of the bound states of length $a$.  Let us denote
\begin{align}
n=\sum_a n_a\,a,\qquad k=\sum_a n_a
\end{align}
so that $k$ is the initial number of bound states and $n$ is the
length of the resulting bound state.  The big bound state is obtained
by clustering the following product
\begin{align}
\prod_a\prod_{j=1}^{n_a}\frac{\rT_a(z_j^{a})}{a}\, .
\end{align}
The clustering rule for the bound states of length $a$ and $b$ is
given, according to \eqref{eq:deltanorm}, by
\begin{align}
\frac{\rT_a(z_j)}{a}\times\frac{\rT_b(z_k)}{b}\rightarrow -
\frac{\rT_a(z_j)\rT_b(z_j)}{a+b}\, .
\end{align}
Therefore, the clustering of $k$ bound states with the initial
configuration $\vec{n}$ gives
\begin{align}
\prod_a\prod_{j=1}^{n_a}\frac{\rT_a(z_j^{a})}{a}\rightarrow
(-1)^{k-1}\frac{(k-1)!}{n}\prod_a \rT_a(z_1)^{n_a}
\end{align}
where the factorial $(k-1)!$ takes into account different orders of
clustering.  The last expression, when summed over all the
configurations of initial bound states with weight $n$, gives, upon
using (\ref{eq:tn}),
\begin{align}
\label{eq:oneclus}
\sum_{\vec{n}:\sum_a n_a\,a=n} (-1)^{k-1}\frac{(k-1)!}{n}\prod_a
\frac{\rT_a(z)^{n_a}}{n_a!}=\frac{\rt_n}{n^2}\;.
\end{align}
We recognise here the factor $1/n^2$ which is necessary to reconstruct
the sought-off dilogarithm, and which appears as a non-trivial
combination of the $1/n$ factor from the measure of integration
$\mu_n(z)$ and the $1/n$ factor in the $\rt-\rT$ relations
\eqref{eq:tn}.

In order to complete the proof, we have to also consider the generic
case when the set of bound states $\vec{n}=\{n_1,n_2,\cdots\}$ cluster
into bound states $\vec{d}=\{d_1,d_2,\cdots\}$ where $d_l$ is the
number of bound state of length $l$.  Again, as in (\ref{eq:vec}), we
find it helpful to use alternatively the non-decreasing sequence
$\{q_1,q_2,\cdots,q_m\}$ to caracterise $\vec{d}$.  The total number
of cluster in the final state is denoted by $m=\sum_l d_l$.  Each
bound state $q_j$ in the final set is obtained from fusing a subset
$\vec{n}^{(j)}=\{n_1^{(j)},n_2^{(j)},\cdots\}$ of the initial bound
states $\vec{n}=\{n_1,n_2,\cdots\}$.  To prepare the clustering we
split the initial factor as
  \begin{align}
\rT_1^{n_1}\rT_2^{n_2}\rT_3^{n_3}\ldots = \prod_{j=1}^m
\left(\rT_1^{n_1^{(j)}}\rT_2^{n_2^{(j)}}\rT_3^{n_3^{(j)}}\ldots\right)\;,
  \end{align}
  such that
\begin{align}
\sum_a n^{(j)}_a\, a=q_j\,,\qquad \sum_a n_a^{(j)}=k_j\,, \qquad
\sum_{j=1}^m n_a^{(j)}=n_a\,.
\end{align}
Let us now count the symmetry factors.  First, there is the factor
from the definition (\ref{eq:Bottom}).  Second, a permutation of the
bound states of the same length in $\{q_1,q_2,\cdots,q_m\}$ leads to
the same representation $\vec{d}$ and we have to take care of this
redundancy as well.  Together, these symmetry factors are given by
\begin{align}
\label{eq:da}
\prod_a\frac{1}{n_a!}\prod_{l}\frac{1}{d_l!}\,.
\end{align}
%
%
Next, there is a factor
\begin{align}
\label{eq:naj}
\frac{n_a!}{\prod_{j=1}^k n_a^{(j)}!}={n_a\choose
n_a^{(1)}}{n_a-n_a^{(1)}\choose n_a^{(2)}}\ldots
\end{align}
coming from distributing $n_a$ bound states of type $a$ into the
different sets $n_a^{(j)}$.  By clustering $\vec{n}^{(j)}$ into a
single bound state of length $q_j$ one gets the same factor as in
(\ref{eq:oneclus}) for each $j$.  We have then
\begin{align}
\label{eq:cluster}
\rB_{\vec{d}}= \prod_l\frac{1}{d_l!}\times \prod_{j=1}^m \int
\frac{dz_j}{2\pi\ep} &\sum_{\vec{n}^{(j)}:\sum_a n_a^{(j)}a\,
=q_j}(-1)^{k_j-1}\frac{(k_j-1)!}{q_j}\prod_a\frac{\rT^{n^{(j)}_a}_a(z_j)}{n_a^{(j)}!}\\\no
&=\prod_l\frac{1}{d_l!}\times \prod_{j=1}^m \int \frac{dz_j}{2\pi\ep}
\frac{\rt_{q_j}(z_j)}{q_j^2}
\end{align}
where we have used (\ref{eq:tn}) repeatedly.  $\rB_{\vec{d}}$ is the
contribution of the bound states after clustering ${\vec{d}}$, not to
be confused with the contribution before clustering, $\rB[\vec{n}]$.
Using the same argument as in section \ref{sec:combas} we can write
the final answer as
\begin{align}
[\CC^{\bullet\circ\circ}]^{\rm bottom}=\sum_{\vec{d}}\rB_{\vec{d}}
=\exp\int\frac{dz}{2\pi\ep}\sum_n\frac{\rt_n(z)}{n^2}
\end{align}

\subsection{The $\su(2)$ bottom mirror contribution}

The result after clustering is remarkably similar to the result for
the asymptotic case (\ref{eq:exactC}), with the exception that we are
now in the zero shift limit and that $\rt_n(z)$ is composed from four
terms.  In the $\sutwo$ sector\footnote{The combination of these four
terms for $n=1$ in the $\sl(2)$ sector was considered in appendix M of
\cite{BKV}.  }
\begin{align}
\rt_n^\sutwo =\tilde g^n(1+1-f^n-\bar{f}^n)\;.
\end{align}
 Here we obtain the full expansion of the dilogarithm
\begin{align}
\log [\CC^{\bullet\circ\circ}]^{\rm
bottom}&=\int_{-1}^1\frac{dz}{2\pi\ep}\left(\Li \left[e^{i(\hat
p^{(2)}(1/x)+\hat p^{(3)}(1/x)-\hat p^{(1)}(1/x))}\right]-\Li
\left[e^{i(\tilde p^{(2)}(1/x)+\tilde p^{(3)}(1/x)-\tilde
p^{(1)}(1/x))}\right]\right)\no \\
&-\int_{-1}^1\frac{dz}{2\pi\ep}\left(\Li \left[e^{i(\hat
p^{(2)}(x)+\hat p^{(3)}(x)-\hat p^{(1)}(x))}\right]-\Li
\left[e^{i(\tilde p^{(2)}(x)+\tilde p^{(3)}(x)-\tilde
p^{(1)}(x))}\right]\right)\;,
\label{eq:mirint}
\end{align}
%
where $\hat p_i(z),\ \tilde p_i(z)$ with $i=1,2,3$ are the AdS and the
sphere parts of the quasi-momenta of the three operators respectively.
Contrary to \cite{BKV}, here it is the sphere part of the
quasi-momenta which is non-trivial,
  \begin{align}
 \tilde p^{(1)} (x) = {\textstyle{1\over 2}} { \Delta }\, p(x) -
 \CG_\uu(x) , \no \quad \tilde p^{(k)}(x) = {\textstyle{1\over 2}}
 {L_k }\, p(x), & \qquad k=2,3\,, \\
 \hat p^{(1)} (x) = {\textstyle{1\over 2}} { \Delta }\, p(x)\,, \qquad
 \hat p^{(k)}(x) = {\textstyle{1\over 2}} {L_k }\, p(x), &\qquad
 k=2,3\,,
  \end{align}
  where $ \ p(x) = { \, x /g( x^2- 1)}$.  In writing (\ref{eq:mirint})
  we also have used $p(1/x)=-p(x)$, as well as the functional equation
  of the dilogarithm.  The integration path in \eqref{eq:mirint} is
  understood as follows: choosing the determination $1/x(z)$ for the
  \Zh variable in the first line means that we integrate below the
  cut, on the contour $U_-$, while choosing the determination $x(z)$
  means we are integrating on the contour $U_+$ above the cut.  The
  integral \eqref{eq:mirint} can be recast as a contour integral on
  the contour $U$ surrounding the cut counterclockwise,
\begin{align}
\!\!\!\!\log [\CC^{\bullet\circ\circ}]^{\rm
bottom}&=\oint_U\frac{dz}{2\pi\ep}\left(\Li \left[e^{i(\hat
p^{(2)}(x)+\hat p^{(3)}(x)-\hat p^{(1)}(x))}\right]-\Li
\left[e^{i(\tilde p^{(2)}(x)+\tilde p^{(3)}(x)-\tilde
p^{(1)}(x))}\right]\right)\;.
\label{eq:mircontint}
\end{align}
This result agrees with the second line of the strong coupling string
result \eqref{strongsu2wr} obtained in \cite{Kazama:2016cfl}.

\subsection{The $\sl(2)$ bottom mirror contribution}

The computation of the wrapping corrections in the $\sl(2)$ sector
goes along the lines of the computation in the $\su(2)$ sector, the
only difference being that the hexagon amplitude $h(u,v)$ and transfer
matrix get replaced with their $\sl(2)$ counterparts.  The measure
$\mu(u)$ and the factor $H_{a,b}(u,v)$ are exactly the same as in the
$\su(2)$ case, while $h_{\sl(2)}(u,v)$ is defined in \eqref{hsl2}, so
that
\begin{align}
\label{defgsl2}
g_a^{\;\sl(2)}(u)\equiv\frac{e^{ip_a (u)\ell_B
}}{\hh^{\;\sl(2)}_{a1}(u,\mathbf{u})}=\frac{R^{(-)[-a]}}{R^{(-)[+a]}}\;\tilde
g_a^{\;\su(2)}(u)\;.
\end{align}
We deduce that the analyticity properties of $g_a^{\;\sl(2)}(u)$ are
the same as the ones for $\tilde g_a^{\;\su(2)}(u)$.  The transfer
matrices $T_a^{\;\sl(2)}(u)$ are defined in terms of the generating
functional
 \begin{align}
 \label{genfnlsl}
\sum _{a=0}^\infty
(-1)^a {T}_{a} ^{[a-1]}(u)
\ID^{2a}=\left(1-Y_{0,1}\ID^2\right)^{-1}\left(1-X_{1,1}\ID^2\right)
\left(1-X_{2,1}\ID^2\right)\left(1-Y_{2,2}\ID^2\right)^{-1}\,,\\
\no \text{with}\quad X_{2,1}=X_{1,1}=\frac{R^{(+)-}}{R^{(-)-}} = f^-,
\quad Y_{2,2}= 1, \quad Y_{0,1}= \frac{B^{(+)+}}{B^{(-)+}}
\frac{R^{(+)-}}{R^{(-)-}} ={f^- \over \bar f^+},
 \quad \ID^2=e^{i\ep \partial_u}\,.
\end{align}
The explicit expression of the $\sl(2)$ transfer matrices, correctly
normalised, is more complicated than \eqref{newnorm} and it is given
by equation (H1) in \cite{BKV}.  Even if it's not immediately obvious,
the matrices ${T}_{a} (u)$ have only two cuts situated at distance
$ia\ep$.  The remaining cuts vanish by virtue of the symmetry of
${T}_{a} (u)$ with respect to the exchange $x^{[k]}\leftrightarrow
1/x^{[k]}$ for $k\neq \pm a$.  Again, we can redefine the quantities
\begin{align}
\label{Tnormsltwo}
\rT_a ^{\;\sl(2)}(u)\equiv\left[g_a(u)T_a(u)\right]^{\;\sl(2)}
\end{align}
which have the required analytical properties to allow continuation to
the mirror dynamics.  In the strong coupling limit, the matrices
$\rT_a(u)$ defined as above obey again the $\rt $-$\rT$ relations
\eqref{eq:recursive}, this time with
\begin{align}
\rt_n ^{\;\sl(2)}=(-1)^{n+1}(g/f)^n(2-f^n-\bar f ^n)\;.
\end{align}
The sign $(-1)^n$ compensates the one in \eqref{eq:intb}, while the
extra minus sign accounts for the exchange of the sector carrying the
non trivial excitations.  All in all, the result of summing out the
mirror excitation gives an expression identical to \eqref{eq:mirint}
and \eqref{eq:mircontint}, with the quasi-momenta being now
  \begin{align}
  \label{semipsltwo}
 \hat p^{(1)} (x) = {\textstyle{1\over 2}} { \Delta }\, & p(x) +
 \CG_\uu(x) , \no \quad \hat p^{(k)}(x) = {\textstyle{1\over 2}} {L_k
 }\, p(x), \qquad k=2,3 \\
 & \tilde p^{(k)}(x) = {\textstyle{1\over 2}} {L_k }\, p(x), \qquad
 k=1,2,3\,,
  \end{align}
Again, the result agrees with the second line of the strong coupling
string result \eqref{strongsl2wr} obtained in \cite{Kazama:2016cfl}.

  \section{Fredholm determinants and free fermions
  \label{subsec:fermitree}}
  
  As it was already remarked in \cite{BKV}, at strong coupling the
  bi-local scalar factors become the same as those at tree level.
  This allows to recycle the techniques developed for the tree-level
  structure constant in \cite{GSV:Tailoring3,3pf-prl,sz,SL,Bettelheim:Semi}.  The
  techniques in question, which we will refer to as determinant
  methods, reformulate the problem in terms of several equivalent
  objects, Fredholm determinants, free fermions and chiral Toda
  theory, which are related to the fermionic system by bosonization.

Although we already obtained the solution of our problem by the
combinatorics of clusters, it is potentially useful to give an
alternative derivation, which will recast the solution in a nice
operator form and potentially give us intuition about how to adjust
our method to more general class of correlation functions.  To our
surprise, the determinant methods work with remarkable efficiency for
the resummation of the mirror channel in section
\ref{sect:clustmirpart}, where they led naturally to an object which
could be named ``quantum spectral determinant".  The quantum spectral
determinant is the determinant in the functional space of a
finite-difference operator, obtained by replacing the spectral
parameter in the spectral (super)determinant with the shift operator
$\ID^2$.  The finite-difference operator in question already appeared
in the literature \cite{Tsuboi-SmatGF, Kazakov:2007fy,
Zabrodin:2007rq} in the guise of exact generating function for the
transfer matrices, {\it cf.} (\ref{genfnl}) and (\ref{defgsl2}).  Our
analysis shows that this operator is not just a formal expansion, but
has a deeper meaning and will certainly play important role in
building analytic methods to study the correlation functions.

It is worth mentioning that the fermionic formalism allows to reveal a
hidden integrable structure of the structure constant and identify it
(at strong coupling) as a $\tau$-function of the KP integrable
hierarchy\footnote{Such a statement has been made in
\cite{Foda:2012wf}; we evoke a different realization of the
$\tau$-function.}.  This means that the structure constant satisfies
an infinite series of non-linear PDE as a function of the conserved
quantities characterizing the operators $\CO_1, \CO_2$ and $\CO_3$.
This integrable structure is not really necessary in the context of
this work, because we have already found the explicit solution, but it
could be useful for studying more complicated objects as the  
I-I-I type structure constant.

In this section we will focus on the description {\it via} Fredholm
determinants.  We will show that not only the leading term in the
semiclassical limit, but the whole semiclassical series can be given a
condensed formulation in terms of a Fredholm determinant.  We will
only briefly mention the fermionic and the bosonic QFT realizations of
the structure constant.  In particular, we will give an interpretation
of the semiclassical result (\ref{weak-semiclassical}) as the grand
canonical partition function of free fermions living on a line.  More
details about the representation in terms of chiral fermions can be
found in \cite{Bettelheim:Semi,generalSU2scpr}.
 Our goal in this section is mainly to give an intuitive explanation
 of the combinatorial factors obtained in sections \ref{sec:asympA}
 and \ref{sec:bottommirror}.

  \subsection{ Multiple contour integrals as a Fredholm determinant
  \label{sec:Fredh}}

  With the help of the Cauchy identity
    \be
    \label{Cauccyid}
 {1\over \ep^n} \prod_{j<k}^n {(z_j-z_k)^2\over (z_j - z_k)^2 + \ep^2}
 = \det_{j,k} \left(\frac{i}{z_j-z_k+i\epsilon} \right) \ee
the $n$-th term in the expansion (\ref{eq:multiC}) can be put in the
form
 \begin{align}\label{fermigastree}
 I_n=\frac{(-1)^n}{n!}\int\limits _\CC\prod_{j=1}^{n}\frac{dz_j}{2\pi
 i} \, \det \left(\frac{ \ff(z_j)}{z_j-z_k+i\epsilon} \right) .
 \end{align}
   Then the sum $\caA = 1+I_1+I_2 + \dots$ in (\ref{eq:multiC}) takes
   the form of the expansion of a Fredholm determinant,
  \be
  \label{defFredhdet}
  \caA=\Det (1-K) \equiv \sum_{n=0}^\infty {(-1)^n\over n!}\,
  \oint_{\CC^{\times n}} {d z_1\dots dz_n } \ \det_{jk}K(z_j, z_k)
  \ee
with the integral operator $K$ defined as \footnote{In the particular
case when $\ff $ is a constant in the interval $[-L/2, L/2]$ and
vanishes outside this interval, the Fredholm determinant has been
computed for by Michel Gaudin \cite{Gaudin1966545}.  }
\begin{align}
\label{FKernel}
K\psi(u) = \oint_\CC {dz} \ K(u,z)\, \psi(z) , \qquad K(u, v) = \
{1\over 2\pi i}\, \frac{\ff(u)}{u-v+i\epsilon }\,.
\end{align}
 The logarithm of the Fredholm determinant is a series of multiple
 integrals
\begin{align}
\label{tracelogK}
 \log \Det(1-K) &= - \oint _\CC dz\, K(z, z) - {1\over 2} \oint
 _{\CC^{\times 2}} dz_1 dz_2 \, K(z_1, z_2)K(z_2, z_1) \no
\\
& - {1\over 3} \oint _{\CC^{\times 3}} dz_1 dz_2 dz_3 \, K(z_1,
z_2)K(z_2, z_3)K(z_3, z_1)+\dots
\end{align}
 The Fredholm operator in (\ref{FKernel}) can be represented in the
 factorised form
\be
\label{operK}
K = \ff \, \ID^2\, \IPplus \ee
where $\ID^2= e^{i\ep \p}$ is the shift operator and $\IPplus$ is the
Cauchy transform
\be
\label{actPplus}
\IPplus\, \psi(u) = \oint _\CC{dz} \ {1 \over 2\pi i}\, {1\over u-z+
i0}\ \psi(z)
. \ee
%

The semiclassical limit of (\ref{tracelogK}) has been studied in
\cite{Bettelheim:Semi}, where the first two terms of the semiclassical
expansion
\be \log \caA = {1\over \ep} \caF_0 + \caF_1 + \ep\, \caF_2 +\dots \ee
 have been obtained.  Compared to the analysis in
 \cite{Bettelheim:Semi}, the method of clustering described in section
 \ref{subsec:deformation} is more powerful because it allows to
 obtain, after some combinatorics, the whole semiclassical expansion.
%
%
%

\subsection{The leading term by Fredholm determinant}

The operator formalism provides an alternative (shorter, but not
rigorous) derivation of the leading term of the semiclassical
expansion.  It is based on the following approximation for the $n$-th
power of the operator $K$,
 \be
 \label{Kfindiff}
 K ^n \equiv (F\ID^2 \IPplus)^n = (F\ID^2)^n \IPplus + \
 \text{subleading}.
 \label{KffD}
 \ee
 Replacing $\IPplus $ with the identity\footnote{Imagine that the
 contour $\CC$ is deformed so that the new contour obtained from $\CC$
 by translation $z\to z+i\ep$ is inside $\CC$.  For that the contour
 should be stretched to $+i\infty$.  Then the operator $\IPplus$ is
 the identity operator in the space $\CH_+$ of the functions analytic
 inside $\CC$.  } on the rhs is equivalent to retaining only the
 contribution of the poles at $z_{j+1} = z_j + i\ep$.  The operator
 $K^n$ in this approximation is an integral operator with kernel
  \be K^n(z_1, z_2) ={1\over 2\pi i}\, {F_n(z_1)\over z_1-z_2+ i n
  \ep}, \ee
 where the function $F_n (u)$ is defined in (\ref{eq:wavef}).  Now we
 can evaluate the sum in (\ref{tracelogK}) as
 \be
 \label{classleadingK}
 \begin{aligned}
 \Tr \log (1- K) = - \sum_{n=1}^\infty \int dz\, K^n(z,z) = {1\over
 \ep} \oint_\CC {dz\over 2\pi} \sum _{n\ge 1} {F_n(u)\over n^2} \, ,
 \end{aligned}
 \ee
 which coincides with the expression (\ref{leadingA}) obtained by the
 clustering method.  This operator representation will be particularly
 useful when computing the wrapping corrections.

\subsection{The leading term as the free energy of a Fermi gas}

The above result can be given an intuitive explanation by the analogy
with a grand ensemble of one-dimensional fermions living on the
contour $\CC$.  We place for simplicity the integration contour $\CC$
along the real axis, from left to right, and close it at infinity in
the upper half-plane.  This simplifying assumption does not change the
short distance behavior, in particular the mechanism for clustering.
The Cauchy transform (\ref{actPplus}) acts as the projection operator
to the functions analytic in the upper half-plane, which we choose as
our Hilbert space.  Such are the wave functions of the right-moving
fermions.  The Hilbert space $\CH_+$ of the right-movinf fermions is
spanned by the plane waves
  \be
  \label{planewaves}
  \psi_k(z) \equiv \langle z|k\rangle=e^{i k z/\ep } \quad \text{with}
  \ k> 0, \ee
while the dual Hilbert space $\CH_-$ of the left-moving fermions is
spanned by the plane waves
  \be
  \label{planewaves1}
  \bar \psi_k(z) \equiv \langle k|z\rangle=e^{-i k z/\ep } \quad
  \text{with} \ k> 0.  \ee
  The plane wave decomposition of the identity operator in $\CH_+$ is
\be
\label{defPplus}
\int_0^\infty |k\rangle \langle k| = \IPplus, \quad \langle z_1|
\IPplus|z_2\rangle \equiv \int_0^\infty {dk\over 2\pi\ep} \, \langle
z_1 |k\rangle \langle k| z_2\rangle = {1\over 2\pi i}\ {1 \over
z_1-z_2+ i0} .  \ee
  The fermions are allowed to have only positive energy $k$ and their
  wave functions are the plane waves (\ref{planewaves}).  The grand
  canonical partition function of the fermionic ensemble is given by
  the Fredholm determinant (\ref{defFredhdet}), with the Fredholm
  kernel (\ref{FKernel}) playing the role of a density matrix in
  coordinate representation:
      \be\label{rhocoordcoord} K(z, u)
    = \int_0^\infty {dk \over 2\pi \ep}\ \, \ff(z)\, \langle z
    |k\rangle \, e^{-k} \, \langle k|u\rangle .  \ee
     The coordinate $u$ and the momentum $\hat k = -i \ep \p/\p u$ are
     canonically canonically conjugate operators
\begin{align}\label{canonical}
[u, \hat k]=i \epsilon\,
\end{align}
and the limit $\epsilon\to 0$ corresponds precisely to the
semiclassical limit of this free fermion system.  In the semiclassical
limit the logarithm of the function $F$ can be considered as an
external potential which adds to the kinetic energy $k$ of the
fermion:
 \begin{align}
 \label{Eclass}
E_{\rm cl} (u,k) = k - \log \ff(u) .
\end{align}
 As is well-known, the grand potential in the semiclassical limit is
 given by the integral over the phase space of the right-moving
 fermions,
\begin{align}
\label{fermienergy}
\ln \caA\simeq -\frac{1}{ \epsilon}\oint\limits _{\CC}{du\over 2\pi}
\int_0^{\infty} dk \,\ln \left(1- e^{-E_{\rm cl}(u,k)} \right) \, .
\end{align}
Performing the integration over $k$, we reproduce the leading term
in  (\ref{weak-semiclassical-all}).   Note that the discussion here is in complete
parallel with the Fermi gas approach to the ABJM matrix model
\cite{M-P-ABJM}.  In that context, the subleading corrections can be
determined by the Wigner-Kirkwood expansion \cite{Wigner-K,
W-Kirkwood}, which can be carried out {\it e.g.} by the method of
co-adjoint orbits \cite{Dhar:1992hr}.  It would be an interesting
future problem to apply it to the three-point function and try to
compute the subleading corrections
systematically\footnote{Conversely, we can apply the clustering method
to the AB JM partition function and obtain the leading term in the
M-theory regime.  This is demonstrated in Appendix \ref{ap:abjm}.  }.

 \subsection{Semiclassical expansion by nested Fredholm determinant }
 
 It happens that the exact result (\ref{eq:exactC}) obtained by the nesting method 
can be expressed in terms of a (different) Fredholm determinant.
Applying the Cauchy identity as in (\ref{Cauccyid}), we can write
the integrand/summand in the $n$-th term of the series
(\ref{eq:exactC}) as 
 \begin{align}
 \label{gendetK}
 \prod_{ j=1}^n {F_{q_j}(z_j)\over \ep\, q_j^2}
 \prod_{j<l}^n\Delta_{q_j,q_l}(z_j,z_l) =(-1)^n \det_{j,l} K_{q_j,
 q_l}(z_j,z_l),
   \end{align}
 where the matrix kernel $\hat K= \{K_{q_1,q_2}(z_1,z_2)\} _{q_1,q_2 \ge 1}$ is defined by
 \be
 \label{matrixkernel}
K_{q_1,q_2}(z_1,z_2) ={1\over 2\pi i }\, {F_{q_1} (z_1) \over
q_1}{1\over z_1- z_2 + i\ep q_1}
. \ee
This turns  (\ref{eq:exactC}) into the expansion of a generalised
Fredholm determinant $\Det(1-\hat K)$.
In addition, this is a "nested Fredholm determinant" in the sense that
the integration for the $n$-th term of the expansion is performed for
a nested configuration of contours    $\CC_1\insde\CC_2\insde \dots\insde 
\CC_n$ associated with the contour $\CC$.  The relation
$\CC_1\insde \CC_2$ means that $\CC_1$ is inside $\CC_2$.\footnote{This
also means that the two contours remain separated at finite distance
in the limit $\ep\to 0$.} All contours $\CC_k$ from the nested
configuration are obtained from the contour $\CC$ by a continuous
deformation without crossing poles or other singularities of the
function $F(z)$.

It is important that   the matrix elements of
the generalised Fredholm kernel (\ref{matrixkernel}) depend only on
the first index $q_1$.  This allows to replace the matrix kernel $\hat K$
with a scalar kernel, but of different functional form.
 Indeed, in the expansion 
 \begin{align}
 \log\Det(I- \hat K )_{\nested} =& -\sum_{q\ge 1} \oint _\CC dz \,
 K_{q,q}(z,z) \no \\
 &- {1\over 2} \sum_{q_1,q_2\ge 1} \oint_{\CC_1\insde  \CC_2} \!\!\!\!
 dz_1\, dz_2\ K_{q_1,q_2}(z_1,z_2) K_{q_2,q_1}(z_2,z_1)-\dots
 \end{align}
  the sums over $q_i$ decouple and the matrix kernel $\hat K$ can be
  replaced with a scalar kernel
 \be
 \label{defIKK}
 \qquad \IIK (z_1,z_2) = {1\over 2\pi i}\ \sum_{q =1}^{\infty}
 {F_q(z_1) \over q} \ {1\over z_1- z_2 + i q \ep}.  \ee
As a consequence, the quantity $\caA$ is given by a nested Fredholm
determinant of the scalar kernel $\IIK$:
 \begin{align}
\label{generalFa}
 \caA \ = \ \ \Det ( \mathrm{I} -\IIK)_\nested\, .
  \end{align}
%
A convenient operator expression for $\caA$ is obtained by
representing the kernel $\IIK$ in a factorised form, as we did for $K$
in (\ref{operK}),
\be \IIK = \sum _{q=1}^\infty {F_q\over q} \ \ID^{2q} \, \IPplus =\sum
_{q =1}^\infty { (F \ID ^2)^{q}\over q}\, \IPplus = - \log(1- F \,
\ID^2)\cdot \IPplus \, .  \ee

\subsection{CFT  representation}

  The representation as a determinant enables us to use the formalism
 of quantum field theory and 2D conformal field theory as in the case
 of random matrix models \cite{Kostov:1999xi}.  To formulate the
 semiclassical expansion in  QFT terms, we first identify the
 fermionic system associated with the determinant (\ref{generalFa})
 and transform it into a bosonic collective field theory.
Let us first remind the representation, given in \cite{KKN}, of the
original ``unnested'' Fredholm determinant (\ref{defFredhdet}) as an
expectation value of two-dimensional chiral fermion $ \psi, \bar\psi$
with with two-point function $ \langle 0| \bar \psi(z) \psi(u)|0
\rangle = {1/( z-u)}$ and interacting with a common external potential
determined by the function $F$.  We have (for the details we refer to
\cite{Bettelheim:Semi, KKN} )
\begin{align}
\label{eq:exactCopb}
 \caA=& \langle 0| \mathrm{exp} \oint_{\CC } {dz\over 2\pi i} { F(z)\
 \bar \psi(z) \ID^2 \psi(z ) }|0 \rangle \, .
\end{align}

The nested Fredholm determinant (\ref{generalFa}) has the
Fock space representation
\begin{align}
\label{eq:exactCop}
 \caA=& \langle 0| \, \left( \exp \int_{\CC } {dz\over 2\pi i} { \bar
 \psi(z) \log(1-F(z)\ID^2)\psi(z ) } \right)_\nested |0 \rangle \, ,
\end{align}
where the ``nested exponent'' is defined as
\be \left( \exp \int_{\CC } {dz } f(z) \right)_\nested = \sum_{k\ge 0}
{1\over k!} \ \oint\limits_{\CC_1\insd  \dots \insd  \CC_k} {dz_1\dots
dz_k } \ f(z_1) \dots f(z_k).  \ee
This simple fermionic system can be transformed into a chiral
Toda-like theory by two-dimensional bosonization, $\psi\to e^{\phi},
\bar\psi\to e^{-\phi}$.  In the bosonic theory, the $q$-clusters are
represented by the vertex operators
\be V_q(z) = e^{-\phi(z) } e^{\phi(z+ i q\ep)} \ee
which are electrically neutral but have dipole charge $q\ep$.  The
function $\Delta_{mn}(u,v)$ defined by (\ref{eq:delta}) is the
correlation function of the dipoles $V_m(u)$ and $V_n(v)$.  The
leading term is the free energy of the gas of dipoles in the external
potential $- \log F$, in the dilute gas approximation.  Taking into
account the interaction generates a series of connected graphs which
is the QFT formulation of the cumulant, or Mayer, expansion.  For
instance, the subdominant term in (\ref{weak-semiclassical-all}) is
given by the connected correlation function of two vertex operators.

 \subsection{Wrapping corrections   by  Fredholm determinant
 \label{subsec:boundstateFredholm}
 }
  
  The sum over the mirror particles has a form very similar to the
  asymptotic contribution after being reformulated in terms of
  multiple contour integrals.  It is possible that this latter
  formulation is more fundamental than the sum over partitions.
  Remarkably, the strong coupling limit of the contribution of the
  mirror particles associated with the bottom edge, eq.
  (\ref{eq:Bottom}), can also be converted to a Fredholm determinant.
  The derivation is again based on converting the bi-local products
  into determinants with the help of Cauchy identity.  All statements
  and derivations in this subsection are valid only to the leading
  order in the semiclassical strong coupling limit.  It is of course
  desirable to develop these techniques beyoud the leading order, but
  this is far beyond the scope of the research presented in this
  paper.  Since we are interested only in the leading behavior, we can
  rescale $\ep$ so that $2g\ep=1$ and then take the limit $\ep\to 0$.

   Let us consider for definiteness the case $\sutwo$.
  The analytical properties of the T-matrices allow to perform a shift  
  $T_a\to T_a^{[-a]}$ in the original integral along the real axis in  
  the mirror dynamics. 
  This leads to (\ref{eq:intbstrong}) with  $z^a\to z^a +
i a\ep/2$. After the shift, the bi-local factors $\Delta_{ab}$, which give the
asymptotics of $H_{a,b}^\g$, transform to
$$
\Delta_{ab}(z-ia \ep/2, z- i b \ep/2)\to \Delta_{ab}(z- i a \ep, z' -
i b\ep) =\Delta_{ba}(z, z')
$$
and (\ref{eq:intbstrong}) can be written equivalently as\footnote{In
the $\sltwo $ case the factor $(-1)^n$ in (\ref{eq:Bottom}) is not
compensated and should be added in (\ref{eq:Bottomdet}).  This will
lead to a factor $(-1)^a$ in (\ref{defrKa}), (\ref{KTa}) and
(\ref{sumTaC}).  }
\begin{align}
\label{eq:intbstrongbis}
\rB_{sc}[\vec{n}]\to \int
_{U_-}\prod_a\prod_{j=1}^{n_a}\;\frac{dz_j^a}{2\pi a \ep}\;\rT_a^{
[-a]}(z_j^a)\!\!\!\prod_{\substack{a \\ 1\le i<j\le n_a}}
\!\!\!\Delta_{a,a}(z_i^a,z_j^a) \times \!\!\prod_{\substack{a<b \\
1\le i\le n_a\\ 1\le j\le n_b}}\!  \!\Delta_{b,a}(z_i^a ,z_j^b ).
\end{align}
The next step is to write the re-organise the sum of all $B[\vec n]$
in (\ref{eq:Bottom}) as
 a sum over all possible sets $I$ of double indices $\{i, a\}$, with
 $a$ and $i$ being otherwise unrestricted positive integers.
Then for any set $I$ we express the product of the factors $\Delta $
in the rhs of (\ref{eq:intbstrong}) as a Cauchy determinant of size
$k=|I|$.  This allows to write the sum over all $B[\vec n]$  as
\begin{align}
\label{eq:Bottomdet}
 [\mathcal{C}_{123}^{\bullet\circ\circ} ]^{\rm bottom}_{\su(2)} &
\to \sum_{k=0}^\infty \ {1\over k!} \sum_{ |I|=k} \int_{U_-} \prod_{\{
i, a\}\in I}\frac{dz_j^a}{2\pi i}\; \ \det_{k\times k}{\rT^{[-a]}_a
(z_j^a)\over z_i^a - z_j^b - i a \ep }.
\end{align}

 Reasoning in the same way as in the derivation of (\ref{generalFa}),
 we express the right-hand side as a Fredholm determinant
 \begin{align}
\label{generalF}
\text{rhs of } (\ref{eq:Bottomdet}) \ = \ \ \Det \left( \mathrm{I} +
\sum_{a =1}^{\infty} {\rK}_{a} \right),
  \end{align}
where the term $\IK_a$ in the Fredholm kernel is associated with the
transfer matrix $\rT_a$:
  \begin{align}
  \label{defrKa}
 {\rK}_{a} (z_1,z_2) = {1\over 2\pi i} \ \frac{
 \rT_a^{[-a]}(v)}{u-v-ia\epsilon }\, .
  \end{align}

The Fredholm determinant (\ref{generalF}) can be computed with the
help of the explicit expression for the generating function for the
transfer matrices, generating function for the transfer matrices,
(\ref{genfnl}) for $\sutwo$ or (\ref{genfnlsl}) for $\sltwo$.   For that we
represent the integral operator with kernel (\ref{defrKa}) can be written in a
factorised form
  \begin{align}
  \label{KTa}
 {\rK}_{a} = \rT_a ^{[-a]} \ \ID^{-2a} \, \IPminus \,
  ,
  \end{align}
where the operator $\IPminus$, which plays the role of the identity
operator,\footnote{Up to exponentially smal terms the operator
$\IPminus$ is the identity operator in the space $\CH_-$ of functions
analytic in the lower half-plane and decaying exponentially for $
x<-1$ or $x>+1$ on the real axis (in the normalization $2g\ep = 1$).}
acts as
\be
\label{defPm}
\IPminus \psi(u) = \int_{U_-} {dz \over 2\pi i} \ {\psi(z) \over u-z - i
0}.  \ee
Now the sum in (\ref{generalF}) turns out to be exactly the generating
function for the transfer matrices, multiplied from the right by the
operator $\IPminus$,
\be
\label{sumTaC}
 [\mathcal{C}_{123}^{\bullet\circ\circ} ]^{\rm bottom}_{\su(2)}
\to \, \Det \left(\sum_{a=0}^\infty \rT_a^{[-a]}\, \ID ^{-2a} \,
\IPminus \right).  \ee

 Both generating functions are products of operators whose
 determinants have been already computed in the semiclassical limit
 $\ep\to 0$.    This fact allows us to write the determinant
 (\ref{generalF}) as a product of simpler determinants for which we
 already know the semiclassical limit. (Since the operators act in a functional space, 
 the factorisation to a product of
 determinants is not exact, but it is corrected by subleading factors.)
Let us see that this indeed
 reproduces the result obtained by clustering.
 
 {\bf The $\su(2)$ bottom mirror contribution.} To evaluate the sum in
 (\ref{sumTaC}), we take the complex conjugate of the generating
 function (\ref{genfnl}),
 \begin{align}
 \label{genTnormexact}
 \sum_{a=0}^\infty \rT_a^{[-a]} \, \ID^{-2a} & = \left(1- g \,
 \ID^{-2}\right) \left(1- { g f^{-1}} \, \ID^{-2}\right)^{-2} \left(
 1- { g } f^{-1} \, \ID^{-2} \bar f\right) .
\end{align}
 where $f= R^{(+)}/R^{(-)}$, $\bar f = {B^{(-)}/B^{(+)}}$ and the
 function $g(u)$ is defined (\ref{defgsutwo}).  Following the
 prescription of section \ref{sec:boundstrong}, which is valid at
 strong coupling, will first take the semiclassical limit of these
 functions and then continue analytically to the lower edge of the \Zh
 cut.  This amounts to replacing
 \begin{align}
 \sum_{a=0}^\infty \rT_a ^{[-a]}\, \ID^{-2a} & \to\ \left(1-\bar
 Y^{-1} \ID^{-2}\right) \left(1- X^{-1}\ID^{-2}\right)^{-2} \left( 1-
 Y^{-1}\ID^{-2} \right)
\, , \qquad u \in U_-
\label{genTnormexactcl}
 \end{align}
where
%
  \be\begin{aligned}
\label{exps}
  \!\!\!\!  { Y(x)}
   \to e^{i\tilde p_2(x)+i\tilde p_3(x)- i\tilde p_1(x)} , \
  {\bar Y(x)}
  \to e^{i\tilde p_1(1/x)-i\tilde p_2(1/x)- i\tilde p_3(1/x)} ,
  \ X (x)
\to e^{ - i p(x) \, {\Delta _1 -L_2-L_3\over 2} } .  \end{aligned}
\ee
In the limit we are concerned with, the last factor in
(\ref{KTa})  acts as   the identity operator
and the product (\ref{genTnormexactcl}) is a product of operators 
of the type (\ref{Kfindiff}) whose determinants we know how to compute.
Since in the leading approximation the determinant of a product of operators 
 equals the  product of their determinants,  the
contribution of the bottom particles is given by the product 
\begin{align} \label{Bsupdet}
  [\mathcal{C}_{123}^{\bullet\circ\circ} ]^{\rm bottom}_{\su(2)} &
 \to \, \textstyle{ {\Det\big(1- \bar Y^{-1}\, \ID^{-2}\big)\
 \Det\big(1- Y^{-1} \, \ID^{-2}\big) \over \Det\big(1- X^{-1} \,
 \ID^{-2}\big)\ \Det\big(1- X^{-1} \, \ID^{ -2}\big)} } \no \\
 & \simeq \textstyle{ {\Det\big(1- \ID^{2}\, \bar Y\big)\ \Det\big(1-
 \ID^{2}\, Y \big) \over \Det\big(1- \ID^{2} X\big)\ \Det\big(1- \ID^{
 2}X\big)} }.
 \end{align}
Evaluating each of the factors   by
(\ref{classleadingK}),  we write
\begin{align} \label{Bottomqc}
 \log [\mathcal{C}_{123}^{\bullet\circ\circ} ]^{\rm bottom}_{\su(2)} &
 \to \, {1\over \ep} \int_{U_-} {du\over 2\pi }\ \left(\Li[ Y(x)] +
 \Li[ \bar Y(x)]- 2 \Li[ X(x) \right) .
 \end{align}
Using the relations $\bar Y(u-i0)= - 1/ Y(u+i0)$ and $X(u-i0)=
1/X(u+i0)$, this linear integral can be further expressed as a contour
integral around the \Zh cut (or along the unit circle in the
$x$-plane)
\begin{align} \label{Bottomqccontour}
\log [\mathcal{C}_{123}^{\bullet\circ\circ} ]^{\rm bottom}_{\su(2)} &
\to \, {1\over \epsilon } \oint\limits_{ U} {du\over 2\pi }\
\Big(\Li(Y(x)) - \Li(X(x)) \Big)
 \end{align}
which coincides with the semiclassical expression
(\ref{eq:mircontint}) obtained by clustering.
  
We have found that the contribution of the bottom mirror particles is
given in the strong coupling, semiclassical limit by the determinant
the generating function for the transfer matrices considered as a
difference operator:
\be  \label{C123mir-QSC}
 [\mathcal{C}_{123}^{\bullet\circ\circ} ]^{\rm
bottom}_{\su(2)} &= \Det^{-1} \Big( {\rm sdet}\left( 1- G\,
\ID^{2}\right) \Big) , \quad G= \text{diag}({{X}},{{ X}}\, | \, {{Y}},
{{\bar Y}}).  \ee
 When the operator $\ID^2$ is considered as a classical expansion
 parameter as in (\ref{eq:Tt}), the superdeterminant in the
 parenthesis gives the equation of the classical spectral curve $W(z,
 u) \equiv \mathrm{sdet}\left( z- G \right (u) ) =0$.  The fact that
 the spectral curve appears as an operator (sometimes called ``quantum
 spectral curve") is very intriguing and merit to be better
 understood.

  {\bf The $\sl(2)$ bottom mirror contribution.} Proceeding in the
  same way as in the $\sutwo$ case, we first write the $\sltwo$ analog
  of (\ref{eq:intbstrongbis}), which differs only by a sign $(-1)^n$.
  Then we express the determinant in terms of the generating function
  for the T-matrices (\ref{Tnormsltwo}), which can be written in the
  form
\begin{align}
\label{eq:Wshift}
\sum_{a\ge 0} (-1)^a \rT_a^{[-a]}\, \ID^{-2a}&= \left(
1-g^-\ID^{-2}\right)^{-1} (1- g^-\ID^{-2} f)^2 \left(1- g^-\bar f
^{-1} \ID^{-2} f\right)^{-1} .
\end{align}
 Evaluating the determinant as in the $\sutwo$ case and taking into
 account the asymptotics (\ref{asymga}) and (\ref{fbarfmir}), we
 obtain an expression identical to (\ref{eq:mircontint}), with the
 quasimomenta given by (\ref{semipsltwo}).

 There is no principal difficulty in computing the  subleading term 
 on the rhs of  (\ref{Bottomqccontour}), which is of order $\ep^0$.
 The subleading  comes from the  $\ep^2$ correction  to  the  
 approximation  (\ref{Habappr}), as well as from the subleading 
 term in the expansion of the 
 logarithm of the functional determinant (\ref{C123mir-QSC}).
 The latter will contain   diagonal terms of the form
  (\ref{weak-semiclassical-all}) with $F $
 replaced by   $X$ and $ Y$, 
as well as  a cross term, which represent a correction to the 
factorised form  (\ref{Bsupdet}) of the determinant.

\section{Conclusion and outlook
\label{sec:conclusion}}
In this paper, we developed two different methods to take the
semiclassical limit of three-point functions in the $\mathfrak{su}(2)$
and $\mathfrak{sl}(2)$ sectors of $\mathcal{N}=4$ SYM theory.  The
first method is based on contour deformations of the multiple contour
integral representation of the structure constant, which leads to
formation of bound states, a phenomena which we call clustering.  The
second method applies the generalised Fredholm determinant and the
Fermi gas approach.  Using these methods, we analyse in detail the
semiclassical limit of two configurations of three-point
functions.\par

For the type-I-I-II case where all the three operators are non-BPS, we
take the semiclassical limit of the asymptotic structure constant at
any coupling both for $\mathfrak{su}(2)$ and $\mathfrak{sl}(2)$
sectors.  For the simpler case where two operators are BPS and the
third is non-BPS we are able to achieve more.  Apart from the
semiclassical limit of the asymptotic structure constant, which is a
special case of the previous type, we are able to take into account
the contribution from all the bottom mirror excitations.  The multiple
contour integral is considerably more complicated compared to the
asymptotic part and we can only take the semiclassical limit at strong
coupling.  We compare our results of both configurations at strong
coupling with the string theoretic calculation and find that they
match precisely.\par

Our work leaves many interesting open questions to be explored in the
near future.  One immediate problem is to take into account the
contributions from mirror excitations on all the three mirror edges in
order to obtain a complete result of the structure constant.  It is
known that when the rapidities of two mirror excitations on two
different mirror edges coincide, the integrand is divergent.  Once
these divergences are properly resolved, we can further develop our
method and try to reproduce the full result at strong coupling.\par

The results obtained in this work for asymptotic part are valid at any
coupling while for the bottom mirror contribution we only take the
semiclassical limit at strong coupling.  It is desirable to fix this
imbalance by studying in more detail the semiclassical limit for the
mirror parts at finite coupling.  The \Zh cut disappears at finite
coupling since $g'=g\ep \to 0$, but it is not clear whether the
wrapping correction will disappear as well.  We are planning to
address this problem elsewhere.\par

We also restricted our analysis to the simplest set-ups, namely the
type-I-I-II case and the BPS-BPS-non BPS case.  Results both at strong
coupling and weak coupling based on classical integrability techniques
strongly suggest that our method can be generalised to more general
set-ups, in particular the type-I-I-I case.  This is one of the most
interesting and at the same time challenging configuration where all
the three operators are entangled with each other and the result in
general does not factorise into simpler building blocks.  At weak
coupling, the semiclassical limit was obtained very recently in
\cite{Kazama:2016cfl} by using similar methods as those used at strong
coupling.  Reproducing this result directly from the sum over
partitions, both at tree-level and at finite coupling, remains an
important open question.

Another direction of generalisation is to consider three-point
functions with operators in other sectors, especially those higher
rank sectors.  The spectral problem has an elegant solution in the
semiclassical limit in terms of algebraic curves.  Our result in the
semiclassical limit shows that the structure constants in this limit
also takes a compact form and the dependence of three operators are
only through the quasi-momenta.  It is tempting to conjecture that
this compact form survives for higher rank sectors.  If this were the
case, we will have an equally elegant description of semiclassical
three-point functions in terms of the spectral curves.\par

It is also important to take into account the subleading corrections
in a systematic manner.  In our approach, the computation of the
subleading corrections is straightforward and is given by the Mayer
expansion of the gas of clusters.  In the fermionic formalism, the
Mayer expansion becomes the standard semiclassical expansion for the
bosonised system.  It would be interesting to compare our formalism
with the Fermi gas approach developed in \cite{M-P-ABJM}, where the
subleading corrections can be worked out by the Wigner-Kirkwood
expansion.  The efficiency of the two methods can be compared on
models which exhibit the clustering phenomenon, but whose subleading
corrections (at all order) are nonetheless completely determined by
the Wigner-Kirkwood expansion.  An example of of such a model is
considered in Appendix \ref{ap:abjm}.

 We would like to emphasise that we consider the methods developed
 here as more important than the final results, which just confirm
 what has been widely expected.  First, in our formalism the
 asymptotic part and the wrapping corrections look very similarly and
 are evaluated by the same method.  Second, we were successful in
 computing the sum over the mirror particles due to the appearance of
 the ``quantum spectral curve" as finite difference operator whose
 classical limit is the classical spectral curve.  The contribution of
 the mirror particles is given by the determinant of this operator in
 the functional space.  The fact that the final result depends only on
 the eigenvalues of the monodromy matrix could give us a hint how to
 generalise the approach to the whole symmetry group, shortcutting the
 intimidating sums that appear in the hexagon proposal.

Finally, it is worth mentioning that similar integral representation
shows up in the study of stochastic processes.  In particular, the
integral appears in the resolution of identity.  This may provide us
with new insights in understanding the origin of integral
representation of structure constants in the present work.

\section*{Acknowledgements} The authors are highly indebted to
Benjamin Basso for numerous insightful discussions.  D.S. would like
to thank Andrei Belitsky for discussions concerning the strong
coupling limit of the amplitudes, and to Alexei Borodin, Ivan Corwin,
Jan de Gier, Alexander Povolotsky and Tomohiro Sasamoto for
discussions concerning role of integral representations in the context
of stochastic processes.  Y.J. would like to thank Zoltan Bajnok and
Alessandro Sfondrini for discussions on the analytic structure of the
dressing phase.  Research at the Perimeter Institute is supported by
the Government of Canada through NSERC and by the Province of Ontario
through MRI. The research of Y.J. is partially supported by the Swiss
National Science Foundation through the NCCR SwissMap.  The research
of I.K. and D.S. leading to these results has received funding from
the People Programme (Marie Curie Actions) of the European Union's
Seventh Framework Programme FP7/2007-2013/ under REA Grant Agreement
No 317089.  I.K. gratefully acknowledges support from the Simons
Center for Geometry and Physics, Stony Brook University.  D.S. thanks
Arizona State University and KITP Santa Barbara for warm hospitality
and acknowledges support by the National Science Foundation under
Grant No.  NSF PHY11-25915.

\appendix
\section{An example for clustering}
\label{sec:n3}
In this appendix, we give an explicit example of clustering.  Consider
the case $n=3$, we have
\begin{align}
\mathrm{I}_3=\oint_{\mathcal{C}_{\mathbf{u}}}\frac{dz_1 dz_2
dz_3}{(2\pi \ep)^3} \ff(z_1)\ff(z_2)\ff(z_3)
\Delta(z_1,z_2)\Delta(z_1,z_3)\Delta(z_2,z_3)
\end{align}
1.  Deform contour of $z_3$, we obtain
\begin{align}
\label{eq:x3}
\mathrm{I}_3=&\,\oint_{\mathcal{C}_3}\frac{dz_3}{2\pi\ep}\oint_{\mathcal{C}_{\mathbf{u}}}\frac{dz_1
dz_2}{(2\pi\ep)^2}\ff(z_1)\ff(z_2)\ff(z_3)
\Delta(z_1,z_2)\Delta(z_1,z_3)\Delta(z_2,z_3)\\\nonumber
&\,+\frac{1}{2}\oint_{\mathcal{C}_{\mathbf{u}}}\frac{dz_1
dz_2}{(2\pi\ep)^2} \ff(z_1)F_2(z_2)\Delta_{12}(z_1,z_2)\\\nonumber
&\,+\frac{1}{2}\oint_{\mathcal{C}_{\mathbf{u}}}\frac{dz_1
dz_2}{(2\pi\ep)^2} F_2(z_1)\ff(z_2)\Delta_{21}(z_1,z_2).
\end{align}
2.  Deform contour of $x_2$ in (\ref{eq:x3}), we obtain
\begin{align}
\label{eq:x2}
\mathrm{I}_3=&\,\oint_{\mathcal{C}_3}\frac{dz_3}{2\pi\ep}\oint_{\mathcal{C}_2}\frac{dz_2}{2\pi\ep}\oint_{\mathcal{C}_{\mathbf{u}}}\frac{dz_1}{2\pi\ep}\ff(z_1)\ff(z_2)\ff(z_3)
\Delta(z_1,z_2)\Delta(z_1,z_3)\Delta(z_2,z_3)\\\nonumber
&\,+\frac{1}{2}\oint_{\mathcal{C}_3}\frac{dz_3}{2\pi\ep}\oint_{\mathcal{C}_{\mathbf{u}}}\frac{dz_1}{2\pi}F_2(z_1)\ff(z_3)\Delta_{21}(z_1,z_3)\\\nonumber
&\,+\frac{1}{2}\oint_{\mathcal{C}_2}\frac{dz_2}{2\pi\ep}\oint_{\mathcal{C}_{\mathbf{u}}}\frac{dz_1}{2\pi}\ff(z_1)F_2(z_2)\Delta_{12}(z_1,z_2)\\\nonumber
&\,+\frac{1}{2}\oint_{\mathcal{C}_2}\frac{dz_2}{2\pi\ep}\oint_{\mathcal{C}_{\mathbf{u}}}\frac{dz_1}{2\pi
i}F_2(z_1)\ff(z_2)\Delta_{21}(z_1,z_2)\\\nonumber
&\,+2\times\frac{1}{3}\oint_{\mathcal{C}_{\mathbf{u}}}\frac{dz_1}{2\pi\ep}F_3(z_1)
\end{align}
The three terms containing a double integral are equal, so that we
therefore obtain the multiplicities in figure \ref{fig:I3}.  The
procedure can be generalised to any number of integrals.
\section{ Scaling limit of the $\sltwo$ and $\sutwo$ scalar factors
\label{app:hslsu}}
Here we give the crossing transforms of the scalar factor $h_\sltwo =
h_{DD}$ and $\hh_{\sutwo}= h_{YY} $ and their semiclassical limit.  We
recall first the definitions of the various objects, together with
their properties under the crossing transformation,
\begin{align}
\label{hsltwo}
h_{\mathfrak{sl}(2)}(u^{\phantom{2\gamma}},v)= &\,\frac{x^- - y^-}{x^-
- y^+}\frac{1-{1\over x^- y^+}}{1-{1\over x^+
y^+}}\frac{1}{\sigma(u,v)},\\\nonumber
h_{\mathfrak{sl}(2)}(u^{{2\gamma}},v)=&\, {1- {1\over x^+y^-}\over 1-
{1\over x^- y^+}}\ \,\sigma(u,v) = {\s(u,v)\over A(u,v)},
\end{align}
%
\begin{align}
h_{\mathfrak{su}(2)}(u^{\phantom{2\gamma}},v)=&\,\frac{x^- - y^-}{x^+
- y^-}\frac{1-{1\over x^- y^+}}{1-{1\over x^+
y^+}}\frac{1}{\sigma(u,v)},\\
h_{\mathfrak{su}(2)}(u^{{2\gamma}},v)=&\,\frac{y^+}{y^-}\,\sigma(u,v),\
\ \bb_{\sutwo} = e^{-i p(v)} h_{\mathfrak{su}(2)}(u^{{2\gamma}},v) =
\s(u,v).  \no
\end{align}
The other transformation properties used are
\begin{align}
h_{\mathfrak{sl}(2)}(u^{4\gamma},v)&=1/h_{\mathfrak{sl}(2)}(v,u) \\
\no h_{\mathfrak{sl}(2)}(u^{2\gamma},v^{2\gamma})&=
h_{\mathfrak{sl}(2)}(u,v),
\\ \no
h_{\mathfrak{su}(2)}(u^{4\gamma},v)&=1/h_{\mathfrak{su}(2)}(v,u) \\
\no h_{\mathfrak{su}(2)}(u^{2\gamma},v^{2\gamma})&=e^{ip(u)-ip(v)}
h_{\mathfrak{su}(2)}(u,v)\,.
\end{align}
 The semiclassical limit $\ep\to 0$ of the various quantities is
    \begin{align}\label{limithA}
  \begin{aligned}
 \log \hh_\sutwo(u,v) & \to\ -\frac{i \left(x^2 y^2-x^2+x
 y-y^2\right)}{g \left(x^2-1\right) \left(y^2-1\right) (x-y)} \\ &= -
 { i\epsilon\, y'\over x-y}+ {ip(x)\over y^2-1} = \ \
 {i\epsilon\,x'\over y-x}- \, {ip(y)\over x^2-1} ,\\
 \log \bb_\sutwo (u,v)\ \ &= \log\s(u,v) \\
  & \ \to\ \ \log \hh_\sutwo(1/x, y) - i\, p(y) =\log \hh_\sutwo(x,
  1/y) +i p(x) \\
  &= - {i\ep y' \over 1/x- y}- {ip(x)\over y^2-1}- i\, p(y) = \frac{i
  \epsilon x'}{{1/y}-x}+\frac{i p(y)}{x^2-1}+i p(x)
 \end{aligned}
    \end{align}

  \begin{align}\label{limithA1}
  \begin{aligned}
 \log \hh_\sltwo(u,v) & \to\ \frac{i x y (x y-1)}{g \left(x^2-1\right)
 \left(y^2-1\right) (x-y)}
\\
& = { i\ep y' \over x-y} - {ip(y)\over x^2-1} = { i\ep x' \over x-y} +
{ip(x)\over y^2-1} ,\\
 \log \bb_\sltwo (u,v) & \ \to\ \ \log \hh_\sltwo (1/x, y) \\
   &= {i\ep y'\over 1/x- y} - {ip(y)\over 1/x^2-1} = \frac{i \epsilon
   x'}{x-1/y}-\frac{i p(x)}{1- 1/y^2} .
 \end{aligned}
    \end{align}

%
%

\section{Computing the phase factor for the asymptotic
$C_{123}^{\bullet\bullet\bullet}$}
\label{app:phases}
In this appendix, we compute the phase factor denoted by
$\texttt{phase}$ in (\ref{eq:sop112}) in the main text and show
explicitly how to derive (\ref{C123beh}) from (\ref{eq:sop112}).  This
phase factor comes from changing between spin chain and string frames
when we perform crossing transformations.  The phase factor is trivial
for $\mathfrak{sl}(2)$ case but is non-trivial for the
$\mathfrak{su}(2)$ case.  In what follows, it is convenient to write
the hexagon form factor in the form
\begin{align}
\label{eq:hFF}
\rH({ \a_1 }|{ \a_3 }|{ \a_2 })=\langle\mathfrak{h}|{ \a_1 }\rangle|{
\a_2 }\rangle|{ \a_3 }\rangle.
\end{align}
The excitations in the two frames are related by
\begin{align}
\label{eq:frameex}
\mathcal{D}_{\text{string}}=\mathcal{D}_{\text{spin}},\qquad
\Phi_{\text{string}}=\sqrt{Z}\,\Phi_{\text{spin}}\,\sqrt{Z}.
\end{align}
We see that the derivative excitation is the same in both frames which
is the reason that the phase factor is trivial.  A state with $n$
scalar excitations in the two frames are related by
\begin{align}
\label{eq:frames}
|\Phi_1\cdots\Phi_n\rangle_{\text{string}}=&\,|\sqrt{Z}\Phi_1\,Z\,\Phi_2\,Z\cdots
Z\,\Phi_n\,\sqrt{Z}\rangle_{\text{spin}}\\\nonumber
=&\,F_n|Z^n\Phi_1\cdots\Phi_n\rangle_{\text{spin}}\,,
\end{align}
where in the second line we have moved all the $Z$-markers to the
leftmost by the rule
\begin{align}
\label{eq:moveZ}
|\cdots\Phi_k\,Z^a\cdots\rangle=e^{ip_k a}|\cdots
Z^a\,\Phi_k\cdots\rangle\,.
\end{align}
It is straightforward to derive that
\begin{align}
\label{eq:Fn}
F_n=\prod_{j=1}^n\frac{e^{ip_j/2}\,\zeta_j}{\zeta},\qquad
\zeta_k=e^{ip_k}\zeta_{k-1},\quad \zeta_1=\zeta.
\end{align}
We can also write (\ref{eq:frames}) as
\begin{align}
\label{eq:frames2}
|\Phi_1\cdots\Phi_n\rangle_{\text{spin}}=F_n^{-1}|Z^{-n}\Phi_1\cdots\Phi_n\rangle_{\text{string}}.
\end{align}
Substituting (\ref{eq:frames2}) into the hexagon form factor
(\ref{eq:hFF})
\begin{align}
\langle\mathfrak{h}|{ \a_1 }\rangle|{ \a_2 }\rangle|{ \a_3
}\rangle_{\text{spin}}= (F_{ \a_1 }F_{ \a_2 }F_{ \a_3
})^{-1}\langle\mathfrak{h}|Z^{-|{ \a_1 }|}{ \a_1 }\rangle|Z^{-|{ \a_2
}|}{ \a_2 }\rangle|Z^{-|{ \a_3 }|}{ \a_3 }\rangle_{\text{string}}\,.
\end{align}
The next step is to move all the excitations on the same edge and then
pull out the $Z$ makers by the rule
\begin{align}
\label{eq:removeZ}
\langle\mathfrak{h}|Z^n\psi\rangle=z^n\langle\mathfrak{h}|\psi\rangle,\qquad
z=e^{-iP/2}\,,
\end{align}
where $P$ is the total momentum of the state $|\psi\rangle$.  This
leads to
\begin{align}
\label{eq:p1}
\langle\mathfrak{h}|{ \a_1 }\rangle|{ \a_2 }\rangle|{ \a_3
}\rangle_{\text{spin}} =&\,e^{\frac{i}{2}[p( \a_1 )(|{ \a_1 }|+|{ \a_3
}|-|{ \a_2 }|)+p(\a_2 )(|{ \a_2 }|+|{ \a_1 }|-|{ \a_3 }|)+p(\a_3 )(|{
\a_3 }|+|{ \a_2 }|-|{ \a_1 }|)]}\\\nonumber &\,\times(F_{ \a_1 } F_{
\a_2 } F_{ \a_3 })^{-1}\langle\mathfrak{h}|{ \a_1 }\rangle|{ \a_2
}\rangle|{ \a_3 }\rangle_{\text{string}}\,,
\end{align}
where $|{ \a_1 }|$, $|{ \a_2 }|$, $|\alpha_3|$ denote the cardinality
and $p({ \a_1 })$, $p({ \a_2 })$, $p({ \a_3 })$ denote the total
momenta of the sets ${ \a_1 }$, ${ \a_2 }$ and ${ \a_3 }$
respectively.  Now that we are in the string frame, we can perform the
crossing transformation and obtain the fundamental hexagon form factor
where all excitations are on the same edge
\begin{align}
\label{eq:p2}
\langle\mathfrak{h}|{ \a_1 }\rangle|{ \a_2 }\rangle|{ \a_3
}\rangle_{\text{string}}= (-1)^{|{ \a_3
}|}\langle\mathfrak{h}|\a_1^{4\gamma},\a_3^{2\gamma},\a_2\rangle_{\text{string}}.
\end{align}
The factor $(-1)^{|{ \a_3 }|}$ appears due to crossing transformation
$(\Phi^{a\dot{b}})^{2\gamma}\to -\Phi^{b\dot{a}}$.\par

In order to compute the fundamental hexagon form factor, we need to
change back to the spin chain frame.  We apply the rules
(\ref{eq:frameex}) and (\ref{eq:moveZ}) again
\begin{align}
\label{eq:p3}
\langle\mathfrak{h}|\a_1^{4\gamma},\a_3^{2\gamma},\a_2\rangle_{\text{string}}
=&\,F_{ \a_1 } F_{ \a_2 } F^{-1}_{ \a_3 }\langle\mathfrak{h}|Z^{|{
\a_1 }|}\a_1^{4\gamma},Z^{|{ \a_3 }|}\a_3^{2\gamma},Z^{|{ \a_2 }|}{
\a_2 }\rangle_{\text{spin}},\\\nonumber =&\,F_{ \a_1 } F_{ \a_2 }
F^{-1}_{ \a_3 }\,e^{i|{ \a_2 }|(-p( \a_3 )+p( \a_1 ))+i|{ \a_3 }|p(
\a_1 )-\frac{i}{2}(|{ \a_1 }|+|{ \a_2 }|+|{ \a_3 }|)(p( \a_1 )-p( \a_3
)+p( \a_2 ))}\\\nonumber
&\,\times\langle\mathfrak{h}|\a_1^{4\gamma},\a_3^{2\gamma},\a_2\rangle_{\text{spin}}\,,
\end{align}
where in the first line, we move the $Z$-markers corresponds to the
excitations ${ \a_1 },{ \a_2 }$ and ${ \a_3 }$ to the leftmost of the
corresponding set, which gives rise to the factors $F_{ \a_1 },F_{
\a_2 }$ and $F_{{ \a_3 }}^{-1}=F_{\a_3^{2\gamma}}$.  In the second
line, we further move the $Z$-markers $Z^{|{ \a_1 }|}$, $Z^{|{ \a_2
}|}$ and $Z^{|{ \a_3 }|}$ to the leftmost and then pull out the
$Z$-markers by the rule (\ref{eq:removeZ}).  Note that $p({ \a_3
})^{2\gamma}=-p( \a_3 )$.

Combining all the phase factors from (\ref{eq:p1}), (\ref{eq:p2}) and
(\ref{eq:p3})
\begin{align}
\langle\mathfrak{h}|{ \a_1 }\rangle|{ \a_2 }\rangle|{ \a_3
}\rangle_{\text{spin}}=(-1)^{|{ \a_3 }|}F_{ \a_3 }^{-2}e^{i|{ \a_3
}|(p( \a_1 )-p( \a_2 )+p(\a_3 ))}
\langle\mathfrak{h}|\a_1^{4\gamma},\a_3^{2\gamma},{ \a_2
}\rangle_{\text{spin}}.
\end{align}
Therefore the phase factors in (\ref{eq:phase12}) read
\begin{align}
\label{eq:phase}
\texttt{phase}_1=&\,(-1)^{|{ \a_3 }|}F_{ \a_3 }^{-2}e^{i|{ \a_3
}|(p(\a_1)-p(\a_2)+p(\a_3))},\\\nonumber
\texttt{phase}_2=&\,(-1)^{|\bar{ \a}_3 |} F_{\bar{ \a}_3
}^{-2}e^{i|\bar{ \a}_3 |(p(\bar{\a}_2)-p(\bar{\a}_1)+p(\bar{ \a}_3))}.
\end{align}
The total phase factor
$\texttt{phase}=\texttt{phase}_1\times\texttt{phase}_2$.  This ends
the derivation of the phase factor.

In order to obtain the more compact result (\ref{C123beh}), we need to
combine the phase factor (\ref{eq:phase}) with the other phase factors
which comes from the crossing transformations of the scalar factors of
the $\mathfrak{su}(2)$ sector.  There are two sources of the other
phase factors.  The first source comes from passing from
(\ref{eq:sop112}) to (\ref{eq:Chh}), where we extract the global
factor $(-1)^N h^<(\uu_1 ,\uu_1 )h^<(\uu_2 ,\uu_2 )h^<(\uu_3 ,\uu_3)$
and neglect it systematically.  In order to extract this global
factor, we used the crossing relation
$h_{\mathfrak{su}(2)}(u^{2\gamma},v^{2\gamma})=e^{ip(u)-ip(v)}
h_{\mathfrak{su}(2)}(u,v)$ which leads to
\begin{align}
h_{\mathfrak{su}(2)}^<(\a_3^{2\gamma},\a_3^{2\gamma})=&\,\texttt{phase}_{
\a_3 }\, h_{\mathfrak{su}(2)}^<({ \a_3 },{ \a_3 }),\\\nonumber
h_{\mathfrak{su}(2)}^<(\bar{\a}_3^{2\gamma},\bar{\a}_3^{2\gamma})=&\,\texttt{phase}_{\bar{\a}_3}\,
h_{\mathfrak{su}(2)}^<(\bar{\a}_3,\bar{\a}_3)\,,
\end{align}
with
\begin{align}
\label{eq:firstsource}
\texttt{phase}_{ \a_3 }=\prod_{\substack{j,k\in{ \a_3 }\\
j<k}}e^{ip_j-ip_k},\qquad
\texttt{phase}_{\bar{\a}_3}=\prod_{\substack{j,k\in\bar{\a}_3\\
j<k}}e^{ip_j-ip_k}\,.
\end{align}
Notice that for $\mathfrak{sl}(2)$ sector we have
$h(u^{2\gamma},v^{2\gamma})=h(u,v)$ and the phase factor is again
trivial.

The second source of phase factor comes from rewriting the following
terms in (\ref{eq:sop112})
\begin{align}
\label{eq:secondsource}
&h_{\mathfrak{su}(2)}(\a_1^{4\gamma},\a_3^{2\gamma})h_{\mathfrak{su}(2)}(\a_3^{2\gamma},{
\a_2 }) =\frac{h_{\mathfrak{su}(2)}(\a_3^{2\gamma},{ \a_2
})}{h_{\mathfrak{su}(2)}(\a_3^{2\gamma},{ \a_1 })} =e^{i(p( \a_2 )-p(
\a_1 ))|{ \a_3 }|}\frac{\sigma({ \a_3 } ,{ \a_2 })}{\sigma({ \a_3 } ,{
\a_1 })},\\\nonumber &h_{\mathfrak{su}(2)}(\bar{\a}_2^{4\gamma},\bar{
\a}_3^{2\gamma})h_{\mathfrak{su}(2)}(\bar{\a}_3^{2\gamma},\bar{ \a}_1
)=
\frac{h_{\mathfrak{su}(2)}(\bar{\a}_3^{2\gamma},\bar{\a}_1)}{h_{\mathfrak{su}(2)}
(\bar{\a}_3^{2\gamma},\bar{\a}_2)}=
e^{i(p(\bar{\a}_1)-p(\bar{\a}_2))|\bar{\a}_3|} \frac{\sigma(\bar{\a}_3
,\bar{\a}_1)}{\sigma(\bar{\a}_3 ,\bar{\a}_2)}
\end{align}
where we have used the crossing property
$h_{\mathfrak{su}(2)}(u^{2\gamma},v)=e^{ip(v)}\sigma(u,v)$.  The phase
factors in (\ref{eq:firstsource}) and (\ref{eq:secondsource}) combines
nicely with (\ref{eq:phase})
\begin{align}
\texttt{phase}_1\times\texttt{phase}_2\times\texttt{phase}_{{ \a_3
}}\times\texttt{phase}_{\bar{\a}_3}\times e^{i(p( \a_2 )-p( \a_1 ))|{
\a_3 }|}\times e^{i(p(\bar{\a}_1)-p(\bar{\a}_2))|\bar{\a}_3|}=1\,.
\end{align}
The final result can be written neatly
\begin{align}
\mathcal{C}_{123}^{\bullet\bullet\bullet}= \sum_{\substack{{ \a_i
}\cup\bar{\a}_i=\uu_i }}\prod_{i=1}^3(-1)^{|\bar{\a}_i|}
\frac{e^{ip_{\bar{
\a}_i}l_{i-1,i}}}{h_{\mathfrak{su}(2)}(\alpha_i,\bar{\alpha}_i)}
\times \frac{1}{h_{\mathfrak{su}(2)}({ \a_2 },{ \a_1
})h_{\mathfrak{su}(2)}(\bar{\a}_1,\bar{\a}_2)} \frac{\sigma({ \a_3 }
,{ \a_2 })}{\sigma({ \a_3 } ,{ \a_1 })}\frac{\sigma(\bar{\a}_3
,\bar{\a}_1)}{\sigma(\bar{\a}_3 ,\bar{\a}_1)}
\end{align}
which is (\ref{C123beh}).  For completeness, we also give the result
for the $\mathfrak{sl}(2)$ sector which takes a very similar form
\begin{align}
\mathcal{C}_{123}^{\bullet\bullet\bullet}=&\,\sum_{\substack{{ \a_i
}\cup\bar{\a}_i=\uu_i}}\prod_{i=1}^3(-1)^{|\bar{\a}_i|}
\frac{e^{ip_{\bar{ \a}_i}l_{i-1,i}}}{h(\alpha_i,\bar{\alpha}_i)}
\times \frac{1}{h({ \a_2 },{ \a_1 })h(\bar{\alpha}_1,\bar{\alpha}_2)}
\frac{h(\a_3^{2\gamma},{ \a_2 })}{h(\a_3^{2\gamma},{ \a_1
})}\frac{h(\bar{\a}_3^{2\gamma},\bar{
\a}_1)}{h(\bar{\a}_3^{2\gamma},\bar{ \a}_2)}.
\end{align}

 \section{The analytic structure of the transfer matrices in mirror
 kinematics
 \label{app:nopolesinthephysicalstrip}}

 \def\beq#1{\begin{align}#1\end{align}} \def\fn#1{\footnote{#1}}

The purpose of this appendix is to show that the combination of the
dressing phase with the extra factors from \eqref{tildega},
\begin{align}
\label{sigmaprod}
\sigma_{a1}(u,\uu)
\prod_{k=1}^{a-1}\frac{R^{(-)[a-2k]}}{R^{(+)[a-2k]}} (u)\,,
\end{align}
does not have any cut between the cuts $x^{[\pm a]}$ when continued to
the mirror dynamics in the variable $u$.  To do so, we compute the
jump of the logarithm of the function given above, between
$u=u_0-i\ep(a/2-k)+i0$ and $u=u_0-i\ep(a/2-k)-i0$ on the interval
$u_0\in[-1,1]$\fn{We choose $\ep=1/2g$ for simplicity.}.
$x^{[a-2k]}(u)$ has a second order branch cut and we call $x^{[a-2k]}$
the determination $u_0+i0$ and $1/x^{[a-2k]}$ the determination
$u_0-i0$.  It is clear that the jump of the logarithm of the product
part is given by
 \begin{align}
 \label{discprod}
\log\left(\frac{x^{[a-2k]}-{\bf y}^+}{x^{[a-2k]}-{\bf
y}^-}\frac{1/x^{[a-2k]}-{\bf y}^-}{1/x^{[a-2k]}-{\bf y}^+}\right)\,.
\end{align}
 To compute the jump of the dressing phase we use the DHM
 representation \cite{Dorey:2007xn}.

\parskip 5pt plus 1pt \jot = 1.3ex

  {\bf DHM representation.} The dressing phase has the following
  structure: \beq{ \log \sigma (u,v)= i\left( \chi (x^{+},y^{-})-\chi
  (x^{-},y^{-})-\chi (x^{+},y^{+})+\chi (x^{-},y^+)\right)\period }
  For two bound states of size $a$ and $b$, it is given by \beq{ \log
  \sigma_{ab} (u,v)=i\left( \chi (x^{[a]},y^{[-b]})-\chi
  (x^{[-a]},y^{[-b]})-\chi (x^{[a]},y^{[b]})+\chi
  (x^{[-a]},y^{[b]})\right)\period }
The function $\chi (x,y)$ is given by the following integral
expression first obtained by Dorey, Hofman and Maldacena
\cite{Dorey:2007xn} \beq{\label{DHM} \chi
(x,y)=-i\oint_{|z|=1}\frac{dz}{2\pi i}\oint_{|w|=1}\frac{dw}{2\pi i}
\frac{1}{x-z}\frac{1}{y-w}\log \frac{\Gamma
[1+ig(z+\frac{1}{z}-w-\frac{1}{w})]}{\Gamma
[1-ig(z+\frac{1}{z}-w-\frac{1}{w})]}\period } This expression is valid
in the physical kinematics; namely when $|x|>1$ and $|y|>1$.  If we
want to compute the quantities in the mirror kinematics, we need to
analytically continue this expression, for example to the regime
$|x|<1$.  Under the analytic continuation, the poles of the integrand
cross the unit circle and yield extra contribution.  For instance, the
dressing phase $\sigma_{ab}(u^\gamma ,v)$ in the mirror-physical
kinematics has extra contributions coming from the analytic
continuation of $\chi (x^{[a]},y^{[-b]})-\chi (x^{[a]},y^{[b]})$.  The
other terms are unaffected.

{\bf Analytic continuation.} Let us explicitly perform such analytic
continuation.  More precisely, we analytically continue $x$ in
\eqref{DHM} to $|x|<1$.  In the process, the integral picks up a pole
from $1/(x-z)$.  The expression after the analytic continuation is
\beq{\label{ancon} \chi_{\rm mirror} (x,y)&=\chi_0(x,y)+\chi_{\rm
jump}(x,y) \,, \qquad |x|<1 } where $\chi_0$ and $\chi_{\rm jump}$ are
given by
\beq{
\label{chi0}
\chi_0(x,y)&=-i\oint_{|z|=1}\frac{dz}{2\pi
i}\oint_{|w|=1}\frac{dw}{2\pi i} \frac{1}{x-z}\frac{1}{y-w}\log
\frac{\Gamma [1+ig(z+\frac{1}{z}-w-\frac{1}{w})]}{\Gamma
[1-ig(z+\frac{1}{z}-w-\frac{1}{w})]}\,,\\
\chi_{\rm jump}(x,y) &=-i\oint_{|w|=1}\frac{dw}{2\pi i}
\frac{1}{y-w}\log \frac{\Gamma
[1+ig(x+\frac{1}{x}-w-\frac{1}{w})]}{\Gamma
[1-ig(x+\frac{1}{x}-w-\frac{1}{w})]}\period
\label{chipole}
} $\chi_0$ has no singularity as a function of $x$ except the one at
$|x|=1$.  As we show below, the function $\chi_{\rm jump}(x,y)$
contains an infinite arrays of branch cuts, which show up in the
mirror dynamics of the bound states dressing phase.

 { \bf Analytic structure of $\chi_{\rm jump}$.}
\begin{figure}[t]
\begin{center}
\begin{minipage}{0.4\hsize}
\begin{center}
\vspace{15pt}
\includegraphics[scale=0.4]{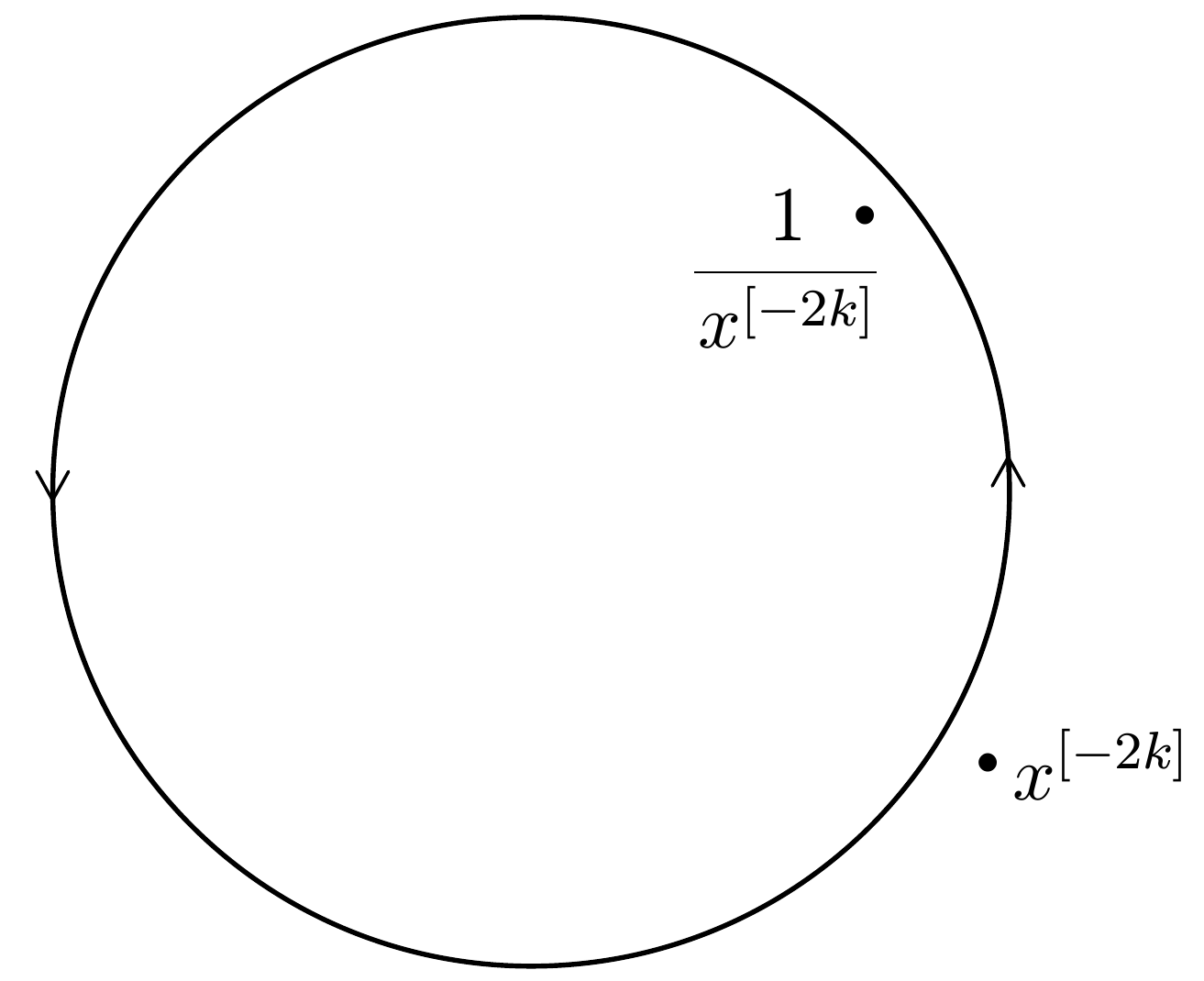}
$\chi^{\uparrow}_{\rm jump}$
\end{center}
\end{minipage}
\begin{minipage}{0.4\hsize}
\begin{center}
\includegraphics[scale=0.4]{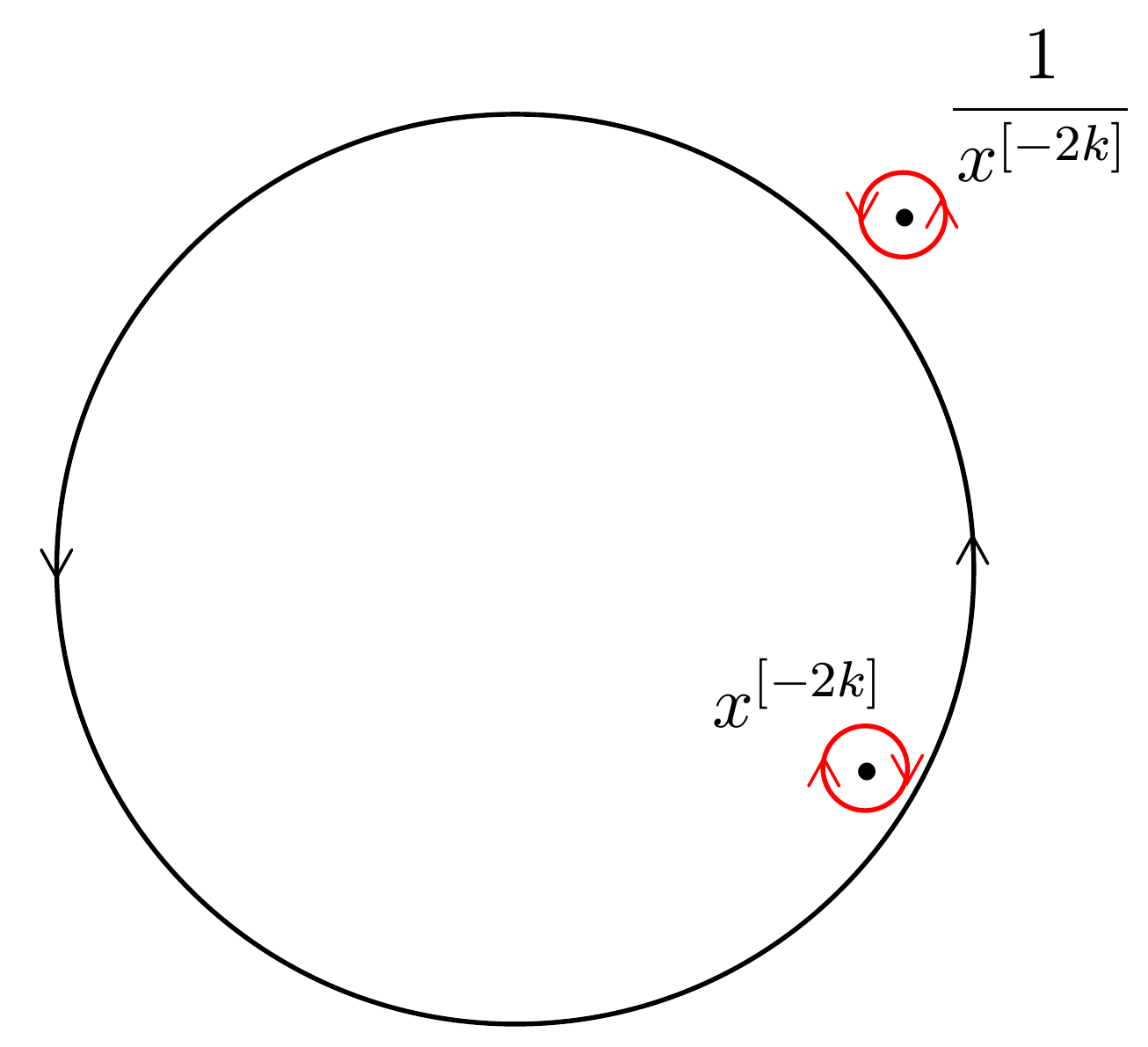}
$\chi^{\downarrow}_{\rm jump}$
\end{center}
\end{minipage}
\end{center}
\caption{The integration contours for $\chi^{\uparrow}_{\rm jump}$ and
$\chi^{\downarrow}_{\rm jump}$ (both denoted in black).  For
$\chi^{\uparrow}_{\rm jump}$, $x^{[-2k]}$ is outside the integration
contour whereas $1/x^{[-2k]}$ is outside.  On the other hand, the
situation is opposite for $\chi^{\downarrow}_{\rm jump}$.  The
discontinuity $\chi^{\uparrow}_{\rm jump}-\chi^{\downarrow}_{\rm
jump}$ is given by the integrations around $x^{[-2k]}$ and
$1/x^{[-2k]}$, which are denoted in red.  \label{fig1}}
\end{figure}
Let us now study the analytic structure of $\chi_{\rm jump}$.  For
this purpose, it is useful to re-express \eqref{chipole} by
integration by parts as \beq{
\label{integparts}
\begin{aligned}\chi_{\rm jump}(x,y)&=-g\oint_{|w|=1}\frac{dw}{2\pi
i}\log (w-y)\left(1-\frac{1}{w^2}\right) \\
&\times\left[\Psi \left(1+ig(x+\frac{1}{x}-w-\frac{1}{w})\right)+\Psi
\left(1-ig(x+\frac{1}{x}-w-\frac{1}{w})\right)\right]\,, \end{aligned}
}
where
\begin{align}
\Psi(z+1)=\frac{d \log \Gamma(z+1)}{dz}=-\sum_{k\geq
1}\left(\frac{1}{z+k}-\frac{1}{k}\right)-\gamma_E\;.
\end{align}
This enables us to write, ignoring the constant in the function
$\Psi(z)$, \beq{
\label{integpartspoles}
\begin{aligned}\chi_{\rm jump}(x,y)&=i\oint_{|w|=1}\frac{dw}{2\pi
i}\log (w-y)\left(1-\frac{1}{w^2}\right) \\
&\times\sum_{k\geq
1}\left[\frac{1}{w+\frac{1}{w}-x^{[-2k]}-\frac{1}{x^{[-2k]}}}-\frac{1}{w+\frac{1}{w}
-x^{[2k]}-\frac{1}{x^{[2k]}}}\right]\\
&=i\oint_{|w|=1}\frac{dw}{2\pi i}\log (w-y)\sum_{k\neq 0}{\rm
sgn}(k)\left[\frac{x^{[-2k]}}{w(w-x^{[-2k]})}+\frac{1}{w-1/x^{[-2k]}}\right]\;.
\end{aligned}
} Obviously, $\chi_{\rm jump}(x,y)$, as a function of $x$, has
discontinuities whenever one of the poles in the sum above hits the
contour of integration $|w|=1$.  This happens whenever $x^{[2k]}$ with
$k$ a non-zero integer hits the unit circle.  These are exactly the
discontinuities we are looking for.  Moreover, we are interested in
the discontinuities of $\chi_{\rm jump}(x^{[a]},y)$ situated between
$u_0+ia\ep/2$ and $u_0-ia\ep/2$.  This means that we restrain
ourselves to the terms with positive $k$, $k=1,\ldots, a-1$ in the
last line of \eqref{integpartspoles}.

Let us now compute the discontinuity $\chi_{\rm jump}^\uparrow -
\chi_{\rm jump}^\downarrow$ when $x^{[-2k]}$ crosses the unit circle
from going from outside to inside (see figure \ref{fig1}).  In the
same time, the pole $1/ x^{[-2k]}$ crosses the unit circle from inside
to outside.  The discontinuity is accounted for by the contribution of
the two poles,
\begin{align}
\label{jumpjump}
&\chi^{\uparrow\downarrow}|_{u+ik\ep}(x,y)\equiv\chi_{\rm
jump}^\uparrow - \chi_{\rm jump}^\downarrow \\ \no &=i\left(\oint_{1/
x^{[-2k]}}-\oint_{x^{[-2k]}}\right)\frac{dw}{2\pi i}\log
(w-y)\left[\frac{x^{[-2k]}}{w(w-x^{[-2k]})}+\frac{1}{w-1/x^{[-2k]}}\right]
=i\log\frac{1/ x^{[-2k]}-y}{x^{[-2k]}-y}\,.
\end{align}
The small subtlety that, after crossing the unit circle, $x^{[-2k]}$
gets exchanged with $1/\bar x^{[-2k]}$ instead of $1/x^{[-2k]}$ is
resolved by noticing that $\chi_{\rm jump}$ only depends on $x+1/x$,
which on the unit circle is the same as its complex conjugate $\bar
x+1/\bar x$.  Combining everything, we obtain the discontinuity of the
mirror dressing factor,
\begin{align}
\label{sigjump}
\log \sigma_{a1}^{\uparrow\downarrow}(x,\yy)|_{u-i(a/2-k)\ep}&
=i(\chi^{\uparrow\downarrow}(x^{[a]},y^-)-\chi^{\uparrow\downarrow}(x^{[a]},y^+))|_{u-i(a/2-k)\ep}\no
\\
&=\log \frac{1/
x^{[a-2k]}-\yy^+}{x^{[a-2k]}-\yy^+}\frac{x^{[a-2k]}-\yy^-}{1/x^{[a-2k]}-\yy^-}
\end{align}
which exactly compensates the one from \eqref{discprod}.  This
completes the proof that the combination in \eqref{sigmaprod} does not
have any discontinuities in the desired region of the mirror regime.

\section{The term $\rB_3$ at strong coupling}
\label{sec:B3strcp}
 In this appendix we compute the terms $\rB_3$
\begin{align}
\label{eq:B3}
\rB_3=&\,\frac{1}{3!}\int\prod_{i=1}^3\frac{dz_i}{2\pi\ep}
\rT_1(z_i)\,\Delta(z_1,z_2)\Delta(z_1,z_3)\Delta(z_2,z_3)\\\nonumber
+&\,\frac{1}{2}\int\frac{dz_1
dz_2}{(2\pi\ep)^2}\rT_1(z_1)\rT_2(z_2)\Delta_{1,2}(z_1,z_2+i\ep/2)+\,
\frac{1}{3}\int\frac{dz}{2\pi\ep}\rT_3(z).
\end{align}
To compute the first term of (\ref{eq:B3}), we can recycle the result
of the asymptotic integral $\rI_3$ in appendix \ref{sec:n3}.  The
combinatorics is the same, with the exception of signs, since in the
present case every fusion of two already existing clusters generates a
minus sign, {\it cf.} (\ref{eq:Deltadelta}) and (\ref{eq:deltanorm}).
This gives
\begin{align}
\label{eq:B31}
&\frac{1}{3!}\left(\int\frac{dz}{2\pi\ep}\rT_1(z)\right)^3
-\frac{1}{4}\left(\int\frac{dz}{2\pi\ep}\rT_1(z)\right)
\left(\int\frac{dz}{2\pi\ep}\rT_1(z)^2\right)+\frac{1}{9}\int\frac{dz}{2\pi\ep}\rT_1(z)^3
\end{align}
For the second term of (\ref{eq:B3}), it is easy to obtain
\begin{align}
\label{eq:B32}
\frac{1}{2}\left(\int\frac{dz}{2\pi\ep}\rT_1(z)\right)
\left(\int\frac{dz}{2\pi\ep}\rT_2(z)\right)-\frac{1}{3}\int\frac{dz}{2\pi\ep}\rT_1(z)\rT_2(z)
\end{align}
Substituting (\ref{eq:B31}) and (\ref{eq:B32}) into (\ref{eq:B3}), and
using (\ref{eq:recursive}) we obtain as announced
\begin{align}
\rB_3=&\frac{1}{3!}\left(\int\frac{dz}{2\pi\ep}\rt_1(z)\right)^3
+\frac{1}{4}\left(\int\frac{dz}{2\pi\ep}\rt_1(z)\right)
\left(\int\frac{dz}{2\pi\ep}\rt_2(z)\right)+\frac{1}{9}\int\frac{dz}{2\pi\ep}\rt_3(z).
\end{align}

\section{The ABJM matrix model and clustering\label{ap:abjm}}
 
In this Appendix, we re-derive the M-theoretic large $N$ limit of the
ABJM matrix model as an extra example of the utility of the clustering
method explained in the main text.

 The partition function of the $U(N) \times U(N)$ ABJM theory on $S^3$
 can be computed by the localisation \cite{KWY}.  The result reads
 \begin{align}
 Z(N) = \frac{1}{N!} \int\prod_i \frac{dx_i}{2\pi k} \frac{1}{2\cosh
 \frac{x_i}{2}} \det\left( \frac{1}{2\cosh
 \frac{x_i-x_j}{2k}}\right)\,,
 \end{align}
 where $k$ is the Chern-Simons level.  To study the M-theory limit
 ($N\to \infty$, $k$ fixed), it is convenient to consider the
 grand-canonical partition function,
 \begin{align}
 \Theta (z) \equiv \sum_{N}z^{N} Z(N) \,,
 \end{align}
 where $z$ is the fugacity.  As explained in \cite{M-P-ABJM}, this can
 be expressed as a Fredholm determinant
 \begin{align}
 \Theta (z) = \det \left(1+ z \hat{\rho} \right)=\exp
 \left(-\sum_{n=1}^{\infty}\frac{(-z)^n}{n}\Tr \left(\hat{\rho}^n
 \right) \right)\,.
 \end{align}
Here the action of the operator $\hat{\rho}$ and the spectral trace
$\Tr \left( \hat{\rho}^n\right)$ are given by
 \begin{align}
 \begin{aligned}
 \hat{\rho}\cdot f (x) &= \int \frac{dy}{2\pi k} \rho (x,y) f(y)\,,\\
 \Tr \left(\hat{\rho}^n \right)&=\int \frac{d^n x}{(2\pi k)^n}\rho
 (x_1,x_2)\cdots \rho (x_{n-1},x_n)\rho(x_n,x_1) \,,
 \end{aligned}
 \end{align}
 with
 \begin{align}
 \rho (x,y) = \frac{1}{2\cosh \frac{x}{2}} \frac{1}{2\cosh
 \frac{x-y}{2k}}\,.
 \end{align}

 To understand the M-theoretic large $N$ limit, we should understand
 the small $k$ expansion of the grand canonical partition function
 $\Theta$.  Just as in the case of the three-point function, the
 kernel $\rho (x,y)$ develops a delta-function singularity in the
 $k\to 0$ limit and the integral exhibits the clustering.  To see
 this, we first express the interaction term in the kernel as a sum
 over rational functions:
 \begin{align}
 \frac{1}{\cosh \frac{x-y}{2k}} =\sum_{a=0}^{\infty} \frac{4k^2 \pi
 (-1)^a (a+1/2)}{(x-y)^2 + 4k^2 \pi^2 (a+1/2)^2}\,.
 \end{align}
 This summation over positive integer is reminiscent of the summation
 over the bound states appearing in the main text.  To make clear the
 relation, we introduce the ``bound-state kernel''
 \begin{align}
 \hat{\rho}_a = \hat{\rho}_{a+} + \hat{\rho}_{a-} \,,
 \end{align}
 with
 \begin{align}
 \begin{aligned}
 \hat{\rho}_{a+}\cdot f &\equiv \int \frac{dy}{4\pi \cosh
 \frac{y}{2}}\frac{i}{(x-y)+2 i k\pi (a+1/2)}f(y)\,,\\
\hat{\rho}_{a-}\cdot f &\equiv \int \frac{dy}{4\pi \cosh
\frac{y}{2}}\frac{-i}{(x-y)-2 i k\pi (a+1/2)}f(y) \,.
 \end{aligned}
 \end{align}
 Then, the grand canonical partition function takes a form similar to
 \eqref{generalF},
 \begin{align}\label{ABJMgrand}
 \ln \Theta =\ln \det \left[1+ z \left( \sum_a (-1)^a \hat{\rho}_a
 \right) \right]=\sum_{n=1}^{\infty}\frac{(-1)^{n-1}}{n}z^n\; \Tr
 \left[\left(\sum_a (-1)^a \hat{\rho}_a\right)^{n}\right]\,.
 \end{align}

 To compute \eqref{ABJMgrand}, we need to evaluate the spectral traces
 \begin{align}
 j_{a_1,\cdots,a_n}\equiv \Tr \left(\hat{\rho}_{a_1}\cdots
 \hat{\rho}_{a_n} \right)\,,
 \end{align}
 which can be further decomposed as
 \begin{align}\label{epsisum}
 j_{a_1,\cdots,a_n} = \sum_{\epsilon_k = \pm}
 \Tr\left(\hat{\rho}_{a_1\epsilon_1}\cdots \hat{\rho}_{a_n\epsilon_n}
 \right)
 \end{align}
 As mentioned above, one needs to take into account the pinching of
 the contour to obtain the correct result.  One important difference
 from the analysis in the main text is that, for most of the choices
 of the signs $\epsilon_k$, in \eqref{epsisum}, there are some
 contours which are not pinched in the limit.  When this happens, the
 contribution coming from such an integral will become subleading.
 The only ases where all the contours are pinched are the ones with
 $\epsilon_k = +$ for all $k $ or $\epsilon_k=-$ for all $k$.  Thus,
 the leading result in the M-theory limit is given by
 \begin{align}
 j_{a_1 ,\cdots, a_n} \sim \Tr \left( \rho_{a_1+}\cdots
 \rho_{a_n+}\right)+\Tr \left( \rho_{a_1-}\cdots \rho_{a_n-}\right)\,.
 \end{align}
  The contributions from these two terms can be determined using the
  clustering method explained in the main text and the result reads
  \begin{align}
  j_{a_1 ,\cdots, a_n} \sim \frac{1}{k\pi \sum_{i=1}^{n}(a_i+1/2)}\int
  \frac{dx}{2\pi } \frac{1}{(2\cosh \frac{x}{2})^n}\,.
  \end{align}
 \begin{align}\label{James1}
 \ln \Theta \sim \sum_{n=1}^{\infty} \frac{(-1)^{n-1}c_n}{k\pi
 n}\int\frac{dx}{2\pi} \left( \frac{z}{2\cosh \frac{x}{2}}\right)^n\,,
 \end{align}
 with the constant $c_n$ given by
 \begin{align}
 c_n = \sum_{a_1=0}^{\infty}\cdots
 \sum_{a_n=0}^{\infty}\frac{(-1)^{\sum_{i=1}^{n}a_i}}{\sum_{i=1}^{n}(a_i+1/2)}\,.
 \end{align}
 To compute this sum, we convert the summand to the following integral
 \begin{align}
 c_n = \sum_{a_1=0}^{\infty}\cdots
 \sum_{a_n=0}^{\infty}(-1)^{\sum_{i=1}^{n}a_i} \int_0^{\infty} dp
 e^{-p \sum_{i=1}^{n}(a_i+1/2)}\,.
 \end{align}
 Then, exchanging the order of the summations and the integration and
 performing the sums explicitly, we arrive at
 \begin{align}\label{James2}
 c_n = \int_0^{\infty} dp \left(\frac{e^{-p/2}}{1+e^{-p}}\right)^n =
 \int^{\infty}_{-\infty} \frac{dp}{2}\frac{1}{(2\cosh
 \frac{p}{2})^n}\,.
 \end{align}
 Now substituting \eqref{James2} into \eqref{James1}, we obtain
 \begin{align}
 \begin{aligned}
 \ln \Theta &\sim \sum_{n=1}^{\infty}\frac{(-1)^{n-1}z^n}{n}\int
 \frac{dxdp}{2\pi \hbar} \left(
 \frac{1}{2\cosh\frac{x}{2}}\frac{1}{2\cosh \frac{p}{2}}\right)\\
 &=\int \frac{dxdp}{2\pi \hbar} \ln \left(1+z e^{-H (p,x)} \right)\,,
 \end{aligned}
 \end{align}
with
\begin{align}
\hbar \equiv 2\pi k \,,\quad H(p,x) \equiv \ln \left(2\cosh
\frac{p}{2} \right) +\log \left( 2\cosh \frac{x}{2}\right)\,.
\end{align}
This is precisely the expression derived in \cite{M-P-ABJM}.

 \section{Separation of variables from clustering\label{ap:SoV}}
In this Appendix, we study the integral representation of the
tree-level three-point function with one non-BPS operator in the
$\mathfrak{su}(2)$ and the $\mathfrak{sl}(2)$ sectors, and show that
they are related to a totally different integral expression, which is
based on the separation of variables (SoV)
\cite{KKN,Korchemsky:sl2sov}.
\subsection*{The $\mathfrak{su}(2)$ sector}
We start from the integral expression\footnote{In this Appendix, we
set $\epsilon=1$.},
\begin{align}\label{sov1}
\mathscr{A}=\sum_{n=0}^{\infty}\frac{\kappa^n}{\,n!}\;
\oint_{\mathcal{C}_{\mathbf{u}}}\;\prod_{j=1}^n\frac{dz_j}{2\pi
}\;f_{\theta}(z_j)\;\prod_{j<k}^n \Delta(z_j,z_k)\,,
\end{align}
where, for later convenience, we introduced the twist $\kappa$ and
deformed the potential term $f(z)$ as
\begin{align}\label{sov2}
f_{\theta}(z)=\frac{1}{h(z,{\bf
u})}\prod_{s=1}^{\ell}\frac{z-\theta_s-i/2}{z-\theta_s+i/2}=
\prod_{i=1}^{M}\frac{z-u_i+i}{z-u_i}\prod_{s=1}^{\ell}\frac{z-\theta_s-i/2}{z-\theta_s+i/2}\,.
\end{align}

As the first step, we rewrite \eqref{sov1} using the Cauchy
determinant identity as
\begin{align}
\mathscr{A}=\sum_{n=0}^{\infty}\frac{\kappa^n}{\,n!}\;
\oint_{\mathcal{C}_{\mathbf{u}}}\;\prod_{j=1}^n\frac{dz_j}{2\pi i
}\;f_{\theta}(z_j)\;\det \left(\frac{1}{z_i-z_j-i} \right)\,.
\end{align}
This expression is similar in form to the Fredholm determinant and
therefore the logarithm of $\mathscr{A}$ admits the expansion:
\begin{align}
\begin{aligned}\label{sov3}
\log \mathscr{A}&=- \sum_{n=1}^{\infty} \frac{(-\kappa)^n}{n} W_n
\,,\\
W_n&=\oint_{\mathcal{C}_{\bf u}} \frac{d^nz}{(2\pi i)^n} \rho
(z_1,z_2)\rho(z_2,z_3)\cdots \rho (z_n,z_1)\,.
\end{aligned}
\end{align}
with
\begin{align}
\rho (u,v)=\frac{f_{\theta}(u)}{u-v-i}\,.
\end{align}
As \eqref{sov3} shows, the integrand of $W_n$ contains poles at $z_j =
\theta_s - i/2$.  The basic idea to make connection with the SoV
representation is to deform the contours and close them around these
poles.  Such deformation yields two additional contributions: The one
comes from poles in the interaction term $\Delta (z_i,z_j)$ and the
other comes from poles at infinity.

Let us first examine the contribution from interaction terms.  Suppose
that we already deformed some of the contours in \eqref{sov3} and are
in a position to deform yet another contour, say the contour of $z_k$.
In the process of deformation, the contour of $z_k$ first hits the
poles at $z_k = z_j +i n$ $(n\in \mathbb{Z})$, where $z_j$ is a
variable whose contour is not yet deformed.  The poles of this type
are precisely the ones considered in section \ref{subsec:deformation}
and, as discussed there, it produces the bound-state contributions.
Now, when the contour of $z_k$ approahes $\theta_s +i/2$, it hits
another set of poles at $z_k = z_m+in$ $(n\in \mathbb{Z})$, where this
time $z_m$ is a variable whose contour is already deformed.
Interestingly, in the case of $\mathfrak{su}(2)$, these two
contributions precisely cancel out\footnote{One can show this by the
combinatorial argument as in section \ref{subsec:deformation}.  Since
the argument is similar, we will not repeat it here.} with each other.
Thus we get
\begin{align}
\log \mathscr{A}=\log \tilde{\mathscr{A}} + \log\mathscr{A}_{\infty}\,,
\end{align}
with
\begin{align}
\begin{aligned}\label{sovatil}
\log \tilde{\mathscr{A}}&=- \sum_{n=1}^{\infty} \frac{\kappa^n}{n}
\tilde{W}_n\\
\tilde{W}_n&=\oint_{\mathcal{C}_{\theta^{-}}} \frac{d^nz}{(2\pi i)^n}
\rho (z_1,z_2)\rho(z_2,z_3)\cdots \rho (z_n,z_1)\,.
\end{aligned}
\end{align}
Here $\mathcal{C}_{\theta^{-}}$ is the contour surrounding
$\theta_s-i/2$ counterclockwise and $\log\mathscr{A}_{\infty}$ denotes
the contribution from infinity.

We next study the contribution from infinity.  To illustrate the basic
mechanism, let us consider $W_2$,
\begin{align}
W_2=\oint_{\mathcal{C}_{\bf u}} \frac{dz_1dz_2}{(2\pi i)^2}
\frac{f_{\theta}(z_1)}{z_1-z_2-i}\frac{f_{\theta}(z_2)}{z_2-z_1-i}.
\end{align}
Since the integrand scales as $1/z_1^2$ when $z_1\sim \infty$, it may
seem that the contribution from infinity is completely absent.
However, this is actually not true: If we first take the pole of
$1/(z_1-z_2-i)$, we arrive at a single integral whose integrand is
given by
\begin{align}
\frac{1}{2i}\oint_{\mathcal{C}_{\bf u}}\frac{dz_2}{2\pi i}
f_{\theta}^{++}(z_2)f_{\theta}(z_2)\period
\end{align}
Then, if we take the residue of this integrand at infinity, we get the
contribution $-(\ell-M)$, where $M$ is the number of magnons.
Repeating such analysis, we can prove the contribution from infinity
for $W_n$ is given by
\begin{align}
-(-1)^n (\ell-M)\,.
\end{align}
Therefore we have
\begin{align}\label{sov4}
\log \mathscr{A}_{\infty}=
(\ell-M)\sum_{n=1}^{\infty}\frac{\kappa^n}{n}=\log
\left((1-\kappa)^{-(\ell-M)} \right)\,.
\end{align}
As \eqref{sov4} shows, $\mathscr{A}_{\infty}$ is divergent in the zero
twist limit $\kappa \to 1$.  As we see later, this divergence is
canceled out by a vanishing factor coming from $\tilde{\mathscr{A}}$.

As the next step, we interpret \eqref{sovatil} as the Fredholm
determinant.  As mentioned before, the form of \eqref{sovatil} is
similar to the expansion of the Fredholm determinant $\det \left(I
+\kappa\hat{G} \right)$.  However, to make the relation more precise,
we need to specify the definition of $\hat{G}$ and the Hilbert space
it acts on\footnote{This was already clarified in
\cite{Bettelheim:Semi} and what we explain below is the minor
modification of their construction.}.  As the Hilbert space, we take a
space of functions with simple poles at $\theta_i^{-}$ $(i=1,\ldots
,\ell)$ and infinity, $\mathcal{H}_{\theta}$.  Obviously it is
$\ell$-dimensional Hilbert space and a general vector belonging to
$\mathcal{H}_{\theta}$ van be expressed as
\begin{align}
F(x)\equiv \sum_j \frac{f_j}{x-\theta_j^{-}}\,.
\end{align}
The dual vector space $\mathcal{H}^{\ast}$ is given by a space of
cycles spanned by $t_j$ $(j=1,\ldots,\ell)$, where $t_j$ is the
contour around $\theta_j^{-}$.  The inner product is then defined by
\begin{align}
\langle \tilde{F}| F\rangle = \oint_{\tilde{F}} \frac{dx}{2\pi i} F(x)
\,.
\end{align}
When $F$ and $\tilde{F}$ are given by
\begin{align}
F(x)= \sum_j \frac{f_j}{x-\theta_j^{-}} \,, \quad \tilde{F}=\sum_k
\tilde{f}_k t_k\,,
\end{align}
the inner product coincides with the standard inner product between
two vectors $(f_1,\ldots,f_{\ell})^t$ and
$(\tilde{f}_1,\ldots,\tilde{f}_{\ell})$.  Now, consider the operator
$\hat{G}$ acting on this function space as
\begin{align}
\hat{G}\cdot F (x)=\oint_{\mathcal{C}_{\theta^{-}}} \frac{dy}{2\pi i}
\frac{f_{\theta}(y)}{x-y} F(y+i)\,.
\end{align}
Then we can check that $\log \det (I+\kappa\hat{G} )= \tr \log
(I+\kappa\hat{G} )$ indeed yields the series \eqref{sovatil}.  Thus in
the end the structure constant can be expressed as
\begin{align}\label{sov5}
\mathscr{A}= \frac{\det (I+\kappa\hat{G} )}{(1-\kappa )^{\ell-M}}\,.
\end{align}

  Now, we relate \eqref{sov5} to the SoV integral formula.  For this
  purpose, we first decompose the operator $I +\kappa\hat{G}$ as
  \begin{align}
  I+\kappa\hat{G} = \hat{K} \hat{L}\,,
  \end{align}
  where $\hat{K}$ and $\hat{L}$ are given by
  \begin{align}
  \begin{aligned}
  \hat{K}\cdot F(x) &= \oint_{\mathcal{C}_{\theta^{-}}}\frac{dy}{2\pi
  i} \frac{e^{-\phi y}}{x-y}\frac{Q_{\theta}^{-}(y)}{Q_{\bf u}(y)}
  F(y)\,,\\
  \hat{L}\cdot F(x)&=\oint_{\mathcal{C}_{\theta^{-}}}\frac{dy}{2\pi i}
  \frac{e^{\phi y}}{x-y}\frac{Q_{\bf u}(y)}{Q_{\theta}^{-}(y)}
  F(y)+\oint_{\mathcal{C}_{\theta^{-}}}\frac{dy}{2\pi } \frac{e^{\phi
  (y+i)}}{x-y}\frac{Q_{\bf u}^{++}(y)}{Q_{\theta}^{+}(y)}F(y+i)
  \end{aligned}
  \end{align}
  Here $Q_{\bf u}$ and $Q_{\theta}$ are the Baxter polynomials defined
  by
  \begin{align}
  Q_{\bf u}(u)= \prod_{i=1}^{M}(u-u_i)\,,\qquad Q_{\theta}(u)=
  \prod_{s=1}^{\ell}(u-\theta_s)\,,
  \end{align}
  and $\phi$ is related to the twist by $e^{i\phi}=-\kappa$.  Since
  the operator $\hat{K}$ acts diagonally on the basis
  $\{1/(x-\theta_1^{-}),\ldots ,1/(x-\theta_{\ell}^{-}) \}$, its
  determinant can be easily computed as
  \begin{align}
  \det \hat{K}=e^{-\phi \sum_s \theta_s^{-}}\frac{\prod_{s,
  t}(\theta_s-\theta_t-i)}{\prod_{s,i}(\theta_s-u_i -i/2)}
  \end{align}
  To compute the determinant of $\hat{L}$, we first express the matrix
  element of $\hat{L}$ in the basis
  \begin{align}\label{sovsimplebasis}
 \left\{\frac{1}{Q_{\theta}^{+}(x)},\frac{x}{Q_{\theta}^{+}(x)},\ldots,
 \frac{x^{\ell-1}}{Q_{\theta}^{+}(x)}\right\}\,.
  \end{align}
  Then we get
  \begin{align}
  \begin{aligned}
  \left( \hat{L}\right)_{n,m}&=\oint_{\theta_n^{-}}\frac{dx}{2\pi i}
  \frac{e^{\phi x}Q_{\bf
  u}(x)x^{m-1}}{Q_{\theta}^{+}(x)Q_{\theta}^{-}(x)}
  +\oint_{\theta_n^{-}}\frac{dx}{2\pi i} \frac{e^{\phi (x+i)}Q_{\bf
  u}^{++}(x)(x+i)^{m-1}}{Q_{\theta}^{+++}(x)Q_{\theta}^{+}(x)}\\
  &=\oint_{\theta_n^{-}\cup \theta_n^{+}}\frac{dx}{2\pi i}
  \frac{e^{\phi x}Q_{\bf
  u}(x)x^{m-1}}{Q_{\theta}^{+}(x)Q_{\theta}^{-}(x)} \,.
  \end{aligned}
  \end{align}
  Now using the Vandermonde determinant formula, we can express the
  determinant of $\hat{L}$ as
  \begin{align}\label{sovlike1}
  \det \hat{L}= J \;\left(\prod_{n=1}^{\ell}\oint_{\theta_n^{-}\cup
  \theta_n^{+}}\frac{dx_n}{2\pi
  i}\right)\prod_{m=1}^{\ell}\frac{e^{\phi x_m}Q_{\bf
  u}(x_m)}{Q_{\theta}^{+}(x_m)Q_{\theta}^{-}(x_m)} \prod_{i<j}
  (x_j-x_i)\,.
  \end{align}
  where $J$ is the Jacobian for the change of basis, which is given by
  \begin{align}
  J^{-1} = \det \left( \oint_{\theta_n^{-}}\frac{dx}{2\pi
  i}\frac{x^m}{Q_{\theta}^{+}(x)}\right)_{n,m}=\frac{1}{\prod_{i<j}(\theta_j-\theta_i)}\,.
  \end{align}

  The expression \eqref{sovlike1} is already very similar to the SoV
  integral formula.  However, there are two important differences.
  First, in the SoV formula derived in \cite{KKN}, the twist $\phi$ is
  set to zero.  Second, the number of integration variables in the SoV
  formula is $\ell-1$ whereas here we have $\ell$.  Thus, to obtain
  the SoV formula, we need to integrate out one of the variables
  sending $\phi \to 0$.  For this purpose, we first change the
  integration contour of \eqref{sovlike1} to
  $\mathcal{C}_{\theta^{\pm}}$, which surround all the $\theta_s \pm
  i/2$, at the cost of introducing an extra factor in the
  integrand\footnote{See \cite{KKN} for details.}:
  \begin{align}
  \det \hat{L}= \frac{J}{\prod_{i<j}(e^{2\pi \theta_j}-e^{2\pi
  \theta_i})}
  \;\oint_{\mathcal{C}_{\theta^{-}}}\prod_{n=1}^{\ell}\frac{dx_n}{2\pi
  i}\frac{e^{\phi x_n}Q_{\bf
  u}(x_n)}{Q_{\theta}^{+}(x_n)Q_{\theta}^{-}(x_n)} \prod_{i<j}
  (e^{2\pi x_j}-e^{2\pi x_i})(x_j-x_i)
  \end{align}
  Now using the Vandermonde determinant formula again, we can convert
  this multiple integral to the following determinant:
  \begin{align}
  \begin{aligned}
  \det \hat{L}&= \frac{J}{\prod_{i<j}(e^{2\pi \theta_j}-e^{2\pi
  \theta_i})}\det M_{n,m}\,,\\
  M_{n,m}&=\oint_{\mathcal{C}_{\theta^{-}}}\frac{dx}{2\pi
  i}\frac{e^{\phi x}Q_{\bf
  u}(x)}{Q_{\theta}^{+}(x)Q_{\theta}^{-}(x)}x^{m-1}e^{2\pi (n-1) x}\,.
  \end{aligned}
  \end{align}
 We then perform the integral of $M_{1,m}$ in the regime $\phi \sim 0$
 to get
 \begin{align}
 M_{1,m}= (-\phi)^{2\ell -M -m} +\cdots\,.
 \end{align}
 Thus the leading contribution comes from $M_{1,\ell}$ and this
 precisely cancels the divergent factor $\mathscr{A}_{\infty}$ given
 by \eqref{sov4}.  As a result, the remaining contribution is given by
 the subdeterminant where the first row and the $\ell$-th column are
 omitted.  Converting this sub-determinant back to the integral, we
 arrive at the following expression for the structure constant,
 \begin{align}
 \begin{aligned}
 \mathscr{A}&=
 C\oint_{\mathcal{C}_{\theta^{\pm}}}\prod_{n}\frac{dx_n}{2\pi
 i}\frac{Q_{\bf u}(x_n)e^{2\pi
 x_n}}{Q_{\theta}^{+}(x_n)Q_{\theta}^{-}(x_n)}\prod_{k<l}(x_k-x_l)(e^{2\pi
 x_k}-e^{2\pi x_l})\,,\\
 C&=e^{-\phi \sum_s \theta_s^{-}}\frac{\prod_{s,
 t}(\theta_s-\theta_t-i)}{\prod_{s,i}(\theta_s-u_i
 -i/2)}\prod_{i<j}\frac{(\theta_i-\theta_j)}{(e^{2\pi
 \theta_i}-e^{2\pi \theta_j})}\,.
 \end{aligned}
 \end{align}
 which coincides with the formula in \cite{KKN}.

 \subsection*{The $\mathfrak{sl}(2)$ sector}
 We next study the $\mathfrak{sl}(2)$ sector,
 \begin{align}
 \mathscr{A}=\sum_{n=0}^{\infty}\frac{(-\kappa)^n}{\,n!}\;
 \oint_{\mathcal{C}_{\mathbf{u}}}\;\prod_{j=1}^n\frac{dz_j}{2\pi
 }\;f^{\mathfrak{sl}(2)}_{\theta}(z_j)\;\prod_{j<k}^n
 \Delta(z_j,z_k)\,,
 \end{align}
 where $f^{\mathfrak{sl}(2)}_{\theta}$ in this case is given by
 \begin{align}
 f^{\mathfrak{sl}(2)}_{\theta}(z)=\frac{1}{h_{\mathfrak{sl}(2)}(z,{\bf
 u})}\prod_{s=1}^{\ell}\frac{z-\theta_s-i/2}{z-\theta_s+i/2}=\prod_{i=1}^{M}
 \frac{z-u_i-i}{z-u_i}\prod_{s=1}^{\ell}\frac{z-\theta_s-i/2}{z-\theta_s+i/2}\,.
 \end{align}

 The basic strategy is the same as in the $\mathfrak{su}(2)$ sector.
 Namely, we write down the expansion for $\log \mathscr{A}$ as in
 \eqref{sov3} and deform the contours.  The important difference is
 that unlike the $\mathfrak{su}(2)$ sector, the contributions from
 poles in the interaction term do not cancel out each other but they
 add up.  By the straightforward (but complicated and tedious)
 computation, we arrive at the following expression:
 \begin{align}
 \begin{aligned}\label{sovsl21}
  \mathscr{A}&=(1-\kappa)^{\ell +M}
  \sum_{\vec{n}}\frac{(-\kappa)^{\sum_a
  n_a}\tilde{B}[\vec{n}]}{\prod_a n_a!}
 \end{aligned}
 \end{align}
 Here the factor $(1-\kappa)^{\ell +M}$ comes from poles at infinity
 as in the $\mathfrak{su}(2)$ case and the contribution from each
 configurartion $\vec{n}$ is given by
 \begin{align}
 \begin{aligned}
 \tilde{B}[\vec{n}]=&\oint_{\mathcal{C}_{\theta^{-}}}\prod_{a}\prod_{j=1}^{n_a}\frac{dz_j^a}{2\pi
 a i}
 f^{\mathfrak{sl}(2)}_{\theta;a}(z_j^{a})\prod_{\substack{a\\a\leq
 i<j\leq n_a}}\Delta_{a,a}(z_i^{a},z_j^{a})\prod_{\substack{a<b\\1\leq
 i\leq n_a\\1\leq j\leq n_b}}\Delta_{a,b}(z_i^{a},z_j^{b})\,,
 \end{aligned}
 \end{align}
 where $\Delta_{a,b}$ are defined by \eqref{eq:delta} and
 $f_{\theta;a}$ is given by
 \begin{align}
 \begin{aligned}
 f^{\mathfrak{sl}(2)}_{\theta;a}(z)&=\frac{Q_{\bf
 u}^{[-2a]}(z)}{Q_{\bf
 u}(z)}\frac{Q_{\theta}^{[1-2a]}(z)}{Q_{\theta}^{+}(z)}\,.
 \end{aligned}
 \end{align}
 Now, as in section \ref{subsec:boundstateFredholm}, we can convert
 this to the following generalised Fredholm determinant\footnote{This
 can be verified by expanding the generalised Fredholm determinant and
 compareing with the series \eqref{sovsl21}.}:
 \begin{align}\label{sovsl2det}
 \mathscr{A}=(1-\kappa)^{\ell +M} \sum_{\vec{n}} \det \left(
 I+\sum_{a=1}^{\infty}(-\kappa)^a\hat{G}_a\right)\,,
 \end{align}
 Here $\hat{G}_a$ is the operator acting on the space $\mathcal{H}$ as
 \begin{align}
 \hat{G}_a \cdot F(x)= \oint_{\mathcal{C}_{\theta}^{-}} \frac{dy}{2\pi
 i}\frac{f^{\mathfrak{sl}(2)}_{\theta;a}(y)}{x-y}F(y-ai)\,.
 \end{align}

 As in the $\mathfrak{su}(2)$ sector, we can decompose the operator $I
 + \sum_a \kappa^a \hat{G}_a$ as
 \begin{align}
 I + \sum_a \kappa^a
 \hat{G}_a=\hat{K}_{\mathfrak{sl}(2)}\hat{L}_{\mathfrak{sl}(2)}\,,
 \end{align}
 where $\hat{K}_{\mathfrak{sl}(2)}$ and $\hat{L}_{\mathfrak{sl}(2)}$
 are given by
 \begin{align}
 \begin{aligned}
 \hat{K}_{\mathfrak{sl}(2)}\cdot
 F(x)&=\oint_{\mathcal{C}_{\theta^{-}}}\frac{dy}{2\pi
 i}\frac{e^{(\phi-\pi) y}}{x-y}\frac{\cosh_ {Q_{\theta}}(y)}{Q_{\bf
 u}(y)Q_{\theta}^{+}(y)}\,,\\
 \hat{L}_{\mathfrak{sl}(2)}\cdot
 F(x)&=\oint_{\mathcal{C}_{\theta^{-}}}\frac{dy}{2\pi i}
 \frac{e^{-(\phi-\pi) y}}{x-y}\frac{Q_{\bf
 u}(y)Q_{\theta}^{+}(y)}{\cosh_ {Q_{\theta}}(y)} F(y)\\&\quad
 +\sum_{a=1}^{\infty} \oint_{\mathcal{C}_{\theta^{-}}}\frac{dy}{2\pi
 i} \frac{e^{-(\phi-\pi) (y-ai)}}{x-y}\frac{Q_{\bf
 u}^{[2a]}(y)Q_{\theta}^{[1-2a]}(y)}{\cosh_{Q_{\theta}}(y)}F(y-ai)
 \end{aligned}
 \end{align}
 where $\cosh_ {Q_{\theta}}(z)$ denotes
 \begin{align}
 \cosh_ {Q_{\theta}}(z) = \prod_{s} \cosh \pi(z-\theta_s)\,.
 \end{align}

\begin{figure}[t]
\begin{center}
\includegraphics[clip, height=4.5cm]{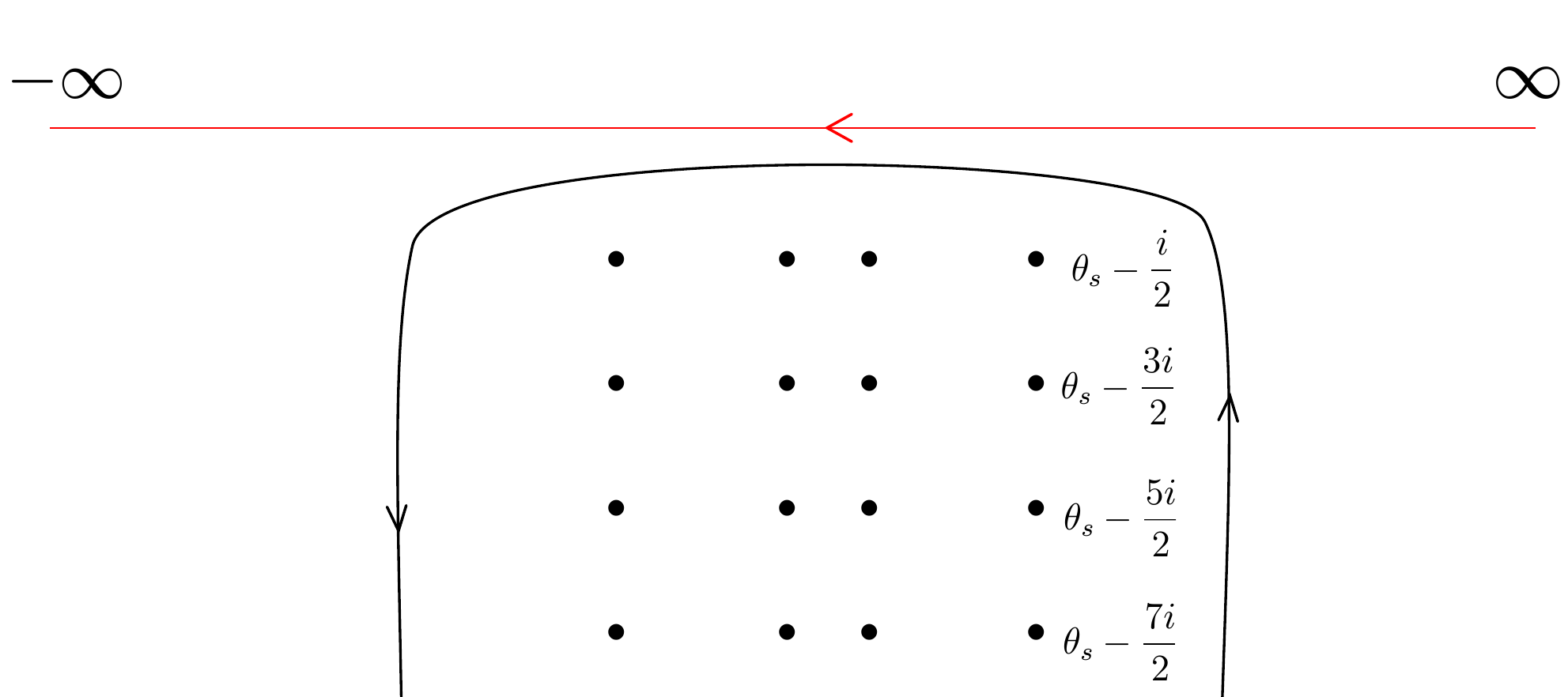}
\end{center}\vspace{-0.5cm}
\caption{The deformation of the contour.  Initially, the contour
encircles counterclockwise $\theta_s-i(a+1/2)$ with $a\geq 0$.  The
contour after the deformation is denoted in red and it runs along the
real axis from $\infty$ to $-\infty$.  $(-1)^{\ell}$ in the formula
comes from the change of the direction of the
contour.}\label{deformcontoursl2}
\end{figure}

 As in the previous case, the determinant of
 $\hat{K}_{\mathfrak{sl}(2)}$ can be computed straightforwardly as
 follows since it acts diagonally on $\mathcal{H}_{\theta}$:
 \begin{align}
 \det \hat{K}_{\mathfrak{sl}(2)} = e^{(\phi-\pi) \sum_s
 \theta_s^{-}}\frac{1}{\prod_{i,s}(\theta_s^{-}-u_i)}\prod_{s\neq
 t}\frac{\cosh \pi(\theta_s-\theta_t)}{(\theta_s-\theta_t)}\,.
 \end{align}
 To compute the determinant of $\hat{L}_{\mathfrak{sl}(2)}$, we again
 use the basis \eqref{sovsimplebasis}.  In this basis, the matrix
 element reads
 \begin{align}
 \begin{aligned}
 \left(\hat{L}_{\mathfrak{sl}(2)}\right)_{n,m}&= \sum_{a=0}^{\infty}
 \oint_{\theta_n^{[-2a-1]}}\frac{dx}{2\pi i}\frac{Q_{\bf
 u}(x)e^{-(\phi-\pi) x}x^{m-1}}{\cosh_{Q_{\theta}}(x)}\,.
 \end{aligned}
 \end{align}
 Converting this to the multiple integral using the Vandermode
 determinant formula and changing the integration contours by
 introducing the factor $\sinh (\pi (x_i- x_j))$, we obtain
\begin{align}
\begin{aligned}
\det \hat{L}_{\mathfrak{sl}(2)}= \frac{J}{\prod_{i<j}\sinh(\pi
(\theta_j-\theta_i))}
\;\oint_{\tilde{\mathcal{C}}_{\theta}}\prod_{n=1}^{\ell}\frac{dx_n}{2\pi
i}\frac{e^{-(\phi-\pi) x_n}Q_{\bf u}(x_n)}{\cosh_{Q_{\theta}}(x_n)}
\prod_{i<j} \sinh (\pi (x_i- x_j))(x_j-x_i)\,,\nonumber
\end{aligned}
    \end{align}
 where $\tilde{C}_{\theta}$ is the contour which encircles all the
 $\theta_s -i (a+1/2)$ with $a>0$.  Now, assuming that $\phi$ has a
 small negative imaginary part, we can deform the contour
 $\tilde{\mathcal{C}}_{\theta}$ to the one along the real axis (see
 figure \ref{deformcontoursl2}).  Then we get
\begin{align}
\begin{aligned}
\det \hat{L}_{\mathfrak{sl}(2)}= \frac{(-1)^{\ell}
J}{\prod_{i<j}\sinh(\pi (\theta_j-\theta_i))}
\;\int_{-\infty}^{\infty}\prod_{n=1}^{\ell}\frac{dx_n}{2\pi
i}\frac{e^{-(\phi-\pi) x_n}Q_{\bf u}(x_n)}{\cosh_{Q_{\theta}}(x_n)}
\prod_{i<j} \sinh (\pi (x_i-x_j))(x_j-x_i)\,, \end{aligned}\nonumber
    \end{align}

To take the limit $\phi\to 0$, we rewrite it as
   \begin{align}
 \begin{aligned}
  \det \hat{L}_{\mathfrak{sl}(2)}&=
  \frac{(-1)^{\ell}J}{\prod_{i<j}(e^{2\pi \theta_j}-e^{2\pi
  \theta_i})}\det M^{\mathfrak{sl}(2)}_{n,m}\,,\\
  M^{\mathfrak{sl}(2)}_{n,m}&=\int_{-\infty}^{\infty}\frac{dx}{2\pi
  i}\frac{e^{-\phi x}Q_{\bf
  u}(x)}{\cosh_{Q_{\theta}}(x)}\frac{x^{m-1}e^{\pi (2n-\ell)
  x}}{2^{\ell-1}}\,.
  \end{aligned}
 \end{align}
 By studying the asymptotic behavior of the integrand, it is easy to
 verify that only when $n=\ell$ does $M_{n,m}^{\mathfrak{sl}(2)}$
 becomes singular in the limit $\phi\to 0$.  Since the divergence
 comes from $x\sim \infty$, we can approximate the integrand as
 \begin{align}
 \begin{aligned}
 M^{\mathfrak{sl}(2)}_{\ell,m}&\sim\int^{\infty}_{0}\frac{dx}{2\pi
 i}\frac{Q_{\bf u}(x)}{\cosh_{Q_{\theta}}(x)}\frac{x^{m-1}e^{(\pi
 \ell-\phi) x}}{2^{\ell-1}}\\
 &\sim \int^{\infty}_{0}\frac{dx}{4\pi i}Q_{\bf u}(x)x^{m-1} e^{-\phi
 x}\\
 &\sim \frac{(M+m-1)!}{4\pi i\phi^{M+m}}
 \end{aligned}
 \end{align}
 Thus, $M_{\ell ,\ell}^{\mathfrak{sl}(2)}$ cancels the prefactor in
 \eqref{sovsl2det}, when $\ell \to 0$.  As a result, the remaining
 contribution is given by the subdeterminant where the $\ell$-th row
 and the $\ell$-th column are omitted.

Converting it back to the integral, we finally arrive at the following
expression,
 \begin{align}
  \mathscr{A}\propto
  \int_{-\infty}^{\infty}\prod_{n=1}^{\ell-1}\frac{dx_n}{2\pi}\frac{Q_{\bf
  u}(x_n)}{\cosh_{Q_{\theta}}(x_n)} \prod_{i<j} \sinh (\pi
  (x_i-x_j))(x_j-x_i)\,.
 \end{align}
 Upon setting $\theta_s=0$, this reduces to the SoV-integral
 expression for the spin $1/2$ $\mathfrak{sl}(2)$ chain obtained in
 \cite{Korchemsky:sl2sov}.

 \bibliographystyle{utphys}
\bibliography{yunfeng}

\end{document}